\pdfoutput=1
\documentclass[11pt,letterpaper]{article}

\usepackage[margin=1in]{geometry}
\usepackage{amsmath,amssymb}
\usepackage{graphicx}
\usepackage{xcolor}
\usepackage{booktabs}
\usepackage{array}
\usepackage[expansion=false]{microtype}
\usepackage[numbers,square,sort&compress]{natbib}
\usepackage{caption}
\usepackage[section]{placeins}

\usepackage{xurl}
\usepackage[hidelinks]{hyperref}
\hypersetup{
  pdftitle    = {A Calculus Gradient: Child Poverty, Mathematics
                 Preparation, and First-Year College Performance at a
                 Hispanic-Serving University},
  pdfsubject  = {Descriptive pre-2020 baseline of student success
                 correlates at a broad-access university},
  pdfkeywords = {gateway mathematics, educational equity, child poverty,
                 record linkage, Hispanic-Serving Institution,
                 reproducible research}
}

\graphicspath{{figures/}}

\IfFileExists{macros/newcommands.tex}{\providecommand{\WarningRunPipeline}{\textcolor{red}{\textbf{[RUN THE PIPELINE]}}}
\newcommand{\MacroTotal}{\WarningRunPipeline}
\newcommand{\MacroModulesWord}{\WarningRunPipeline}
\newcommand{\CalcDistricts}{\WarningRunPipeline}
\newcommand{\CalcDistrictsWithIncome}{\WarningRunPipeline}
\newcommand{\CalcDistrictsZero}{\WarningRunPipeline}
\newcommand{\CalcDistrictsAboveDisplay}{\WarningRunPipeline}
\newcommand{\CalcBexarDistricts}{\WarningRunPipeline}
\newcommand{\CalcMeanPerHundred}{\WarningRunPipeline}
\newcommand{\CalcSlope}{\WarningRunPipeline}
\newcommand{\CalcSlopeSE}{\WarningRunPipeline}
\newcommand{\CalcSeatsPerTenPoints}{\WarningRunPipeline}
\newcommand{\CalcRelDecline}{\WarningRunPipeline}
\newcommand{\CalcRelSigned}{\WarningRunPipeline}
\newcommand{\CalcRelDeclineRounded}{\WarningRunPipeline}
\newcommand{\CalcSlopeUnweighted}{\WarningRunPipeline}
\newcommand{\CalcSlopeUnweightedSE}{\WarningRunPipeline}
\newcommand{\CalcTexasDistricts}{\WarningRunPipeline}
\newcommand{\CalcTexasSlope}{\WarningRunPipeline}
\newcommand{\CalcTexasSlopeSE}{\WarningRunPipeline}
\newcommand{\CalcTexasRelDecline}{\WarningRunPipeline}
\newcommand{\CalcAlamoPovertyPct}{\WarningRunPipeline}
\newcommand{\CalcEdgewoodPovertyPct}{\WarningRunPipeline}
\newcommand{\CalcAlamoPerHundred}{\WarningRunPipeline}
\newcommand{\CalcEdgewoodPerHundred}{\WarningRunPipeline}
\newcommand{\CalcAlamoRatio}{\WarningRunPipeline}
\newcommand{\CalcAlamoRatioTwo}{\WarningRunPipeline}
\newcommand{\CalcEdgewoodRatio}{\WarningRunPipeline}
\newcommand{\CalcLargestTwoRatio}{\WarningRunPipeline}
\newcommand{\CalcSanAntonioRatio}{\WarningRunPipeline}
\newcommand{\CalcIncomeLowNear}{\WarningRunPipeline}
\newcommand{\CalcIncomeHighAbove}{\WarningRunPipeline}
\newcommand{\CalcIncomeCorr}{\WarningRunPipeline}
\newcommand{\CalcIncomeSpearman}{\WarningRunPipeline}
\newcommand{\CalcIncomePovertyCorr}{\WarningRunPipeline}

\newcommand{\DataSaipeVintages}{\WarningRunPipeline}
\newcommand{\DataSaipeDistrictsEver}{\WarningRunPipeline}
\newcommand{\DataSaipeDistrictsEveryVintage}{\WarningRunPipeline}
\newcommand{\DataSaipeDistrictsEveryVintagePct}{\WarningRunPipeline}
\newcommand{\DataCrdcCycles}{\WarningRunPipeline}
\newcommand{\DataCourseFamilies}{\WarningRunPipeline}
\newcommand{\DataCourseFamiliesEarliestCycle}{\WarningRunPipeline}
\newcommand{\DataSaipeRows}{\WarningRunPipeline}
\newcommand{\DataCrdcSchoolCycleRows}{\WarningRunPipeline}
\newcommand{\DataCrdcDistrictCycleRows}{\WarningRunPipeline}
\newcommand{\DataCourseFamilyRows}{\WarningRunPipeline}
\newcommand{\DataEdgeSchoolRows}{\WarningRunPipeline}
\newcommand{\DataAtlasTracts}{\WarningRunPipeline}
\newcommand{\DataAtlasZctas}{\WarningRunPipeline}
\newcommand{\DataZctaTractWeights}{\WarningRunPipeline}
\newcommand{\DataDistrictTractShares}{\WarningRunPipeline}
\newcommand{\DataSourceFiles}{\WarningRunPipeline}
\newcommand{\DataManifestArchivesWord}{\WarningRunPipeline}
\newcommand{\DataAlgTwoZeroPctEarliestCycle}{\WarningRunPipeline}
\newcommand{\DataAlgTwoZeroPctLaterMin}{\WarningRunPipeline}
\newcommand{\DataAlgTwoZeroPctLaterMax}{\WarningRunPipeline}
\newcommand{\DataOfferAgreePctMin}{\WarningRunPipeline}
\newcommand{\DataOfferAgreePctMax}{\WarningRunPipeline}
\newcommand{\DataOfferKappa}{\WarningRunPipeline}
\newcommand{\DataReportingRateExcessPct}{\WarningRunPipeline}
\newcommand{\DataAlgTwoSkipCodeSchoolsRounded}{\WarningRunPipeline}
\newcommand{\DataAlgTwoSchoolsWithCountPct}{\WarningRunPipeline}
\newcommand{\DataAlgTwoSchoolsWithCount}{\WarningRunPipeline}
\newcommand{\DataAlgTwoEnrolleesRebuiltKey}{\WarningRunPipeline}
\newcommand{\DataStoredKeySchools}{\WarningRunPipeline}
\newcommand{\DataStoredKeySchoolsLost}{\WarningRunPipeline}
\newcommand{\DataAlgTwoEnrolleesStoredKey}{\WarningRunPipeline}
\newcommand{\DataStoredKeyEnrollmentLostPct}{\WarningRunPipeline}
\newcommand{\DataSlopeSeStateClustered}{\WarningRunPipeline}
\newcommand{\DataSlopeSeNaive}{\WarningRunPipeline}
\newcommand{\DataSlopeSeRatio}{\WarningRunPipeline}
\newcommand{\DataMinChildren}{\WarningRunPipeline}
\newcommand{\DataAlgTwoDistrictsNoFloor}{\WarningRunPipeline}
\newcommand{\DataAlgTwoPerChildMax}{\WarningRunPipeline}
\newcommand{\DataAlgTwoPerChildMaxFloor}{\WarningRunPipeline}
\newcommand{\DataUnweightedSlopeNoFloor}{\WarningRunPipeline}
\newcommand{\DataUnweightedSlopeFloor}{\WarningRunPipeline}
\newcommand{\DataWeightedSlope}{\WarningRunPipeline}
\newcommand{\DataSlopeGapFloor}{\WarningRunPipeline}
\newcommand{\DataWeightedSlopeLadderFlattest}{\WarningRunPipeline}
\newcommand{\DataWeightedSlopeLadderSteepest}{\WarningRunPipeline}
\newcommand{\DataUnweightedSlopeWithoutMismatch}{\WarningRunPipeline}
\newcommand{\DataMismatchMedianChildren}{\WarningRunPipeline}
\newcommand{\DataNeutralFrameDistricts}{\WarningRunPipeline}
\newcommand{\DataNoAlgTwoDistricts}{\WarningRunPipeline}
\newcommand{\DataNoAlgTwoChildren}{\WarningRunPipeline}
\newcommand{\DataNoCoreDistricts}{\WarningRunPipeline}
\newcommand{\DataNoCoreChildren}{\WarningRunPipeline}
\newcommand{\DataCrdcCyclesWord}{\WarningRunPipeline}
\newcommand{\DataMismatchDistrictsWord}{\WarningRunPipeline}
\newcommand{\DataMismatchDistrictsCapitalWord}{\WarningRunPipeline}
\newcommand{\DataIncompleteDistrictsCapitalWord}{\WarningRunPipeline}
\newcommand{\DataIncompleteDistrictsEarlierWord}{\WarningRunPipeline}
\newcommand{\DataIncompleteDistrictsLaterWord}{\WarningRunPipeline}
\newcommand{\DataCoreCoursesWord}{\WarningRunPipeline}

\newcommand{\GradMinChildren}{\WarningRunPipeline}
\newcommand{\GradBootstrapResamples}{\WarningRunPipeline}
\newcommand{\GradIncompleteDistricts}{\WarningRunPipeline}
\newcommand{\GradDistrictsBeforeExclusion}{\WarningRunPipeline}
\newcommand{\GradBiologyRel}{\WarningRunPipeline}
\newcommand{\GradBiologyRelLow}{\WarningRunPipeline}
\newcommand{\GradBiologyRelHigh}{\WarningRunPipeline}
\newcommand{\GradAlgebraTwoRel}{\WarningRunPipeline}
\newcommand{\GradAlgebraTwoRelLow}{\WarningRunPipeline}
\newcommand{\GradAlgebraTwoRelHigh}{\WarningRunPipeline}
\newcommand{\GradGeometryRel}{\WarningRunPipeline}
\newcommand{\GradGeometryRelLow}{\WarningRunPipeline}
\newcommand{\GradGeometryRelHigh}{\WarningRunPipeline}
\newcommand{\GradAlgebraOneNineTenRel}{\WarningRunPipeline}
\newcommand{\GradAlgebraOneNineTenRelLow}{\WarningRunPipeline}
\newcommand{\GradAlgebraOneNineTenRelHigh}{\WarningRunPipeline}
\newcommand{\GradApMathRelDecline}{\WarningRunPipeline}
\newcommand{\GradApScienceRelDecline}{\WarningRunPipeline}
\newcommand{\GradApAnyRelDecline}{\WarningRunPipeline}
\newcommand{\GradAdvMathRelDecline}{\WarningRunPipeline}
\newcommand{\GradPhysicsRelDecline}{\WarningRunPipeline}
\newcommand{\GradCalculusRelDeclineLow}{\WarningRunPipeline}
\newcommand{\GradCalculusRelDeclineHigh}{\WarningRunPipeline}
\newcommand{\GradEarlyAlgebraRelDecline}{\WarningRunPipeline}
\newcommand{\GradEarlyAlgebraRelDeclineLow}{\WarningRunPipeline}
\newcommand{\GradEarlyAlgebraRelDeclineHigh}{\WarningRunPipeline}
\newcommand{\GradEarlyAlgebraFirstCycleRelDecline}{\WarningRunPipeline}
\newcommand{\GradEarlyAlgebraFirstCycleRelDeclineLow}{\WarningRunPipeline}
\newcommand{\GradEarlyAlgebraFirstCycleRelDeclineHigh}{\WarningRunPipeline}
\newcommand{\GradAdvancedRelDeclineMin}{\WarningRunPipeline}
\newcommand{\GradAdvancedRelDeclineMax}{\WarningRunPipeline}
\newcommand{\GradLowestTenthPovertyMax}{\WarningRunPipeline}
\newcommand{\GradHighestTenthPovertyMin}{\WarningRunPipeline}
\newcommand{\GradCalculusLowestTenthRatio}{\WarningRunPipeline}
\newcommand{\GradCalculusHighestTenthRatio}{\WarningRunPipeline}
\newcommand{\GradBiologyLowestTenthRatio}{\WarningRunPipeline}
\newcommand{\GradBiologyHighestTenthRatio}{\WarningRunPipeline}
\newcommand{\GradBexarMinSharePct}{\WarningRunPipeline}
\newcommand{\GradBexarGroupOnePovertyLow}{\WarningRunPipeline}
\newcommand{\GradBexarGroupOnePovertyHigh}{\WarningRunPipeline}
\newcommand{\GradBexarGroupOneCalculusRatio}{\WarningRunPipeline}
\newcommand{\GradBexarGroupTwoPoverty}{\WarningRunPipeline}
\newcommand{\GradBexarGroupTwoCalculusRatio}{\WarningRunPipeline}
\newcommand{\GradBexarGroupThreePovertyLow}{\WarningRunPipeline}
\newcommand{\GradBexarGroupThreePovertyHigh}{\WarningRunPipeline}
\newcommand{\GradBexarGroupThreeCalculusRatio}{\WarningRunPipeline}
\newcommand{\GradBexarGroupFourPovertyLow}{\WarningRunPipeline}
\newcommand{\GradBexarGroupFourPovertyHigh}{\WarningRunPipeline}
\newcommand{\GradBexarGroupFourCalculusRatio}{\WarningRunPipeline}
\newcommand{\GradBexarGroupFivePovertyLow}{\WarningRunPipeline}
\newcommand{\GradBexarGroupFivePovertyHigh}{\WarningRunPipeline}
\newcommand{\GradBexarGroupFiveCalculusRatio}{\WarningRunPipeline}
\newcommand{\GradBexarCoreRatioMin}{\WarningRunPipeline}
\newcommand{\GradBexarCoreRatioMax}{\WarningRunPipeline}
\newcommand{\GradBexarCalculusRelDecline}{\WarningRunPipeline}
\newcommand{\GradBexarCalculusRelDeclineLow}{\WarningRunPipeline}
\newcommand{\GradBexarCalculusRelDeclineHigh}{\WarningRunPipeline}
\newcommand{\GradBexarApMathRelDecline}{\WarningRunPipeline}
\newcommand{\GradBexarApScienceRelDecline}{\WarningRunPipeline}
\newcommand{\GradBexarAdvMathRel}{\WarningRunPipeline}
\newcommand{\GradBexarAdvMathRelLow}{\WarningRunPipeline}
\newcommand{\GradBexarAdvMathRelHigh}{\WarningRunPipeline}
\newcommand{\GradBexarCalculusRel}{\WarningRunPipeline}
\newcommand{\GradBexarCalculusRelLow}{\WarningRunPipeline}
\newcommand{\GradBexarCalculusRelHigh}{\WarningRunPipeline}
\newcommand{\GradBexarUpperGroupsCalculusRatioMin}{\WarningRunPipeline}
\newcommand{\GradBexarUpperGroupsCalculusRatioMax}{\WarningRunPipeline}
\newcommand{\GradBexarUpperGroupsPovertyAbove}{\WarningRunPipeline}
\newcommand{\GradCalculusFirstCycleRelDecline}{\WarningRunPipeline}
\newcommand{\GradCalculusLastCycleRelDecline}{\WarningRunPipeline}
\newcommand{\GradCoreCyclesRelMin}{\WarningRunPipeline}
\newcommand{\GradCoreCyclesRelMax}{\WarningRunPipeline}
\newcommand{\GradCalculusNeutralFrameRelDecline}{\WarningRunPipeline}
\newcommand{\GradCalculusStateFeRelDecline}{\WarningRunPipeline}
\newcommand{\GradCalculusExactSpanRelDecline}{\WarningRunPipeline}
\newcommand{\GradCalculusExactSpanRelDeclineLow}{\WarningRunPipeline}
\newcommand{\GradCalculusExactSpanRelDeclineHigh}{\WarningRunPipeline}
\newcommand{\GradCalculusChecksRelDeclineMin}{\WarningRunPipeline}
\newcommand{\GradCalculusChecksRelDeclineMax}{\WarningRunPipeline}
\newcommand{\GradCoreNeutralFrameRelMin}{\WarningRunPipeline}
\newcommand{\GradCoreNeutralFrameRelMax}{\WarningRunPipeline}
\newcommand{\GradExactSpanSchools}{\WarningRunPipeline}
\newcommand{\GradTexasApMathRelDecline}{\WarningRunPipeline}
\newcommand{\GradTexasEarlyAlgebraRelDecline}{\WarningRunPipeline}
\newcommand{\GradTexasAlgebraTwoRel}{\WarningRunPipeline}
\newcommand{\GradTexasAlgebraTwoZ}{\WarningRunPipeline}
\newcommand{\GradApAnySeats}{\WarningRunPipeline}
\newcommand{\GradEarlyAlgebraSeats}{\WarningRunPipeline}
\newcommand{\GradAdvMathSeats}{\WarningRunPipeline}
\newcommand{\GradApScienceSeats}{\WarningRunPipeline}
\newcommand{\GradApMathSeats}{\WarningRunPipeline}
\newcommand{\GradCoreSeatsMax}{\WarningRunPipeline}
\newcommand{\GradIncomeCoreCorrMin}{\WarningRunPipeline}
\newcommand{\GradIncomeCoreCorrMax}{\WarningRunPipeline}
\newcommand{\GradIncomeAdvMathCorr}{\WarningRunPipeline}
\newcommand{\GradIncomeApCorr}{\WarningRunPipeline}
\newcommand{\GradHsEnrollmentPerChildRelDecline}{\WarningRunPipeline}
\newcommand{\GradHsEnrollmentPerChildRelDeclineLow}{\WarningRunPipeline}
\newcommand{\GradHsEnrollmentPerChildRelDeclineHigh}{\WarningRunPipeline}
\newcommand{\GradCorePerChildRelDeclineMin}{\WarningRunPipeline}
\newcommand{\GradCorePerChildRelDeclineMax}{\WarningRunPipeline}
\newcommand{\GradAdvancedPerChildRelDeclineMin}{\WarningRunPipeline}
\newcommand{\GradAdvancedPerChildRelDeclineMax}{\WarningRunPipeline}
\newcommand{\GradApAnySeatsLow}{\WarningRunPipeline}
\newcommand{\GradApAnySeatsHigh}{\WarningRunPipeline}
\newcommand{\GradEarlyAlgebraSeatsLow}{\WarningRunPipeline}
\newcommand{\GradEarlyAlgebraSeatsHigh}{\WarningRunPipeline}
\newcommand{\GradAdvancedMathSeatsLow}{\WarningRunPipeline}
\newcommand{\GradAdvancedMathSeatsHigh}{\WarningRunPipeline}
\newcommand{\GradCourseFamiliesWord}{\WarningRunPipeline}
\newcommand{\GradCyclesWord}{\WarningRunPipeline}
\newcommand{\GradIncompleteDistrictsWord}{\WarningRunPipeline}
\newcommand{\GradBexarDistrictsWord}{\WarningRunPipeline}
\newcommand{\GradBexarDistrictsCapitalWord}{\WarningRunPipeline}
\newcommand{\GradBexarGroupsWord}{\WarningRunPipeline}
\newcommand{\GradBexarLowerGroupsWord}{\WarningRunPipeline}
\newcommand{\GradBexarUpperGroupsWord}{\WarningRunPipeline}
\newcommand{\GradFrameMismatchDistrictsWord}{\WarningRunPipeline}

\newcommand{\InstCohortN}{\WarningRunPipeline}
\newcommand{\InstCohortFirstFallN}{\WarningRunPipeline}
\newcommand{\InstCohortSecondFallN}{\WarningRunPipeline}
\newcommand{\InstCohortSpringSummerN}{\WarningRunPipeline}
\newcommand{\InstCohortFallPct}{\WarningRunPipeline}
\newcommand{\InstCapN}{\WarningRunPipeline}
\newcommand{\InstCapSeniorPct}{\WarningRunPipeline}
\newcommand{\InstNoCapN}{\WarningRunPipeline}
\newcommand{\InstNoCapSeniorPct}{\WarningRunPipeline}
\newcommand{\InstNoCapOddsMat}{\WarningRunPipeline}
\newcommand{\InstSeniorPct}{\WarningRunPipeline}
\newcommand{\InstSeniorStrictN}{\WarningRunPipeline}
\newcommand{\InstSeniorStrictPct}{\WarningRunPipeline}
\newcommand{\InstOddsMat}{\WarningRunPipeline}
\newcommand{\InstOddsMatStrict}{\WarningRunPipeline}
\newcommand{\InstMatTakersPct}{\WarningRunPipeline}
\newcommand{\InstOddsWrc}{\WarningRunPipeline}
\newcommand{\InstOddsWrcStrict}{\WarningRunPipeline}
\newcommand{\InstWrcTakersPct}{\WarningRunPipeline}
\newcommand{\InstOddsAis}{\WarningRunPipeline}
\newcommand{\InstOddsAisStrict}{\WarningRunPipeline}
\newcommand{\InstAisTakersPct}{\WarningRunPipeline}
\newcommand{\InstMatNoSeniorPct}{\WarningRunPipeline}
\newcommand{\InstNoMatNoSeniorPct}{\WarningRunPipeline}
\newcommand{\InstGraduateNumberedN}{\WarningRunPipeline}
\newcommand{\InstPreEntryRecordsN}{\WarningRunPipeline}
\newcommand{\InstMissingMajorN}{\WarningRunPipeline}
\newcommand{\InstMajorChangePct}{\WarningRunPipeline}
\newcommand{\InstConfirmedChangePct}{\WarningRunPipeline}
\newcommand{\InstMajorCompleteN}{\WarningRunPipeline}
\newcommand{\InstConfirmedChangeCompletePct}{\WarningRunPipeline}
\newcommand{\InstBoundedConfirmedN}{\WarningRunPipeline}
\newcommand{\InstBoundedConfirmedChangePct}{\WarningRunPipeline}
\newcommand{\InstLaterObservedN}{\WarningRunPipeline}
\newcommand{\InstLaterObservedChangePct}{\WarningRunPipeline}
\newcommand{\InstNotObservedN}{\WarningRunPipeline}
\newcommand{\InstNotObservedChangePct}{\WarningRunPipeline}
\newcommand{\InstIntroBelow}{\WarningRunPipeline}
\newcommand{\InstDfwIntervalLevel}{\WarningRunPipeline}
\newcommand{\InstDfwMatLaterN}{\WarningRunPipeline}
\newcommand{\InstDfwMatLaterWithN}{\WarningRunPipeline}
\newcommand{\InstDfwMatLaterWithoutN}{\WarningRunPipeline}
\newcommand{\InstDfwMatLaterChangedWithPct}{\WarningRunPipeline}
\newcommand{\InstDfwMatLaterChangedWithoutPct}{\WarningRunPipeline}
\newcommand{\InstDfwMatLaterDiff}{\WarningRunPipeline}
\newcommand{\InstDfwMatLaterDiffLo}{\WarningRunPipeline}
\newcommand{\InstDfwMatLaterDiffHi}{\WarningRunPipeline}
\newcommand{\InstDfwMatLaterOdds}{\WarningRunPipeline}
\newcommand{\InstDfwMatLaterOddsLo}{\WarningRunPipeline}
\newcommand{\InstDfwMatLaterOddsHi}{\WarningRunPipeline}
\newcommand{\InstDfwMatAllN}{\WarningRunPipeline}
\newcommand{\InstDfwMatAllChangedWithPct}{\WarningRunPipeline}
\newcommand{\InstDfwMatAllChangedWithoutPct}{\WarningRunPipeline}
\newcommand{\InstDfwMatAllDiff}{\WarningRunPipeline}
\newcommand{\InstDfwMatAllDiffLo}{\WarningRunPipeline}
\newcommand{\InstDfwMatAllDiffHi}{\WarningRunPipeline}
\newcommand{\InstDfwMatAllOdds}{\WarningRunPipeline}
\newcommand{\InstDfwMatAllOddsLo}{\WarningRunPipeline}
\newcommand{\InstDfwMatAllOddsHi}{\WarningRunPipeline}
\newcommand{\InstDfwPhyLaterWithN}{\WarningRunPipeline}
\newcommand{\InstDfwCsLaterDiff}{\WarningRunPipeline}
\newcommand{\InstDfwIsLaterDiff}{\WarningRunPipeline}
\newcommand{\InstDfwPhyLaterDiff}{\WarningRunPipeline}
\newcommand{\InstDfwCheLaterDiff}{\WarningRunPipeline}
\newcommand{\InstDfwCheLaterDiffLo}{\WarningRunPipeline}
\newcommand{\InstDfwCheLaterDiffHi}{\WarningRunPipeline}
\newcommand{\InstDfwCheAllDiff}{\WarningRunPipeline}
\newcommand{\InstDfwCheAllDiffLo}{\WarningRunPipeline}
\newcommand{\InstDfwCheAllDiffHi}{\WarningRunPipeline}
\newcommand{\InstDfwWrcAllDiff}{\WarningRunPipeline}
\newcommand{\InstDfwHisAllDiff}{\WarningRunPipeline}
\newcommand{\InstDfwAisAllDiff}{\WarningRunPipeline}
\newcommand{\InstDfwMatDeclaredN}{\WarningRunPipeline}
\newcommand{\InstDfwMatDeclaredDiff}{\WarningRunPipeline}
\newcommand{\InstDfwMatDeclaredOdds}{\WarningRunPipeline}
\newcommand{\InstDfwMatDeclaredOddsLo}{\WarningRunPipeline}
\newcommand{\InstDfwMatDeclaredOddsHi}{\WarningRunPipeline}
\newcommand{\InstDfwMatBoundedLaterDiff}{\WarningRunPipeline}
\newcommand{\InstDfwMatBoundedLaterDiffLo}{\WarningRunPipeline}
\newcommand{\InstDfwMatBoundedLaterDiffHi}{\WarningRunPipeline}
\newcommand{\InstDfwMatBoundedAllDiff}{\WarningRunPipeline}
\newcommand{\InstDfwMatBoundedAllDiffLo}{\WarningRunPipeline}
\newcommand{\InstDfwMatBoundedAllDiffHi}{\WarningRunPipeline}
\newcommand{\InstHeatMinCell}{\WarningRunPipeline}
\newcommand{\InstHeatTopKept}{\WarningRunPipeline}
\newcommand{\InstHeatChePerMover}{\WarningRunPipeline}
\newcommand{\InstHeatCheMovers}{\WarningRunPipeline}
\newcommand{\InstHeatCheCells}{\WarningRunPipeline}
\newcommand{\InstHeatMatPolPerMover}{\WarningRunPipeline}
\newcommand{\InstHeatBlwEcoBioPerMoverLow}{\WarningRunPipeline}
\newcommand{\InstHeatBlwEcoBioPerMoverHigh}{\WarningRunPipeline}
\newcommand{\InstHeatMatPerMoverBelow}{\WarningRunPipeline}
\newcommand{\InstHeatChePerMoverBelow}{\WarningRunPipeline}
\newcommand{\InstHeatMatTotal}{\WarningRunPipeline}
\newcommand{\InstHeatPolTotal}{\WarningRunPipeline}
\newcommand{\InstHeatMatMovers}{\WarningRunPipeline}
\newcommand{\InstHeatPolMovers}{\WarningRunPipeline}
\newcommand{\InstHeatDonorMajorColumns}{\WarningRunPipeline}
\newcommand{\InstHeatCheTotal}{\WarningRunPipeline}
\newcommand{\InstHeatIsPerMover}{\WarningRunPipeline}
\newcommand{\InstHeatArtPerMover}{\WarningRunPipeline}
\newcommand{\InstHeatAisTotal}{\WarningRunPipeline}
\newcommand{\InstHeatAisMovers}{\WarningRunPipeline}
\newcommand{\InstHeatAisPerMover}{\WarningRunPipeline}
\newcommand{\InstHeatIsTotal}{\WarningRunPipeline}
\newcommand{\InstNaiveOddsMat}{\WarningRunPipeline}
\newcommand{\InstNaiveOddsWrc}{\WarningRunPipeline}
\newcommand{\InstNaiveOddsAis}{\WarningRunPipeline}
\newcommand{\InstChordStudents}{\WarningRunPipeline}
\newcommand{\InstChordMinFlowWord}{\WarningRunPipeline}
\newcommand{\InstChordDrawnFlows}{\WarningRunPipeline}
\newcommand{\InstChordDrawnStudents}{\WarningRunPipeline}
\newcommand{\InstChordDrawnPct}{\WarningRunPipeline}
\newcommand{\InstChordWithheldFlows}{\WarningRunPipeline}
\newcommand{\InstChordWithheldStudents}{\WarningRunPipeline}
\newcommand{\InstChordSectors}{\WarningRunPipeline}
\newcommand{\InstChordAnchorDeg}{\WarningRunPipeline}
\newcommand{\InstFdrQ}{\WarningRunPipeline}
\newcommand{\InstSectionSizeBin}{\WarningRunPipeline}
\newcommand{\InstHpLambda}{\WarningRunPipeline}
\newcommand{\InstHorizonYearsWord}{\WarningRunPipeline}
\newcommand{\InstExactBoundaryCourseNText}{\WarningRunPipeline}
\newcommand{\InstDfwSubjectsWord}{\WarningRunPipeline}
\newcommand{\InstChordOtherMajorsWord}{\WarningRunPipeline}
\newcommand{\InstDisclosureMinCellWord}{\WarningRunPipeline}

\newcommand{\LinkEntrants}{\WarningRunPipeline}
\newcommand{\LinkSample}{\WarningRunPipeline}
\newcommand{\LinkSampleScored}{\WarningRunPipeline}
\newcommand{\LinkSampleGradeCohorts}{\WarningRunPipeline}
\newcommand{\LinkNonApplicantPct}{\WarningRunPipeline}
\newcommand{\LinkSubjects}{\WarningRunPipeline}
\newcommand{\LinkQuantSubjects}{\WarningRunPipeline}
\newcommand{\LinkOtherSubjects}{\WarningRunPipeline}
\newcommand{\LinkSubjectMinimum}{\WarningRunPipeline}
\newcommand{\LinkFdrLevel}{\WarningRunPipeline}
\newcommand{\LinkScoreSD}{\WarningRunPipeline}
\newcommand{\LinkVarDistrictPct}{\WarningRunPipeline}
\newcommand{\LinkVarSchoolPct}{\WarningRunPipeline}
\newcommand{\LinkVarBetweenPct}{\WarningRunPipeline}
\newcommand{\LinkVarWithinPct}{\WarningRunPipeline}
\newcommand{\LinkVarWithinPctRound}{\WarningRunPipeline}
\newcommand{\LinkGapVarBetweenPct}{\WarningRunPipeline}
\newcommand{\LinkGapVarWithinPct}{\WarningRunPipeline}
\newcommand{\LinkVarDistrictReducedPct}{\WarningRunPipeline}
\newcommand{\LinkVarSchoolReducedPct}{\WarningRunPipeline}
\newcommand{\LinkVarSchoolReducedWithSchoolPct}{\WarningRunPipeline}
\newcommand{\LinkIndexPoints}{\WarningRunPipeline}
\newcommand{\LinkIndexPointsSigned}{\WarningRunPipeline}
\newcommand{\LinkPovertyPoints}{\WarningRunPipeline}
\newcommand{\LinkPovertyPointsSigned}{\WarningRunPipeline}
\newcommand{\LinkIncomeRankPoints}{\WarningRunPipeline}
\newcommand{\LinkFirstGenPoints}{\WarningRunPipeline}
\newcommand{\LinkIncomeBetween}{\WarningRunPipeline}
\newcommand{\LinkIncomeWithin}{\WarningRunPipeline}
\newcommand{\LinkIncomeBetweenMinusWithin}{\WarningRunPipeline}
\newcommand{\LinkIncomeBetweenMinusWithinZ}{\WarningRunPipeline}
\newcommand{\LinkHOne}{\WarningRunPipeline}
\newcommand{\LinkHOneLow}{\WarningRunPipeline}
\newcommand{\LinkHOneHigh}{\WarningRunPipeline}
\newcommand{\LinkHOnePoints}{\WarningRunPipeline}
\newcommand{\LinkGapIncomeWithin}{\WarningRunPipeline}
\newcommand{\LinkGapMale}{\WarningRunPipeline}
\newcommand{\LinkGapInternational}{\WarningRunPipeline}
\newcommand{\LinkGapInternationalSE}{\WarningRunPipeline}
\newcommand{\LinkGapAsian}{\WarningRunPipeline}
\newcommand{\LinkIndexEntryLevel}{\WarningRunPipeline}
\newcommand{\LinkMathHeldOutPct}{\WarningRunPipeline}
\newcommand{\LinkGapHeldOutPct}{\WarningRunPipeline}
\newcommand{\LinkNoAidIndexPoints}{\WarningRunPipeline}
\newcommand{\LinkNoAidPovertyPoints}{\WarningRunPipeline}
\newcommand{\LinkNoAidHOne}{\WarningRunPipeline}
\newcommand{\LinkNoAidHOneLow}{\WarningRunPipeline}
\newcommand{\LinkNoAidHOneHigh}{\WarningRunPipeline}
\newcommand{\LinkNoAidPooledMathOther}{\WarningRunPipeline}
\newcommand{\LinkNoAidSubjectsPassing}{\WarningRunPipeline}
\newcommand{\LinkMatMathGrade}{\WarningRunPipeline}
\newcommand{\LinkMatVerbalGrade}{\WarningRunPipeline}
\newcommand{\LinkMatMathDfwPoints}{\WarningRunPipeline}
\newcommand{\LinkMatMathDfwPointsSigned}{\WarningRunPipeline}
\newcommand{\LinkSubjectsPassing}{\WarningRunPipeline}
\newcommand{\LinkOtherSubjectsPassing}{\WarningRunPipeline}
\newcommand{\LinkPassingAttemptsPct}{\WarningRunPipeline}
\newcommand{\LinkPooledMathOther}{\WarningRunPipeline}
\newcommand{\LinkPooledMathOtherTwo}{\WarningRunPipeline}
\newcommand{\LinkPooledMathOtherLow}{\WarningRunPipeline}
\newcommand{\LinkPooledMathOtherHigh}{\WarningRunPipeline}
\newcommand{\LinkPooledMathOtherLowIndep}{\WarningRunPipeline}
\newcommand{\LinkPooledMathOtherHighIndep}{\WarningRunPipeline}
\newcommand{\LinkPooledVerbalOther}{\WarningRunPipeline}
\newcommand{\LinkPooledDiffOther}{\WarningRunPipeline}
\newcommand{\LinkPooledDiffOtherLow}{\WarningRunPipeline}
\newcommand{\LinkPooledDiffOtherHigh}{\WarningRunPipeline}
\newcommand{\LinkPooledMathQuant}{\WarningRunPipeline}
\newcommand{\LinkPooledMathQuantTwo}{\WarningRunPipeline}
\newcommand{\LinkPooledMathQuantLow}{\WarningRunPipeline}
\newcommand{\LinkPooledMathQuantHigh}{\WarningRunPipeline}
\newcommand{\LinkPooledVerbalQuant}{\WarningRunPipeline}
\newcommand{\LinkPooledVerbalOtherTwo}{\WarningRunPipeline}
\newcommand{\LinkPooledVerbalQuantTwo}{\WarningRunPipeline}
\newcommand{\LinkWeightedMathOther}{\WarningRunPipeline}
\newcommand{\LinkWeightedMathQuant}{\WarningRunPipeline}
\newcommand{\LinkWeightedDiffOtherLowIndep}{\WarningRunPipeline}
\newcommand{\LinkWeightedDiffOtherHighIndep}{\WarningRunPipeline}
\newcommand{\LinkWeightedDiffOtherLow}{\WarningRunPipeline}
\newcommand{\LinkWeightedDiffOtherHigh}{\WarningRunPipeline}
\newcommand{\LinkSubjectCorrMean}{\WarningRunPipeline}
\newcommand{\LinkSubjectCorrMax}{\WarningRunPipeline}
\newcommand{\LinkSkillMedian}{\WarningRunPipeline}
\newcommand{\LinkSkillMedianPct}{\WarningRunPipeline}
\newcommand{\LinkSkillMinPct}{\WarningRunPipeline}
\newcommand{\LinkSkillMaxPct}{\WarningRunPipeline}
\newcommand{\LinkSkillWeightedPct}{\WarningRunPipeline}
\newcommand{\LinkPrepAdded}{\WarningRunPipeline}
\newcommand{\LinkPrepAddedLow}{\WarningRunPipeline}
\newcommand{\LinkPrepAddedHigh}{\WarningRunPipeline}
\newcommand{\LinkFamilyAdded}{\WarningRunPipeline}
\newcommand{\LinkDemoAdded}{\WarningRunPipeline}
\newcommand{\LinkPovertyAdded}{\WarningRunPipeline}
\newcommand{\LinkPovertyAddedSigned}{\WarningRunPipeline}
\newcommand{\LinkPovertyAddedLow}{\WarningRunPipeline}
\newcommand{\LinkPovertyAddedHigh}{\WarningRunPipeline}
\newcommand{\LinkPovertyAddedMin}{\WarningRunPipeline}
\newcommand{\LinkPovertyAddedMax}{\WarningRunPipeline}
\newcommand{\LinkPovertyAddedPositive}{\WarningRunPipeline}
\newcommand{\LinkSchoolAdded}{\WarningRunPipeline}
\newcommand{\LinkSchoolAddedLow}{\WarningRunPipeline}
\newcommand{\LinkSchoolAddedHigh}{\WarningRunPipeline}
\newcommand{\LinkSchoolAddedMin}{\WarningRunPipeline}
\newcommand{\LinkSchoolAddedMax}{\WarningRunPipeline}
\newcommand{\LinkSchoolAddedPositive}{\WarningRunPipeline}
\newcommand{\LinkPlanSkillMedianPct}{\WarningRunPipeline}
\newcommand{\LinkPlanPrepAdded}{\WarningRunPipeline}
\newcommand{\LinkPlanPovertyAdded}{\WarningRunPipeline}
\newcommand{\LinkPlanSchoolAdded}{\WarningRunPipeline}
\newcommand{\LinkLaterRidgeMedianPct}{\WarningRunPipeline}
\newcommand{\LinkLaterRidgeMinPct}{\WarningRunPipeline}
\newcommand{\LinkLaterRidgeMaxPct}{\WarningRunPipeline}
\newcommand{\LinkLaterLinearMedianPct}{\WarningRunPipeline}
\newcommand{\LinkLaterPovertyAdded}{\WarningRunPipeline}
\newcommand{\LinkLaterSchoolAdded}{\WarningRunPipeline}
\newcommand{\LinkAucHeldOut}{\WarningRunPipeline}
\newcommand{\LinkAucLater}{\WarningRunPipeline}
\newcommand{\LinkAucHeldOutBaseline}{\WarningRunPipeline}
\newcommand{\LinkAucLaterBaseline}{\WarningRunPipeline}
\newcommand{\LinkForestGainHeldOut}{\WarningRunPipeline}
\newcommand{\LinkForestGainLater}{\WarningRunPipeline}
\newcommand{\LinkHeldOutFallbackPct}{\WarningRunPipeline}
\newcommand{\LinkLaterFallbackPct}{\WarningRunPipeline}
\newcommand{\LinkCapGradeMin}{\WarningRunPipeline}
\newcommand{\LinkCapGradeMax}{\WarningRunPipeline}
\newcommand{\LinkCapSubjects}{\WarningRunPipeline}
\newcommand{\LinkExamCreditSubjects}{\WarningRunPipeline}
\newcommand{\LinkMatDualCredit}{\WarningRunPipeline}
\newcommand{\LinkMatExamCredit}{\WarningRunPipeline}
\newcommand{\LinkFirstGenGradeMin}{\WarningRunPipeline}
\newcommand{\LinkFirstGenGradeMax}{\WarningRunPipeline}
\newcommand{\LinkFirstGenSubjects}{\WarningRunPipeline}
\newcommand{\LinkQuantSubjectsWord}{\WarningRunPipeline}
\newcommand{\LinkQuantSubjectsCapitalWord}{\WarningRunPipeline}
\newcommand{\LinkPlanQuantSubjectsCapitalWord}{\WarningRunPipeline}
\newcommand{\LinkVerbalAboveMathSubjectsWord}{\WarningRunPipeline}
\newcommand{\LinkVerbalAboveMathPassingWord}{\WarningRunPipeline}
\newcommand{\LinkVarBetweenOnePct}{\WarningRunPipeline}
\newcommand{\LinkSubjectsGradeBelowMinimumWord}{\WarningRunPipeline}
\newcommand{\LinkPlanCohortsWord}{\WarningRunPipeline}
\newcommand{\LinkSnapshotMaxLagWord}{\WarningRunPipeline}
\newcommand{\LinkSnapshotMaxLagEarlierWord}{\WarningRunPipeline}
\newcommand{\LinkDualSubstituteMaxLagWord}{\WarningRunPipeline}
\newcommand{\LinkDualSubstituteLinked}{\WarningRunPipeline}
\newcommand{\LinkSensitivityRunsWord}{\WarningRunPipeline}
\newcommand{\LinkSensitivityReachingZeroWord}{\WarningRunPipeline}
\newcommand{\LinkFoldsWord}{\WarningRunPipeline}
\newcommand{\LinkEntryLevelsWord}{\WarningRunPipeline}

\newcommand{\PersistPopulationN}{\WarningRunPipeline}
\newcommand{\PersistLeftN}{\WarningRunPipeline}
\newcommand{\PersistLeftPct}{\WarningRunPipeline}
\newcommand{\PersistCapN}{\WarningRunPipeline}
\newcommand{\PersistCapLeftPct}{\WarningRunPipeline}
\newcommand{\PersistMatN}{\WarningRunPipeline}
\newcommand{\PersistMatDfwN}{\WarningRunPipeline}
\newcommand{\PersistMatDfwPct}{\WarningRunPipeline}
\newcommand{\PersistMatLeftWithPct}{\WarningRunPipeline}
\newcommand{\PersistMatLeftWithoutPct}{\WarningRunPipeline}
\newcommand{\PersistMatDiff}{\WarningRunPipeline}
\newcommand{\PersistMatDiffLo}{\WarningRunPipeline}
\newcommand{\PersistMatDiffHi}{\WarningRunPipeline}
\newcommand{\PersistMatOdds}{\WarningRunPipeline}
\newcommand{\PersistMatFailN}{\WarningRunPipeline}
\newcommand{\PersistMatFailLeftWithPct}{\WarningRunPipeline}
\newcommand{\PersistMatFailLeftWithoutPct}{\WarningRunPipeline}
\newcommand{\PersistMatFailDiff}{\WarningRunPipeline}
\newcommand{\PersistMatFailDiffLo}{\WarningRunPipeline}
\newcommand{\PersistMatFailDiffHi}{\WarningRunPipeline}
\newcommand{\PersistMatWithdrewN}{\WarningRunPipeline}
\newcommand{\PersistMatWithdrewLeftPct}{\WarningRunPipeline}
\newcommand{\PersistMatGoneAfterTermPct}{\WarningRunPipeline}
\newcommand{\PersistMatBothN}{\WarningRunPipeline}
\newcommand{\PersistMatOnlyN}{\WarningRunPipeline}
\newcommand{\PersistMatWithElsewhereLeftWithPct}{\WarningRunPipeline}
\newcommand{\PersistMatWithElsewhereLeftWithoutPct}{\WarningRunPipeline}
\newcommand{\PersistMatWithElsewhereDiff}{\WarningRunPipeline}
\newcommand{\PersistMatWithElsewhereDiffLo}{\WarningRunPipeline}
\newcommand{\PersistMatWithElsewhereDiffHi}{\WarningRunPipeline}
\newcommand{\PersistMatAloneLeftWithPct}{\WarningRunPipeline}
\newcommand{\PersistMatAloneLeftWithoutPct}{\WarningRunPipeline}
\newcommand{\PersistMatAloneDiff}{\WarningRunPipeline}
\newcommand{\PersistMatAloneDiffLo}{\WarningRunPipeline}
\newcommand{\PersistMatAloneDiffHi}{\WarningRunPipeline}
\newcommand{\PersistLeftMatDfwN}{\WarningRunPipeline}
\newcommand{\PersistLeftMatDfwPct}{\WarningRunPipeline}
\newcommand{\PersistStayN}{\WarningRunPipeline}
\newcommand{\PersistStayMatDfwPct}{\WarningRunPipeline}
\newcommand{\PersistSubjects}{\WarningRunPipeline}
\newcommand{\PersistSubjectsWord}{\WarningRunPipeline}
\newcommand{\PersistSecondSubjectDfwN}{\WarningRunPipeline}
\newcommand{\PersistDiffMin}{\WarningRunPipeline}
\newcommand{\PersistDiffMax}{\WarningRunPipeline}
\newcommand{\PersistSubjectsAboveMatWord}{\WarningRunPipeline}
\newcommand{\PersistMinTakers}{\WarningRunPipeline}
\newcommand{\PersistAnyN}{\WarningRunPipeline}
\newcommand{\PersistAnyDfwN}{\WarningRunPipeline}
\newcommand{\PersistAnyDfwPct}{\WarningRunPipeline}
\newcommand{\PersistAnyLeftWithPct}{\WarningRunPipeline}
\newcommand{\PersistAnyLeftWithoutPct}{\WarningRunPipeline}
\newcommand{\PersistAnyDiff}{\WarningRunPipeline}
\newcommand{\PersistAnyDiffLo}{\WarningRunPipeline}
\newcommand{\PersistAnyDiffHi}{\WarningRunPipeline}
\newcommand{\PersistMatWholeCohortLeftWithPct}{\WarningRunPipeline}
\newcommand{\PersistMatWholeCohortLeftWithoutPct}{\WarningRunPipeline}
\newcommand{\PersistMatWholeCohortDiff}{\WarningRunPipeline}
\newcommand{\PersistMatWholeCohortDiffLo}{\WarningRunPipeline}
\newcommand{\PersistMatWholeCohortDiffHi}{\WarningRunPipeline}
\newcommand{\PersistOutcomes}{\WarningRunPipeline}
\newcommand{\PersistFirstAttempts}{\WarningRunPipeline}
\newcommand{\PersistOutcomeRatePct}{\WarningRunPipeline}
\newcommand{\PersistMatOutcomes}{\WarningRunPipeline}
\newcommand{\PersistMatOutcomeSharePct}{\WarningRunPipeline}
\newcommand{\PersistSecondSubjectOutcomeSharePct}{\WarningRunPipeline}
\newcommand{\PersistMatOutcomeShareWithRepeatsPct}{\WarningRunPipeline}
\newcommand{\PersistMatOutcomeRatePct}{\WarningRunPipeline}
\newcommand{\PersistMatFirstAttempts}{\WarningRunPipeline}
\newcommand{\PersistLeftWithDfwN}{\WarningRunPipeline}
\newcommand{\PersistLeftWithDfwPct}{\WarningRunPipeline}
\newcommand{\PersistLeftOutcomes}{\WarningRunPipeline}
\newcommand{\PersistLeftMatOutcomes}{\WarningRunPipeline}
\newcommand{\PersistLeftMatOutcomeSharePct}{\WarningRunPipeline}
\newcommand{\PersistLeftSecondSubjectOutcomeSharePct}{\WarningRunPipeline}
\newcommand{\PersistLeftWithDfwHoldingMatPct}{\WarningRunPipeline}
\newcommand{\PersistStayWithDfwPct}{\WarningRunPipeline}
\newcommand{\PersistStayMatOutcomeSharePct}{\WarningRunPipeline}
\newcommand{\PersistLeftPctWithNone}{\WarningRunPipeline}
\newcommand{\PersistLeftPctWithOne}{\WarningRunPipeline}
\newcommand{\PersistLeftPctWithTwo}{\WarningRunPipeline}
\newcommand{\PersistLeftPctWithThree}{\WarningRunPipeline}
\newcommand{\PersistLeftPctWithFourPlus}{\WarningRunPipeline}
\newcommand{\PersistFourPlusN}{\WarningRunPipeline}

\newcommand{\PopExtractStudents}{\WarningRunPipeline}
\newcommand{\PopExtractRecords}{\WarningRunPipeline}
\newcommand{\PopExtractTableStudents}{\WarningRunPipeline}
\newcommand{\PopActiveStudents}{\WarningRunPipeline}
\newcommand{\PopActiveRecords}{\WarningRunPipeline}
\newcommand{\PopActiveWithNamePct}{\WarningRunPipeline}
\newcommand{\PopActiveWithZipPct}{\WarningRunPipeline}
\newcommand{\PopActiveTexasName}{\WarningRunPipeline}
\newcommand{\PopActiveTexasNamePct}{\WarningRunPipeline}
\newcommand{\PopActiveTableStudents}{\WarningRunPipeline}
\newcommand{\PopTableShortfall}{\WarningRunPipeline}

\newcommand{\ScoreBaselineInstitutions}{\WarningRunPipeline}
\newcommand{\ScoreBaselineUndergraduates}{\WarningRunPipeline}
\newcommand{\ScoreBaselineCompletion}{\WarningRunPipeline}
\newcommand{\ScoreBaselinePellShare}{\WarningRunPipeline}
\newcommand{\ScoreBaselineExpectedCompletion}{\WarningRunPipeline}
\newcommand{\ScoreBaselineResidual}{\WarningRunPipeline}
\newcommand{\ScoreCurrentInstitutions}{\WarningRunPipeline}
\newcommand{\ScoreCurrentUndergraduates}{\WarningRunPipeline}
\newcommand{\ScoreCurrentCompletion}{\WarningRunPipeline}
\newcommand{\ScoreCurrentPellShare}{\WarningRunPipeline}
\newcommand{\ScoreCurrentExpectedCompletion}{\WarningRunPipeline}
\newcommand{\ScoreCurrentResidual}{\WarningRunPipeline}
\newcommand{\ScoreBaselinePointsBelow}{\WarningRunPipeline}
\newcommand{\ScoreCurrentPointsAbove}{\WarningRunPipeline}
\newcommand{\ScoreLowessFraction}{\WarningRunPipeline}
\newcommand{\ScoreVintagesWord}{\WarningRunPipeline}

\newcommand{\SizeCap}{\WarningRunPipeline}
\newcommand{\SizeMinSection}{\WarningRunPipeline}
\newcommand{\SizeEntryMinStudents}{\WarningRunPipeline}
\newcommand{\SizeEntryMinSdLogSize}{\WarningRunPipeline}
\newcommand{\SizeBootstrapDraws}{\WarningRunPipeline}
\newcommand{\SizeFirstAttempts}{\WarningRunPipeline}
\newcommand{\SizeEntrants}{\WarningRunPipeline}
\newcommand{\SizeSubjectsModeled}{\WarningRunPipeline}
\newcommand{\SizeSubjectsConsidered}{\WarningRunPipeline}
\newcommand{\SizeSameStudentInstructors}{\WarningRunPipeline}
\newcommand{\SizeCalculusMedianSize}{\WarningRunPipeline}
\newcommand{\SizeCapCalculusPercentile}{\WarningRunPipeline}
\newcommand{\SizeCalculusSizeBound}{\WarningRunPipeline}
\newcommand{\SizeLargeSectionBound}{\WarningRunPipeline}
\newcommand{\SizeCurveGridMax}{\WarningRunPipeline}
\newcommand{\SizeCurveBin}{\WarningRunPipeline}
\newcommand{\SizeCurveSmoothing}{\WarningRunPipeline}
\newcommand{\SizeMatAllSectionsGrade}{\WarningRunPipeline}
\newcommand{\SizeMatAllSectionsGradeLow}{\WarningRunPipeline}
\newcommand{\SizeMatAllSectionsGradeHigh}{\WarningRunPipeline}
\newcommand{\SizeMatCourseTermGrade}{\WarningRunPipeline}
\newcommand{\SizeMatPrimaryGrade}{\WarningRunPipeline}
\newcommand{\SizeMatPrimaryGradeLow}{\WarningRunPipeline}
\newcommand{\SizeMatPrimaryGradeHigh}{\WarningRunPipeline}
\newcommand{\SizeMatPrimaryGradeP}{\WarningRunPipeline}
\newcommand{\SizeMatPrimaryDfw}{\WarningRunPipeline}
\newcommand{\SizeMatPrimaryDfwLow}{\WarningRunPipeline}
\newcommand{\SizeMatPrimaryDfwHigh}{\WarningRunPipeline}
\newcommand{\SizeMatSameStudentGrade}{\WarningRunPipeline}
\newcommand{\SizeMatBalanceMathScore}{\WarningRunPipeline}
\newcommand{\SizeMatInteractionGrade}{\WarningRunPipeline}
\newcommand{\SizeMatInteractionGradeLow}{\WarningRunPipeline}
\newcommand{\SizeMatInteractionGradeHigh}{\WarningRunPipeline}
\newcommand{\SizeMatInteractionGradeP}{\WarningRunPipeline}
\newcommand{\SizeMatInteractionDfwFall}{\WarningRunPipeline}
\newcommand{\SizeMatInteractionDfwFallLow}{\WarningRunPipeline}
\newcommand{\SizeMatInteractionDfwFallHigh}{\WarningRunPipeline}
\newcommand{\SizeMatInteractionDfwP}{\WarningRunPipeline}
\newcommand{\SizeMatInteractionGradeSigned}{\WarningRunPipeline}
\newcommand{\SizeMatSlopeLowScoreDfw}{\WarningRunPipeline}
\newcommand{\SizeMatSlopeLowScoreDfwLow}{\WarningRunPipeline}
\newcommand{\SizeMatSlopeLowScoreDfwHigh}{\WarningRunPipeline}
\newcommand{\SizeMatSlopeLowScoreDfwP}{\WarningRunPipeline}
\newcommand{\SizeMatSlopeLowScoreGradeLower}{\WarningRunPipeline}
\newcommand{\SizeMatSlopeLowScoreGradeLow}{\WarningRunPipeline}
\newcommand{\SizeMatSlopeLowScoreGradeHigh}{\WarningRunPipeline}
\newcommand{\SizeMatSlopeLowScoreGradeP}{\WarningRunPipeline}
\newcommand{\SizeMatSlopeHighScoreDfw}{\WarningRunPipeline}
\newcommand{\SizeMatSlopeHighScoreDfwLow}{\WarningRunPipeline}
\newcommand{\SizeMatSlopeHighScoreDfwHigh}{\WarningRunPipeline}
\newcommand{\SizeMatSlopeHighScoreGrade}{\WarningRunPipeline}
\newcommand{\SizeMatSlopeHighScoreGradeLow}{\WarningRunPipeline}
\newcommand{\SizeMatSlopeHighScoreGradeHigh}{\WarningRunPipeline}
\newcommand{\SizePooledGrade}{\WarningRunPipeline}
\newcommand{\SizePooledGradeLow}{\WarningRunPipeline}
\newcommand{\SizePooledGradeHigh}{\WarningRunPipeline}
\newcommand{\SizeAisPrimaryGrade}{\WarningRunPipeline}
\newcommand{\SizeAisSizeMedian}{\WarningRunPipeline}
\newcommand{\SizeAisSizeNinetieth}{\WarningRunPipeline}
\newcommand{\SizeMatGradeFirstAttempts}{\WarningRunPipeline}
\newcommand{\SizeFewestGradeFirstAttempts}{\WarningRunPipeline}
\newcommand{\SizeMusPrimaryGrade}{\WarningRunPipeline}
\newcommand{\SizeSameStudentGrade}{\WarningRunPipeline}
\newcommand{\SizeSameStudentGradeLow}{\WarningRunPipeline}
\newcommand{\SizeSameStudentGradeHigh}{\WarningRunPipeline}
\newcommand{\SizeHisDfwPerCohort}{\WarningRunPipeline}
\newcommand{\SizeHisDfwPerCohortLow}{\WarningRunPipeline}
\newcommand{\SizeHisDfwPerCohortHigh}{\WarningRunPipeline}
\newcommand{\SizeMatDfwPerCohort}{\WarningRunPipeline}
\newcommand{\SizeMatDfwPerCohortLow}{\WarningRunPipeline}
\newcommand{\SizeMatDfwPerCohortHigh}{\WarningRunPipeline}
\newcommand{\SizeCheDfwPerCohort}{\WarningRunPipeline}
\newcommand{\SizeCheDfwPerCohortLow}{\WarningRunPipeline}
\newcommand{\SizeCheDfwPerCohortHigh}{\WarningRunPipeline}
\newcommand{\SizeAisDfwPerCohort}{\WarningRunPipeline}
\newcommand{\SizeAisDfwPerCohortLow}{\WarningRunPipeline}
\newcommand{\SizeAisDfwPerCohortHigh}{\WarningRunPipeline}
\newcommand{\SizeMusDfwPerCohort}{\WarningRunPipeline}
\newcommand{\SizeMusDfwPerCohortLow}{\WarningRunPipeline}
\newcommand{\SizeMusDfwPerCohortHigh}{\WarningRunPipeline}
\newcommand{\SizeMatDfwPerCohortFewer}{\WarningRunPipeline}
\newcommand{\SizeMatDfwPerCohortLowFewer}{\WarningRunPipeline}
\newcommand{\SizeMatDfwPerCohortHighMore}{\WarningRunPipeline}
\newcommand{\SizeMatAffectedPerCohort}{\WarningRunPipeline}
\newcommand{\SizeMatScenarioDfw}{\WarningRunPipeline}
\newcommand{\SizeMatScenarioDfwLow}{\WarningRunPipeline}
\newcommand{\SizeMatScenarioDfwHigh}{\WarningRunPipeline}
\newcommand{\SizeMatSectionsPerYear}{\WarningRunPipeline}
\newcommand{\SizeMatSectionsAddedPerYear}{\WarningRunPipeline}
\newcommand{\SizeMatSectionsAddedPct}{\WarningRunPipeline}
\newcommand{\SizeLowThirdDfwFewer}{\WarningRunPipeline}
\newcommand{\SizeLowThirdDfwLow}{\WarningRunPipeline}
\newcommand{\SizeLowThirdDfwHigh}{\WarningRunPipeline}
\newcommand{\SizeLowThirdDfwP}{\WarningRunPipeline}
\newcommand{\SizeLowThirdGrade}{\WarningRunPipeline}
\newcommand{\SizeLowThirdGradeLow}{\WarningRunPipeline}
\newcommand{\SizeLowThirdGradeHigh}{\WarningRunPipeline}
\newcommand{\SizeLowThirdGradeP}{\WarningRunPipeline}
\newcommand{\SizeLowThirdDfwInvertedLow}{\WarningRunPipeline}
\newcommand{\SizeLowThirdDfwInvertedHigh}{\WarningRunPipeline}
\newcommand{\SizeLowThirdDfwClusterLow}{\WarningRunPipeline}
\newcommand{\SizeLowThirdDfwClusterHigh}{\WarningRunPipeline}
\newcommand{\SizeMidThirdDfwFewer}{\WarningRunPipeline}
\newcommand{\SizeMidThirdDfwLow}{\WarningRunPipeline}
\newcommand{\SizeMidThirdDfwHigh}{\WarningRunPipeline}
\newcommand{\SizeHighThirdDfwMore}{\WarningRunPipeline}
\newcommand{\SizeHighThirdDfwLow}{\WarningRunPipeline}
\newcommand{\SizeHighThirdDfwHigh}{\WarningRunPipeline}
\newcommand{\SizeAffectedMedianSize}{\WarningRunPipeline}
\newcommand{\SizeAffectedMedianSizeAfterSplit}{\WarningRunPipeline}
\newcommand{\SizeAffectedShareWithSmallSection}{\WarningRunPipeline}
\newcommand{\SizeAllSectionsPerYear}{\WarningRunPipeline}
\newcommand{\SizeAllSectionsAddedPerYear}{\WarningRunPipeline}
\newcommand{\SizeAllDfwPerCohort}{\WarningRunPipeline}
\newcommand{\SizeAllDfwPerCohortLow}{\WarningRunPipeline}
\newcommand{\SizeAllDfwPerCohortHigh}{\WarningRunPipeline}
\newcommand{\SizeNextOriginFirstAttempts}{\WarningRunPipeline}
\newcommand{\SizeNextContinuers}{\WarningRunPipeline}
\newcommand{\SizeNextBothGrades}{\WarningRunPipeline}
\newcommand{\SizeNextDifferentInstructor}{\WarningRunPipeline}
\newcommand{\SizeNextGrade}{\WarningRunPipeline}
\newcommand{\SizeNextGradeLow}{\WarningRunPipeline}
\newcommand{\SizeNextGradeHigh}{\WarningRunPipeline}
\newcommand{\SizeNextGradeDifferentInstructor}{\WarningRunPipeline}
\newcommand{\SizeNextGradeDifferentInstructorLow}{\WarningRunPipeline}
\newcommand{\SizeNextGradeDifferentInstructorHigh}{\WarningRunPipeline}
\newcommand{\SizeContinuationPerDoubling}{\WarningRunPipeline}
\newcommand{\SizeContinuationPerDoublingLow}{\WarningRunPipeline}
\newcommand{\SizeContinuationPerDoublingHigh}{\WarningRunPipeline}
\newcommand{\SizeBandSmallLow}{\WarningRunPipeline}
\newcommand{\SizeBandSmallHigh}{\WarningRunPipeline}
\newcommand{\SizeBandSmallGrade}{\WarningRunPipeline}
\newcommand{\SizeBandReferenceLow}{\WarningRunPipeline}
\newcommand{\SizeBandReferenceHigh}{\WarningRunPipeline}
\newcommand{\SizeBandMiddleLow}{\WarningRunPipeline}
\newcommand{\SizeBandMiddleHigh}{\WarningRunPipeline}
\newcommand{\SizeBandMiddleGrade}{\WarningRunPipeline}
\newcommand{\SizeBandLargeLow}{\WarningRunPipeline}
\newcommand{\SizeBandLargeHigh}{\WarningRunPipeline}
\newcommand{\SizeBandLargeGrade}{\WarningRunPipeline}
\newcommand{\SizeBandLargestGrade}{\WarningRunPipeline}
\newcommand{\SizeContrastShuffles}{\WarningRunPipeline}
\newcommand{\SizeContrastWildRejectMin}{\WarningRunPipeline}
\newcommand{\SizeContrastWildRejectMax}{\WarningRunPipeline}
\newcommand{\SizeContrastClusterRejectMin}{\WarningRunPipeline}
\newcommand{\SizeContrastClusterRejectMax}{\WarningRunPipeline}
\newcommand{\SizeDryShuffles}{\WarningRunPipeline}
\newcommand{\SizeDrySubjects}{\WarningRunPipeline}
\newcommand{\SizeDryClusterRejectMin}{\WarningRunPipeline}
\newcommand{\SizeDryClusterRejectMax}{\WarningRunPipeline}
\newcommand{\SizeDryWildRejectMin}{\WarningRunPipeline}
\newcommand{\SizeDryWildRejectMax}{\WarningRunPipeline}
\newcommand{\SizeCurveMatGpaSmallest}{\WarningRunPipeline}
\newcommand{\SizeCurveMatGpaReference}{\WarningRunPipeline}
\newcommand{\SizeCurveReferenceSize}{\WarningRunPipeline}
\newcommand{\SizeCurveMinStudents}{\WarningRunPipeline}
\newcommand{\SizeSubjectsModeledCapitalWord}{\WarningRunPipeline}
\newcommand{\SizeSubjectsOutsideWord}{\WarningRunPipeline}
\newcommand{\SizeNestedModelsCapitalWord}{\WarningRunPipeline}
\newcommand{\SizeSubjectsPassingCapitalWord}{\WarningRunPipeline}

\newcommand{\XwalkWeightContain}{\WarningRunPipeline}
\newcommand{\XwalkWeightJaccard}{\WarningRunPipeline}
\newcommand{\XwalkWeightTrigram}{\WarningRunPipeline}
\newcommand{\XwalkAutoScore}{\WarningRunPipeline}
\newcommand{\XwalkAutoMargin}{\WarningRunPipeline}
\newcommand{\XwalkMinCandidateScore}{\WarningRunPipeline}
\newcommand{\XwalkFirstRunScore}{\WarningRunPipeline}
\newcommand{\XwalkRecoveryScore}{\WarningRunPipeline}
\newcommand{\XwalkNameRowsAll}{\WarningRunPipeline}
\newcommand{\XwalkReviewRows}{\WarningRunPipeline}
\newcommand{\XwalkNameRows}{\WarningRunPipeline}
\newcommand{\XwalkStageAutoPct}{\WarningRunPipeline}
\newcommand{\XwalkStageAdjudicatedPct}{\WarningRunPipeline}
\newcommand{\XwalkStageVintagePct}{\WarningRunPipeline}
\newcommand{\XwalkCoveragePct}{\WarningRunPipeline}
\newcommand{\XwalkUsableStudents}{\WarningRunPipeline}
\newcommand{\XwalkPrivateNames}{\WarningRunPipeline}
\newcommand{\XwalkPrivateNamedStudents}{\WarningRunPipeline}
\newcommand{\XwalkPrivateNamedPct}{\WarningRunPipeline}
\newcommand{\XwalkRecoveredInDirectoryPct}{\WarningRunPipeline}
\newcommand{\XwalkMatchedSchools}{\WarningRunPipeline}
\newcommand{\XwalkMatchedDistricts}{\WarningRunPipeline}
\newcommand{\XwalkSchoolsInTablePct}{\WarningRunPipeline}
\newcommand{\XwalkTableCycleSchools}{\WarningRunPipeline}
\newcommand{\XwalkTableCycleEqualPct}{\WarningRunPipeline}
\newcommand{\XwalkTableOtherCyclesLow}{\WarningRunPipeline}
\newcommand{\XwalkTableOtherCyclesHigh}{\WarningRunPipeline}
\newcommand{\XwalkMatched}{\WarningRunPipeline}
\newcommand{\XwalkOutsideTexas}{\WarningRunPipeline}
\newcommand{\XwalkUnresolved}{\WarningRunPipeline}
\newcommand{\XwalkPrivate}{\WarningRunPipeline}
\newcommand{\XwalkNoSchool}{\WarningRunPipeline}
\newcommand{\XwalkRejected}{\WarningRunPipeline}
\newcommand{\XwalkUnresolvedDualPct}{\WarningRunPipeline}
\newcommand{\XwalkMatchedDualPct}{\WarningRunPipeline}
\newcommand{\XwalkDualZ}{\WarningRunPipeline}
\newcommand{\XwalkDualZClustered}{\WarningRunPipeline}
\newcommand{\XwalkUnresolvedApPct}{\WarningRunPipeline}
\newcommand{\XwalkMatchedApPct}{\WarningRunPipeline}
\newcommand{\XwalkApZ}{\WarningRunPipeline}
\newcommand{\XwalkApZClustered}{\WarningRunPipeline}
\newcommand{\XwalkUnresolvedPellPct}{\WarningRunPipeline}
\newcommand{\XwalkMatchedPellPct}{\WarningRunPipeline}
\newcommand{\XwalkPellZ}{\WarningRunPipeline}
\newcommand{\XwalkPellZClustered}{\WarningRunPipeline}
\newcommand{\XwalkUnresolvedFirstCoursePct}{\WarningRunPipeline}
\newcommand{\XwalkMatchedFirstCoursePct}{\WarningRunPipeline}
\newcommand{\XwalkFirstCourseZ}{\WarningRunPipeline}
\newcommand{\XwalkFirstCourseZClustered}{\WarningRunPipeline}
\newcommand{\XwalkUnresolvedScore}{\WarningRunPipeline}
\newcommand{\XwalkMatchedScore}{\WarningRunPipeline}
\newcommand{\XwalkScoreZ}{\WarningRunPipeline}
\newcommand{\XwalkScoreZClustered}{\WarningRunPipeline}
\newcommand{\XwalkMaxZClustered}{\WarningRunPipeline}
\newcommand{\XwalkUnresolvedSchools}{\WarningRunPipeline}
\newcommand{\XwalkMatchedFirstGenPct}{\WarningRunPipeline}
\newcommand{\XwalkOutsideTexasTransferPct}{\WarningRunPipeline}
\newcommand{\XwalkOutsideTexasApPct}{\WarningRunPipeline}
\newcommand{\XwalkOutsideTexasDualPct}{\WarningRunPipeline}
\newcommand{\XwalkNoSchoolTransferPct}{\WarningRunPipeline}
\newcommand{\XwalkNoSchoolApPct}{\WarningRunPipeline}
\newcommand{\XwalkNoSchoolDualPct}{\WarningRunPipeline}
\newcommand{\XwalkAllTransferPct}{\WarningRunPipeline}
\newcommand{\XwalkAllApPct}{\WarningRunPipeline}
\newcommand{\XwalkAllDualPct}{\WarningRunPipeline}
\newcommand{\XwalkPrivatePellPct}{\WarningRunPipeline}
\newcommand{\XwalkPrivateFirstGenPct}{\WarningRunPipeline}
\newcommand{\XwalkProvisionStudents}{\WarningRunPipeline}
\newcommand{\XwalkProvisionSchools}{\WarningRunPipeline}
\newcommand{\XwalkNoApBandApPct}{\WarningRunPipeline}
\newcommand{\XwalkNoApBandDualPct}{\WarningRunPipeline}
\newcommand{\XwalkTopBandApPct}{\WarningRunPipeline}
\newcommand{\XwalkTopBandDualPct}{\WarningRunPipeline}
\newcommand{\XwalkTopBandEdgePct}{\WarningRunPipeline}
\newcommand{\XwalkNoApBandMedianEnrollment}{\WarningRunPipeline}
\newcommand{\XwalkTopBandMedianEnrollment}{\WarningRunPipeline}
\newcommand{\XwalkPovertyStudents}{\WarningRunPipeline}
\newcommand{\XwalkPovertyDistricts}{\WarningRunPipeline}
\newcommand{\XwalkPovertyMedianStudents}{\WarningRunPipeline}
\newcommand{\XwalkPovertyMedianChildren}{\WarningRunPipeline}
\newcommand{\XwalkPovertyMedianGap}{\WarningRunPipeline}
\newcommand{\XwalkLargestBinLow}{\WarningRunPipeline}
\newcommand{\XwalkLargestBinHigh}{\WarningRunPipeline}
\newcommand{\XwalkLargestBinPct}{\WarningRunPipeline}
\newcommand{\XwalkNorthsideStudents}{\WarningRunPipeline}
\newcommand{\XwalkNorthsidePovertyPct}{\WarningRunPipeline}
\newcommand{\XwalkNorthEastStudents}{\WarningRunPipeline}
\newcommand{\XwalkNorthEastPovertyPct}{\WarningRunPipeline}
\newcommand{\XwalkDistanceNinetieth}{\WarningRunPipeline}
\newcommand{\XwalkDistanceNinetyFifth}{\WarningRunPipeline}
\newcommand{\XwalkFarMiles}{\WarningRunPipeline}
\newcommand{\XwalkFarLocalHomePct}{\WarningRunPipeline}
\newcommand{\XwalkPrimaryMiles}{\WarningRunPipeline}
\newcommand{\XwalkStrictMiles}{\WarningRunPipeline}
\newcommand{\XwalkWithinPrimaryPct}{\WarningRunPipeline}
\newcommand{\XwalkPointWithinPrimaryPct}{\WarningRunPipeline}
\newcommand{\XwalkZipZeroPct}{\WarningRunPipeline}
\newcommand{\XwalkPointZeroPct}{\WarningRunPipeline}
\newcommand{\XwalkMisreadZipStudents}{\WarningRunPipeline}
\newcommand{\XwalkBestTractSharePct}{\WarningRunPipeline}
\newcommand{\XwalkTractsPerZip}{\WarningRunPipeline}
\newcommand{\XwalkBestTractShareAllZips}{\WarningRunPipeline}
\newcommand{\XwalkContextCoveragePct}{\WarningRunPipeline}

}{}

\title{A Calculus Gradient: Child Poverty, Mathematics Preparation, and First-Year College Performance at a Hispanic-Serving University}
\author{Juan B. Guti\'errez\\
  Department of Mathematics\\
  University of Texas at San Antonio}
\date{}

\begin{document}
\IfFileExists{macros/renewcommands.tex}{\renewcommand{\MacroTotal}{856}
\renewcommand{\MacroModulesWord}{ten}
\renewcommand{\CalcDistricts}{9{,}790}
\renewcommand{\CalcDistrictsWithIncome}{9{,}775}
\renewcommand{\CalcDistrictsZero}{2{,}321}
\renewcommand{\CalcDistrictsAboveDisplay}{112}
\renewcommand{\CalcBexarDistricts}{15}
\renewcommand{\CalcMeanPerHundred}{4.6}
\renewcommand{\CalcSlope}{-0.171}
\renewcommand{\CalcSlopeSE}{0.020}
\renewcommand{\CalcSeatsPerTenPoints}{1.7}
\renewcommand{\CalcRelDecline}{36.9}
\renewcommand{\CalcRelSigned}{-36.9}
\renewcommand{\CalcRelDeclineRounded}{37}
\renewcommand{\CalcSlopeUnweighted}{-0.179}
\renewcommand{\CalcSlopeUnweightedSE}{0.019}
\renewcommand{\CalcTexasDistricts}{860}
\renewcommand{\CalcTexasSlope}{-0.083}
\renewcommand{\CalcTexasSlopeSE}{0.023}
\renewcommand{\CalcTexasRelDecline}{24.7}
\renewcommand{\CalcAlamoPovertyPct}{7}
\renewcommand{\CalcEdgewoodPovertyPct}{39}
\renewcommand{\CalcAlamoPerHundred}{12.0}
\renewcommand{\CalcEdgewoodPerHundred}{0.7}
\renewcommand{\CalcAlamoRatio}{2.6}
\renewcommand{\CalcAlamoRatioTwo}{2.60}
\renewcommand{\CalcEdgewoodRatio}{0.14}
\renewcommand{\CalcLargestTwoRatio}{0.92}
\renewcommand{\CalcSanAntonioRatio}{0.49}
\renewcommand{\CalcIncomeLowNear}{30{,}000}
\renewcommand{\CalcIncomeHighAbove}{60{,}000}
\renewcommand{\CalcIncomeCorr}{0.45}
\renewcommand{\CalcIncomeSpearman}{0.41}
\renewcommand{\CalcIncomePovertyCorr}{-0.72}

\renewcommand{\DataSaipeVintages}{28}
\renewcommand{\DataSaipeDistrictsEver}{15{,}007}
\renewcommand{\DataSaipeDistrictsEveryVintage}{12{,}659}
\renewcommand{\DataSaipeDistrictsEveryVintagePct}{84.4}
\renewcommand{\DataCrdcCycles}{5}
\renewcommand{\DataCourseFamilies}{15}
\renewcommand{\DataCourseFamiliesEarliestCycle}{7}
\renewcommand{\DataSaipeRows}{382{,}099}
\renewcommand{\DataCrdcSchoolCycleRows}{481{,}347}
\renewcommand{\DataCrdcDistrictCycleRows}{84{,}361}
\renewcommand{\DataCourseFamilyRows}{5{,}030{,}159}
\renewcommand{\DataEdgeSchoolRows}{204{,}467}
\renewcommand{\DataAtlasTracts}{74{,}123}
\renewcommand{\DataAtlasZctas}{28{,}744}
\renewcommand{\DataZctaTractWeights}{125{,}330}
\renewcommand{\DataDistrictTractShares}{112{,}698}
\renewcommand{\DataSourceFiles}{378}
\renewcommand{\DataManifestArchivesWord}{six}
\renewcommand{\DataAlgTwoZeroPctEarliestCycle}{26.25}
\renewcommand{\DataAlgTwoZeroPctLaterMin}{0.01}
\renewcommand{\DataAlgTwoZeroPctLaterMax}{0.41}
\renewcommand{\DataOfferAgreePctMin}{93}
\renewcommand{\DataOfferAgreePctMax}{100}
\renewcommand{\DataOfferKappa}{0}
\renewcommand{\DataReportingRateExcessPct}{40}
\renewcommand{\DataAlgTwoSkipCodeSchoolsRounded}{75{,}000}
\renewcommand{\DataAlgTwoSchoolsWithCountPct}{22}
\renewcommand{\DataAlgTwoSchoolsWithCount}{21{,}419}
\renewcommand{\DataAlgTwoEnrolleesRebuiltKey}{2{,}931{,}194}
\renewcommand{\DataStoredKeySchools}{96{,}360}
\renewcommand{\DataStoredKeySchoolsLost}{66{,}986}
\renewcommand{\DataAlgTwoEnrolleesStoredKey}{807{,}667}
\renewcommand{\DataStoredKeyEnrollmentLostPct}{72}
\renewcommand{\DataSlopeSeStateClustered}{0.0136}
\renewcommand{\DataSlopeSeNaive}{0.0027}
\renewcommand{\DataSlopeSeRatio}{5}
\renewcommand{\DataMinChildren}{200}
\renewcommand{\DataAlgTwoDistrictsNoFloor}{10{,}425}
\renewcommand{\DataAlgTwoPerChildMax}{31.3}
\renewcommand{\DataAlgTwoPerChildMaxFloor}{0.68}
\renewcommand{\DataUnweightedSlopeNoFloor}{+0.030}
\renewcommand{\DataUnweightedSlopeFloor}{-0.049}
\renewcommand{\DataWeightedSlope}{-0.052}
\renewcommand{\DataSlopeGapFloor}{0.003}
\renewcommand{\DataWeightedSlopeLadderFlattest}{-0.0521}
\renewcommand{\DataWeightedSlopeLadderSteepest}{-0.0527}
\renewcommand{\DataUnweightedSlopeWithoutMismatch}{-0.037}
\renewcommand{\DataMismatchMedianChildren}{35}
\renewcommand{\DataNeutralFrameDistricts}{9{,}934}
\renewcommand{\DataNoAlgTwoDistricts}{144}
\renewcommand{\DataNoAlgTwoChildren}{450{,}005}
\renewcommand{\DataNoCoreDistricts}{65}
\renewcommand{\DataNoCoreChildren}{247{,}131}
\renewcommand{\DataCrdcCyclesWord}{five}
\renewcommand{\DataMismatchDistrictsWord}{seven}
\renewcommand{\DataMismatchDistrictsCapitalWord}{Seven}
\renewcommand{\DataIncompleteDistrictsCapitalWord}{Five}
\renewcommand{\DataIncompleteDistrictsEarlierWord}{three}
\renewcommand{\DataIncompleteDistrictsLaterWord}{thirteen}
\renewcommand{\DataCoreCoursesWord}{nine}

\renewcommand{\GradMinChildren}{200}
\renewcommand{\GradBootstrapResamples}{5{,}000}
\renewcommand{\GradIncompleteDistricts}{5}
\renewcommand{\GradDistrictsBeforeExclusion}{9{,}795}
\renewcommand{\GradBiologyRel}{-0.6}
\renewcommand{\GradBiologyRelLow}{-2.3}
\renewcommand{\GradBiologyRelHigh}{+1.7}
\renewcommand{\GradAlgebraTwoRel}{-0.7}
\renewcommand{\GradAlgebraTwoRelLow}{-4.7}
\renewcommand{\GradAlgebraTwoRelHigh}{+1.9}
\renewcommand{\GradGeometryRel}{+1.1}
\renewcommand{\GradGeometryRelLow}{-1.0}
\renewcommand{\GradGeometryRelHigh}{+3.2}
\renewcommand{\GradAlgebraOneNineTenRel}{+6.5}
\renewcommand{\GradAlgebraOneNineTenRelLow}{-2.0}
\renewcommand{\GradAlgebraOneNineTenRelHigh}{+13.8}
\renewcommand{\GradApMathRelDecline}{36.9}
\renewcommand{\GradApScienceRelDecline}{35.9}
\renewcommand{\GradApAnyRelDecline}{24.5}
\renewcommand{\GradAdvMathRelDecline}{14.9}
\renewcommand{\GradPhysicsRelDecline}{13.3}
\renewcommand{\GradCalculusRelDeclineLow}{28.7}
\renewcommand{\GradCalculusRelDeclineHigh}{46.2}
\renewcommand{\GradEarlyAlgebraRelDecline}{16.4}
\renewcommand{\GradEarlyAlgebraRelDeclineLow}{-4.9}
\renewcommand{\GradEarlyAlgebraRelDeclineHigh}{34.0}
\renewcommand{\GradEarlyAlgebraFirstCycleRelDecline}{25.9}
\renewcommand{\GradEarlyAlgebraFirstCycleRelDeclineLow}{20.6}
\renewcommand{\GradEarlyAlgebraFirstCycleRelDeclineHigh}{31.3}
\renewcommand{\GradAdvancedRelDeclineMin}{36}
\renewcommand{\GradAdvancedRelDeclineMax}{37}
\renewcommand{\GradLowestTenthPovertyMax}{5.6}
\renewcommand{\GradHighestTenthPovertyMin}{28.6}
\renewcommand{\GradCalculusLowestTenthRatio}{1.93}
\renewcommand{\GradCalculusHighestTenthRatio}{0.46}
\renewcommand{\GradBiologyLowestTenthRatio}{1.00}
\renewcommand{\GradBiologyHighestTenthRatio}{0.99}
\renewcommand{\GradBexarMinSharePct}{10}
\renewcommand{\GradBexarGroupOnePovertyLow}{7.0}
\renewcommand{\GradBexarGroupOnePovertyHigh}{14.3}
\renewcommand{\GradBexarGroupOneCalculusRatio}{1.02}
\renewcommand{\GradBexarGroupTwoPoverty}{15.2}
\renewcommand{\GradBexarGroupTwoCalculusRatio}{0.92}
\renewcommand{\GradBexarGroupThreePovertyLow}{17.8}
\renewcommand{\GradBexarGroupThreePovertyHigh}{18.9}
\renewcommand{\GradBexarGroupThreeCalculusRatio}{0.38}
\renewcommand{\GradBexarGroupFourPovertyLow}{25.9}
\renewcommand{\GradBexarGroupFourPovertyHigh}{28.6}
\renewcommand{\GradBexarGroupFourCalculusRatio}{0.40}
\renewcommand{\GradBexarGroupFivePovertyLow}{31.1}
\renewcommand{\GradBexarGroupFivePovertyHigh}{39.1}
\renewcommand{\GradBexarGroupFiveCalculusRatio}{0.45}
\renewcommand{\GradBexarCoreRatioMin}{0.81}
\renewcommand{\GradBexarCoreRatioMax}{1.26}
\renewcommand{\GradBexarCalculusRelDecline}{32}
\renewcommand{\GradBexarCalculusRelDeclineLow}{11}
\renewcommand{\GradBexarCalculusRelDeclineHigh}{63}
\renewcommand{\GradBexarApMathRelDecline}{34}
\renewcommand{\GradBexarApScienceRelDecline}{32}
\renewcommand{\GradBexarAdvMathRel}{+16}
\renewcommand{\GradBexarAdvMathRelLow}{0.2}
\renewcommand{\GradBexarAdvMathRelHigh}{35}
\renewcommand{\GradBexarCalculusRel}{-32}
\renewcommand{\GradBexarCalculusRelLow}{-63}
\renewcommand{\GradBexarCalculusRelHigh}{-11}
\renewcommand{\GradBexarUpperGroupsCalculusRatioMin}{0.38}
\renewcommand{\GradBexarUpperGroupsCalculusRatioMax}{0.45}
\renewcommand{\GradBexarUpperGroupsPovertyAbove}{17}
\renewcommand{\GradCalculusFirstCycleRelDecline}{30.7}
\renewcommand{\GradCalculusLastCycleRelDecline}{45.3}
\renewcommand{\GradCoreCyclesRelMin}{-1.4}
\renewcommand{\GradCoreCyclesRelMax}{+3.4}
\renewcommand{\GradCalculusNeutralFrameRelDecline}{37.1}
\renewcommand{\GradCalculusStateFeRelDecline}{37.0}
\renewcommand{\GradCalculusExactSpanRelDecline}{37.5}
\renewcommand{\GradCalculusExactSpanRelDeclineLow}{29.1}
\renewcommand{\GradCalculusExactSpanRelDeclineHigh}{47.0}
\renewcommand{\GradCalculusChecksRelDeclineMin}{37.0}
\renewcommand{\GradCalculusChecksRelDeclineMax}{37.5}
\renewcommand{\GradCoreNeutralFrameRelMin}{-0.6}
\renewcommand{\GradCoreNeutralFrameRelMax}{+1.4}
\renewcommand{\GradExactSpanSchools}{13{,}250}
\renewcommand{\GradTexasApMathRelDecline}{35.7}
\renewcommand{\GradTexasEarlyAlgebraRelDecline}{21.5}
\renewcommand{\GradTexasAlgebraTwoRel}{+0.4}
\renewcommand{\GradTexasAlgebraTwoZ}{0.30}
\renewcommand{\GradApAnySeats}{5.0}
\renewcommand{\GradEarlyAlgebraSeats}{2.5}
\renewcommand{\GradAdvMathSeats}{2.4}
\renewcommand{\GradApScienceSeats}{2.0}
\renewcommand{\GradApMathSeats}{1.7}
\renewcommand{\GradCoreSeatsMax}{0.3}
\renewcommand{\GradIncomeCoreCorrMin}{-0.06}
\renewcommand{\GradIncomeCoreCorrMax}{0.06}
\renewcommand{\GradIncomeAdvMathCorr}{0.28}
\renewcommand{\GradIncomeApCorr}{0.48}
\renewcommand{\GradHsEnrollmentPerChildRelDecline}{8.7}
\renewcommand{\GradHsEnrollmentPerChildRelDeclineLow}{6.1}
\renewcommand{\GradHsEnrollmentPerChildRelDeclineHigh}{11.7}
\renewcommand{\GradCorePerChildRelDeclineMin}{7.7}
\renewcommand{\GradCorePerChildRelDeclineMax}{9.4}
\renewcommand{\GradAdvancedPerChildRelDeclineMin}{42}
\renewcommand{\GradAdvancedPerChildRelDeclineMax}{43}
\renewcommand{\GradApAnySeatsLow}{4.1}
\renewcommand{\GradApAnySeatsHigh}{5.9}
\renewcommand{\GradEarlyAlgebraSeatsLow}{-0.8}
\renewcommand{\GradEarlyAlgebraSeatsHigh}{5.7}
\renewcommand{\GradAdvancedMathSeatsLow}{1.6}
\renewcommand{\GradAdvancedMathSeatsHigh}{3.1}
\renewcommand{\GradCourseFamiliesWord}{fifteen}
\renewcommand{\GradCyclesWord}{three}
\renewcommand{\GradIncompleteDistrictsWord}{five}
\renewcommand{\GradBexarDistrictsWord}{fifteen}
\renewcommand{\GradBexarDistrictsCapitalWord}{Fifteen}
\renewcommand{\GradBexarGroupsWord}{five}
\renewcommand{\GradBexarLowerGroupsWord}{two}
\renewcommand{\GradBexarUpperGroupsWord}{three}
\renewcommand{\GradFrameMismatchDistrictsWord}{seven}

\renewcommand{\InstCohortN}{10{,}027}
\renewcommand{\InstCohortFirstFallN}{4{,}935}
\renewcommand{\InstCohortSecondFallN}{4{,}923}
\renewcommand{\InstCohortSpringSummerN}{169}
\renewcommand{\InstCohortFallPct}{98}
\renewcommand{\InstCapN}{1{,}190}
\renewcommand{\InstCapSeniorPct}{24.5}
\renewcommand{\InstNoCapN}{8{,}837}
\renewcommand{\InstNoCapSeniorPct}{54.0}
\renewcommand{\InstNoCapOddsMat}{1.42}
\renewcommand{\InstSeniorPct}{50.5}
\renewcommand{\InstSeniorStrictN}{4{,}755}
\renewcommand{\InstSeniorStrictPct}{47.4}
\renewcommand{\InstOddsMat}{1.46}
\renewcommand{\InstOddsMatStrict}{1.60}
\renewcommand{\InstMatTakersPct}{91.0}
\renewcommand{\InstOddsWrc}{1.39}
\renewcommand{\InstOddsWrcStrict}{1.45}
\renewcommand{\InstWrcTakersPct}{70.4}
\renewcommand{\InstOddsAis}{0.23}
\renewcommand{\InstOddsAisStrict}{0.26}
\renewcommand{\InstAisTakersPct}{92.6}
\renewcommand{\InstMatNoSeniorPct}{50.4}
\renewcommand{\InstNoMatNoSeniorPct}{41.1}
\renewcommand{\InstGraduateNumberedN}{67}
\renewcommand{\InstPreEntryRecordsN}{34}
\renewcommand{\InstMissingMajorN}{27}
\renewcommand{\InstMajorChangePct}{43.0}
\renewcommand{\InstConfirmedChangePct}{42.7}
\renewcommand{\InstMajorCompleteN}{10{,}000}
\renewcommand{\InstConfirmedChangeCompletePct}{42.8}
\renewcommand{\InstBoundedConfirmedN}{10{,}000}
\renewcommand{\InstBoundedConfirmedChangePct}{42.7}
\renewcommand{\InstLaterObservedN}{7{,}250}
\renewcommand{\InstLaterObservedChangePct}{56.5}
\renewcommand{\InstNotObservedN}{2{,}777}
\renewcommand{\InstNotObservedChangePct}{7.7}
\renewcommand{\InstIntroBelow}{2000}
\renewcommand{\InstDfwIntervalLevel}{95}
\renewcommand{\InstDfwMatLaterN}{6{,}390}
\renewcommand{\InstDfwMatLaterWithN}{2{,}196}
\renewcommand{\InstDfwMatLaterWithoutN}{4{,}194}
\renewcommand{\InstDfwMatLaterChangedWithPct}{65.2}
\renewcommand{\InstDfwMatLaterChangedWithoutPct}{54.6}
\renewcommand{\InstDfwMatLaterDiff}{10.6}
\renewcommand{\InstDfwMatLaterDiffLo}{8.1}
\renewcommand{\InstDfwMatLaterDiffHi}{13.1}
\renewcommand{\InstDfwMatLaterOdds}{1.56}
\renewcommand{\InstDfwMatLaterOddsLo}{1.40}
\renewcommand{\InstDfwMatLaterOddsHi}{1.73}
\renewcommand{\InstDfwMatAllN}{8{,}892}
\renewcommand{\InstDfwMatAllChangedWithPct}{45.8}
\renewcommand{\InstDfwMatAllChangedWithoutPct}{42.9}
\renewcommand{\InstDfwMatAllDiff}{2.8}
\renewcommand{\InstDfwMatAllDiffLo}{0.7}
\renewcommand{\InstDfwMatAllDiffHi}{5.0}
\renewcommand{\InstDfwMatAllOdds}{1.12}
\renewcommand{\InstDfwMatAllOddsLo}{1.03}
\renewcommand{\InstDfwMatAllOddsHi}{1.22}
\renewcommand{\InstDfwPhyLaterWithN}{127}
\renewcommand{\InstDfwCsLaterDiff}{17.3}
\renewcommand{\InstDfwIsLaterDiff}{10.7}
\renewcommand{\InstDfwPhyLaterDiff}{10.7}
\renewcommand{\InstDfwCheLaterDiff}{12.8}
\renewcommand{\InstDfwCheLaterDiffLo}{8.7}
\renewcommand{\InstDfwCheLaterDiffHi}{17.0}
\renewcommand{\InstDfwCheAllDiff}{0.3}
\renewcommand{\InstDfwCheAllDiffLo}{-3.4}
\renewcommand{\InstDfwCheAllDiffHi}{4.1}
\renewcommand{\InstDfwWrcAllDiff}{-10.2}
\renewcommand{\InstDfwHisAllDiff}{-9.6}
\renewcommand{\InstDfwAisAllDiff}{-16.0}
\renewcommand{\InstDfwMatDeclaredN}{5{,}275}
\renewcommand{\InstDfwMatDeclaredDiff}{13.2}
\renewcommand{\InstDfwMatDeclaredOdds}{1.70}
\renewcommand{\InstDfwMatDeclaredOddsLo}{1.52}
\renewcommand{\InstDfwMatDeclaredOddsHi}{1.91}
\renewcommand{\InstDfwMatBoundedLaterDiff}{10.7}
\renewcommand{\InstDfwMatBoundedLaterDiffLo}{8.2}
\renewcommand{\InstDfwMatBoundedLaterDiffHi}{13.2}
\renewcommand{\InstDfwMatBoundedAllDiff}{2.9}
\renewcommand{\InstDfwMatBoundedAllDiffLo}{0.7}
\renewcommand{\InstDfwMatBoundedAllDiffHi}{5.0}
\renewcommand{\InstHeatMinCell}{15}
\renewcommand{\InstHeatTopKept}{30}
\renewcommand{\InstHeatChePerMover}{-0.50}
\renewcommand{\InstHeatCheMovers}{2{,}047}
\renewcommand{\InstHeatCheCells}{14}
\renewcommand{\InstHeatMatPolPerMover}{-0.43}
\renewcommand{\InstHeatBlwEcoBioPerMoverLow}{-0.39}
\renewcommand{\InstHeatBlwEcoBioPerMoverHigh}{-0.38}
\renewcommand{\InstHeatMatPerMoverBelow}{0.43}
\renewcommand{\InstHeatChePerMoverBelow}{0.50}
\renewcommand{\InstHeatMatTotal}{-1{,}662}
\renewcommand{\InstHeatPolTotal}{-1{,}593}
\renewcommand{\InstHeatMatMovers}{3{,}851}
\renewcommand{\InstHeatPolMovers}{3{,}708}
\renewcommand{\InstHeatDonorMajorColumns}{30}
\renewcommand{\InstHeatCheTotal}{-1{,}028}
\renewcommand{\InstHeatIsPerMover}{+0.61}
\renewcommand{\InstHeatArtPerMover}{+0.57}
\renewcommand{\InstHeatAisTotal}{+1{,}487}
\renewcommand{\InstHeatAisMovers}{4{,}036}
\renewcommand{\InstHeatAisPerMover}{+0.37}
\renewcommand{\InstHeatIsTotal}{+925}
\renewcommand{\InstNaiveOddsMat}{1.47}
\renewcommand{\InstNaiveOddsWrc}{1.80}
\renewcommand{\InstNaiveOddsAis}{1.81}
\renewcommand{\InstChordStudents}{19{,}489}
\renewcommand{\InstChordMinFlowWord}{ten}
\renewcommand{\InstChordDrawnFlows}{415}
\renewcommand{\InstChordDrawnStudents}{17{,}164}
\renewcommand{\InstChordDrawnPct}{88.1}
\renewcommand{\InstChordWithheldFlows}{610}
\renewcommand{\InstChordWithheldStudents}{2{,}325}
\renewcommand{\InstChordSectors}{36}
\renewcommand{\InstChordAnchorDeg}{135}
\renewcommand{\InstFdrQ}{0.001}
\renewcommand{\InstSectionSizeBin}{5}
\renewcommand{\InstHpLambda}{10^{4}}
\renewcommand{\InstHorizonYearsWord}{four}
\renewcommand{\InstExactBoundaryCourseNText}{Fewer than ten}
\renewcommand{\InstDfwSubjectsWord}{seventeen}
\renewcommand{\InstChordOtherMajorsWord}{fourteen}
\renewcommand{\InstDisclosureMinCellWord}{ten}

\renewcommand{\LinkEntrants}{28{,}209}
\renewcommand{\LinkSample}{23{,}489}
\renewcommand{\LinkSampleScored}{23{,}478}
\renewcommand{\LinkSampleGradeCohorts}{19{,}874}
\renewcommand{\LinkNonApplicantPct}{14.4}
\renewcommand{\LinkSubjects}{25}
\renewcommand{\LinkQuantSubjects}{7}
\renewcommand{\LinkOtherSubjects}{18}
\renewcommand{\LinkSubjectMinimum}{1{,}000}
\renewcommand{\LinkFdrLevel}{0.05}
\renewcommand{\LinkScoreSD}{71}
\renewcommand{\LinkVarDistrictPct}{7.9}
\renewcommand{\LinkVarSchoolPct}{14.3}
\renewcommand{\LinkVarBetweenPct}{22}
\renewcommand{\LinkVarWithinPct}{77.8}
\renewcommand{\LinkVarWithinPctRound}{78}
\renewcommand{\LinkGapVarBetweenPct}{2.5}
\renewcommand{\LinkGapVarWithinPct}{97.5}
\renewcommand{\LinkVarDistrictReducedPct}{95}
\renewcommand{\LinkVarSchoolReducedPct}{35}
\renewcommand{\LinkVarSchoolReducedWithSchoolPct}{63}
\renewcommand{\LinkIndexPoints}{9.7}
\renewcommand{\LinkIndexPointsSigned}{+9.7}
\renewcommand{\LinkPovertyPoints}{10.4}
\renewcommand{\LinkPovertyPointsSigned}{-10.4}
\renewcommand{\LinkIncomeRankPoints}{+5.9}
\renewcommand{\LinkFirstGenPoints}{10.6}
\renewcommand{\LinkIncomeBetween}{0.436}
\renewcommand{\LinkIncomeWithin}{0.048}
\renewcommand{\LinkIncomeBetweenMinusWithin}{0.39}
\renewcommand{\LinkIncomeBetweenMinusWithinZ}{10.2}
\renewcommand{\LinkHOne}{+0.027}
\renewcommand{\LinkHOneLow}{0.008}
\renewcommand{\LinkHOneHigh}{0.046}
\renewcommand{\LinkHOnePoints}{1.9}
\renewcommand{\LinkGapIncomeWithin}{-0.01}
\renewcommand{\LinkGapMale}{+0.36}
\renewcommand{\LinkGapInternational}{+0.72}
\renewcommand{\LinkGapInternationalSE}{0.17}
\renewcommand{\LinkGapAsian}{+0.39}
\renewcommand{\LinkIndexEntryLevel}{+0.07}
\renewcommand{\LinkMathHeldOutPct}{29}
\renewcommand{\LinkGapHeldOutPct}{6.5}
\renewcommand{\LinkNoAidIndexPoints}{+9.8}
\renewcommand{\LinkNoAidPovertyPoints}{-11.5}
\renewcommand{\LinkNoAidHOne}{+0.027}
\renewcommand{\LinkNoAidHOneLow}{0.008}
\renewcommand{\LinkNoAidHOneHigh}{0.046}
\renewcommand{\LinkNoAidPooledMathOther}{0.145}
\renewcommand{\LinkNoAidSubjectsPassing}{24}
\renewcommand{\LinkMatMathGrade}{0.38}
\renewcommand{\LinkMatVerbalGrade}{-0.03}
\renewcommand{\LinkMatMathDfwPoints}{10.6}
\renewcommand{\LinkMatMathDfwPointsSigned}{-10.6}
\renewcommand{\LinkSubjectsPassing}{24}
\renewcommand{\LinkOtherSubjectsPassing}{17}
\renewcommand{\LinkPassingAttemptsPct}{98}
\renewcommand{\LinkPooledMathOther}{0.140}
\renewcommand{\LinkPooledMathOtherTwo}{0.14}
\renewcommand{\LinkPooledMathOtherLow}{0.109}
\renewcommand{\LinkPooledMathOtherHigh}{0.171}
\renewcommand{\LinkPooledMathOtherLowIndep}{0.111}
\renewcommand{\LinkPooledMathOtherHighIndep}{0.169}
\renewcommand{\LinkPooledVerbalOther}{0.112}
\renewcommand{\LinkPooledDiffOther}{+0.027}
\renewcommand{\LinkPooledDiffOtherLow}{-0.015}
\renewcommand{\LinkPooledDiffOtherHigh}{+0.068}
\renewcommand{\LinkPooledMathQuant}{0.263}
\renewcommand{\LinkPooledMathQuantTwo}{0.26}
\renewcommand{\LinkPooledMathQuantLow}{0.197}
\renewcommand{\LinkPooledMathQuantHigh}{0.328}
\renewcommand{\LinkPooledVerbalQuant}{0.065}
\renewcommand{\LinkPooledVerbalOtherTwo}{0.11}
\renewcommand{\LinkPooledVerbalQuantTwo}{0.06}
\renewcommand{\LinkWeightedMathOther}{0.125}
\renewcommand{\LinkWeightedMathQuant}{0.316}
\renewcommand{\LinkWeightedDiffOtherLowIndep}{0.001}
\renewcommand{\LinkWeightedDiffOtherHighIndep}{0.035}
\renewcommand{\LinkWeightedDiffOtherLow}{-0.008}
\renewcommand{\LinkWeightedDiffOtherHigh}{+0.043}
\renewcommand{\LinkSubjectCorrMean}{0.08}
\renewcommand{\LinkSubjectCorrMax}{0.47}
\renewcommand{\LinkSkillMedian}{0.086}
\renewcommand{\LinkSkillMedianPct}{8.6}
\renewcommand{\LinkSkillMinPct}{2}
\renewcommand{\LinkSkillMaxPct}{19}
\renewcommand{\LinkSkillWeightedPct}{10.3}
\renewcommand{\LinkPrepAdded}{0.056}
\renewcommand{\LinkPrepAddedLow}{0.044}
\renewcommand{\LinkPrepAddedHigh}{0.065}
\renewcommand{\LinkFamilyAdded}{0.003}
\renewcommand{\LinkDemoAdded}{0.008}
\renewcommand{\LinkPovertyAdded}{0.0004}
\renewcommand{\LinkPovertyAddedSigned}{+0.0004}
\renewcommand{\LinkPovertyAddedLow}{-0.0008}
\renewcommand{\LinkPovertyAddedHigh}{+0.0016}
\renewcommand{\LinkPovertyAddedMin}{-0.004}
\renewcommand{\LinkPovertyAddedMax}{+0.003}
\renewcommand{\LinkPovertyAddedPositive}{17}
\renewcommand{\LinkSchoolAdded}{-0.0015}
\renewcommand{\LinkSchoolAddedLow}{-0.0033}
\renewcommand{\LinkSchoolAddedHigh}{0.0000}
\renewcommand{\LinkSchoolAddedMin}{-0.013}
\renewcommand{\LinkSchoolAddedMax}{+0.003}
\renewcommand{\LinkSchoolAddedPositive}{7}
\renewcommand{\LinkPlanSkillMedianPct}{8.5}
\renewcommand{\LinkPlanPrepAdded}{0.057}
\renewcommand{\LinkPlanPovertyAdded}{0.0004}
\renewcommand{\LinkPlanSchoolAdded}{-0.0022}
\renewcommand{\LinkLaterRidgeMedianPct}{8.35}
\renewcommand{\LinkLaterRidgeMinPct}{1}
\renewcommand{\LinkLaterRidgeMaxPct}{17}
\renewcommand{\LinkLaterLinearMedianPct}{6.9}
\renewcommand{\LinkLaterPovertyAdded}{0.001}
\renewcommand{\LinkLaterSchoolAdded}{-0.004}
\renewcommand{\LinkAucHeldOut}{0.67}
\renewcommand{\LinkAucLater}{0.63}
\renewcommand{\LinkAucHeldOutBaseline}{0.58}
\renewcommand{\LinkAucLaterBaseline}{0.52}
\renewcommand{\LinkForestGainHeldOut}{0.009}
\renewcommand{\LinkForestGainLater}{0.008}
\renewcommand{\LinkHeldOutFallbackPct}{0.03}
\renewcommand{\LinkLaterFallbackPct}{1.9}
\renewcommand{\LinkCapGradeMin}{0.1}
\renewcommand{\LinkCapGradeMax}{0.6}
\renewcommand{\LinkCapSubjects}{18}
\renewcommand{\LinkExamCreditSubjects}{17}
\renewcommand{\LinkMatDualCredit}{-0.17}
\renewcommand{\LinkMatExamCredit}{+0.10}
\renewcommand{\LinkFirstGenGradeMin}{0.07}
\renewcommand{\LinkFirstGenGradeMax}{0.19}
\renewcommand{\LinkFirstGenSubjects}{12}
\renewcommand{\LinkQuantSubjectsWord}{seven}
\renewcommand{\LinkQuantSubjectsCapitalWord}{Seven}
\renewcommand{\LinkPlanQuantSubjectsCapitalWord}{Eight}
\renewcommand{\LinkVerbalAboveMathSubjectsWord}{four}
\renewcommand{\LinkVerbalAboveMathPassingWord}{two}
\renewcommand{\LinkVarBetweenOnePct}{22.2}
\renewcommand{\LinkSubjectsGradeBelowMinimumWord}{one}
\renewcommand{\LinkPlanCohortsWord}{eleven}
\renewcommand{\LinkSnapshotMaxLagWord}{one}
\renewcommand{\LinkSnapshotMaxLagEarlierWord}{five}
\renewcommand{\LinkDualSubstituteMaxLagWord}{two}
\renewcommand{\LinkDualSubstituteLinked}{8{,}180}
\renewcommand{\LinkSensitivityRunsWord}{eleven}
\renewcommand{\LinkSensitivityReachingZeroWord}{two}
\renewcommand{\LinkFoldsWord}{ten}
\renewcommand{\LinkEntryLevelsWord}{five}

\renewcommand{\PersistPopulationN}{8{,}837}
\renewcommand{\PersistLeftN}{2{,}006}
\renewcommand{\PersistLeftPct}{22.7}
\renewcommand{\PersistCapN}{1{,}190}
\renewcommand{\PersistCapLeftPct}{64.8}
\renewcommand{\PersistMatN}{7{,}782}
\renewcommand{\PersistMatDfwN}{3{,}164}
\renewcommand{\PersistMatDfwPct}{40.7}
\renewcommand{\PersistMatLeftWithPct}{34.8}
\renewcommand{\PersistMatLeftWithoutPct}{14.8}
\renewcommand{\PersistMatDiff}{20.0}
\renewcommand{\PersistMatDiffLo}{18.0}
\renewcommand{\PersistMatDiffHi}{21.9}
\renewcommand{\PersistMatOdds}{3.07}
\renewcommand{\PersistMatFailN}{2{,}413}
\renewcommand{\PersistMatFailLeftWithPct}{32.5}
\renewcommand{\PersistMatFailLeftWithoutPct}{14.8}
\renewcommand{\PersistMatFailDiff}{17.7}
\renewcommand{\PersistMatFailDiffLo}{15.5}
\renewcommand{\PersistMatFailDiffHi}{19.8}
\renewcommand{\PersistMatWithdrewN}{751}
\renewcommand{\PersistMatWithdrewLeftPct}{42.1}
\renewcommand{\PersistMatGoneAfterTermPct}{15.1}
\renewcommand{\PersistMatBothN}{2{,}490}
\renewcommand{\PersistMatOnlyN}{674}
\renewcommand{\PersistMatWithElsewhereLeftWithPct}{40.3}
\renewcommand{\PersistMatWithElsewhereLeftWithoutPct}{18.7}
\renewcommand{\PersistMatWithElsewhereDiff}{21.6}
\renewcommand{\PersistMatWithElsewhereDiffLo}{18.9}
\renewcommand{\PersistMatWithElsewhereDiffHi}{24.3}
\renewcommand{\PersistMatAloneLeftWithPct}{14.2}
\renewcommand{\PersistMatAloneLeftWithoutPct}{12.7}
\renewcommand{\PersistMatAloneDiff}{1.6}
\renewcommand{\PersistMatAloneDiffLo}{-1.3}
\renewcommand{\PersistMatAloneDiffHi}{4.5}
\renewcommand{\PersistLeftMatDfwN}{1{,}100}
\renewcommand{\PersistLeftMatDfwPct}{54.8}
\renewcommand{\PersistStayN}{6{,}831}
\renewcommand{\PersistStayMatDfwPct}{30.2}
\renewcommand{\PersistSubjects}{19}
\renewcommand{\PersistSubjectsWord}{nineteen}
\renewcommand{\PersistSecondSubjectDfwN}{1{,}673}
\renewcommand{\PersistDiffMin}{7.8}
\renewcommand{\PersistDiffMax}{50.7}
\renewcommand{\PersistSubjectsAboveMatWord}{fourteen}
\renewcommand{\PersistMinTakers}{400}
\renewcommand{\PersistAnyN}{8{,}823}
\renewcommand{\PersistAnyDfwN}{5{,}267}
\renewcommand{\PersistAnyDfwPct}{59.7}
\renewcommand{\PersistAnyLeftWithPct}{29.5}
\renewcommand{\PersistAnyLeftWithoutPct}{12.7}
\renewcommand{\PersistAnyDiff}{16.8}
\renewcommand{\PersistAnyDiffLo}{15.2}
\renewcommand{\PersistAnyDiffHi}{18.5}
\renewcommand{\PersistMatWholeCohortLeftWithPct}{35.1}
\renewcommand{\PersistMatWholeCohortLeftWithoutPct}{23.9}
\renewcommand{\PersistMatWholeCohortDiff}{11.3}
\renewcommand{\PersistMatWholeCohortDiffLo}{9.3}
\renewcommand{\PersistMatWholeCohortDiffHi}{13.2}
\renewcommand{\PersistOutcomes}{14{,}145}
\renewcommand{\PersistFirstAttempts}{67{,}838}
\renewcommand{\PersistOutcomeRatePct}{20.9}
\renewcommand{\PersistMatOutcomes}{3{,}298}
\renewcommand{\PersistMatOutcomeSharePct}{23.3}
\renewcommand{\PersistSecondSubjectOutcomeSharePct}{12.8}
\renewcommand{\PersistMatOutcomeShareWithRepeatsPct}{25.4}
\renewcommand{\PersistMatOutcomeRatePct}{32.5}
\renewcommand{\PersistMatFirstAttempts}{10{,}138}
\renewcommand{\PersistLeftWithDfwN}{1{,}553}
\renewcommand{\PersistLeftWithDfwPct}{77.4}
\renewcommand{\PersistLeftOutcomes}{5{,}584}
\renewcommand{\PersistLeftMatOutcomes}{1{,}154}
\renewcommand{\PersistLeftMatOutcomeSharePct}{20.7}
\renewcommand{\PersistLeftSecondSubjectOutcomeSharePct}{13.9}
\renewcommand{\PersistLeftWithDfwHoldingMatPct}{70.8}
\renewcommand{\PersistStayWithDfwPct}{54.4}
\renewcommand{\PersistStayMatOutcomeSharePct}{25.0}
\renewcommand{\PersistLeftPctWithNone}{12.7}
\renewcommand{\PersistLeftPctWithOne}{14.6}
\renewcommand{\PersistLeftPctWithTwo}{19.4}
\renewcommand{\PersistLeftPctWithThree}{30.9}
\renewcommand{\PersistLeftPctWithFourPlus}{52.6}
\renewcommand{\PersistFourPlusN}{1{,}542}

\renewcommand{\PopExtractStudents}{92{,}769}
\renewcommand{\PopExtractRecords}{1{,}984{,}991}
\renewcommand{\PopExtractTableStudents}{86{,}165}
\renewcommand{\PopActiveStudents}{68{,}558}
\renewcommand{\PopActiveRecords}{1{,}236{,}347}
\renewcommand{\PopActiveWithNamePct}{98.6}
\renewcommand{\PopActiveWithZipPct}{99.96}
\renewcommand{\PopActiveTexasName}{60{,}045}
\renewcommand{\PopActiveTexasNamePct}{87.6}
\renewcommand{\PopActiveTableStudents}{61{,}954}
\renewcommand{\PopTableShortfall}{6{,}604}

\renewcommand{\ScoreBaselineInstitutions}{1{,}869}
\renewcommand{\ScoreBaselineUndergraduates}{27{,}573}
\renewcommand{\ScoreBaselineCompletion}{0.423}
\renewcommand{\ScoreBaselinePellShare}{0.440}
\renewcommand{\ScoreBaselineExpectedCompletion}{0.478}
\renewcommand{\ScoreBaselineResidual}{-0.055}
\renewcommand{\ScoreCurrentInstitutions}{1{,}814}
\renewcommand{\ScoreCurrentUndergraduates}{30{,}580}
\renewcommand{\ScoreCurrentCompletion}{0.526}
\renewcommand{\ScoreCurrentPellShare}{0.441}
\renewcommand{\ScoreCurrentExpectedCompletion}{0.462}
\renewcommand{\ScoreCurrentResidual}{+0.064}
\renewcommand{\ScoreBaselinePointsBelow}{5.5}
\renewcommand{\ScoreCurrentPointsAbove}{6.4}
\renewcommand{\ScoreLowessFraction}{0.4}
\renewcommand{\ScoreVintagesWord}{two}

\renewcommand{\SizeCap}{45}
\renewcommand{\SizeMinSection}{10}
\renewcommand{\SizeEntryMinStudents}{200}
\renewcommand{\SizeEntryMinSdLogSize}{0.20}
\renewcommand{\SizeBootstrapDraws}{999}
\renewcommand{\SizeFirstAttempts}{177{,}289}
\renewcommand{\SizeEntrants}{23{,}766}
\renewcommand{\SizeSubjectsModeled}{20}
\renewcommand{\SizeSubjectsConsidered}{25}
\renewcommand{\SizeSameStudentInstructors}{758}
\renewcommand{\SizeCalculusMedianSize}{36}
\renewcommand{\SizeCapCalculusPercentile}{90}
\renewcommand{\SizeCalculusSizeBound}{50}
\renewcommand{\SizeLargeSectionBound}{100}
\renewcommand{\SizeCurveGridMax}{500}
\renewcommand{\SizeCurveBin}{5}
\renewcommand{\SizeCurveSmoothing}{10000}
\renewcommand{\SizeMatAllSectionsGrade}{-0.255}
\renewcommand{\SizeMatAllSectionsGradeLow}{-0.373}
\renewcommand{\SizeMatAllSectionsGradeHigh}{-0.138}
\renewcommand{\SizeMatCourseTermGrade}{-0.026}
\renewcommand{\SizeMatPrimaryGrade}{-0.043}
\renewcommand{\SizeMatPrimaryGradeLow}{-0.117}
\renewcommand{\SizeMatPrimaryGradeHigh}{+0.031}
\renewcommand{\SizeMatPrimaryGradeP}{0.26}
\renewcommand{\SizeMatPrimaryDfw}{+1.0}
\renewcommand{\SizeMatPrimaryDfwLow}{-1.3}
\renewcommand{\SizeMatPrimaryDfwHigh}{+3.3}
\renewcommand{\SizeMatSameStudentGrade}{-0.026}
\renewcommand{\SizeMatBalanceMathScore}{+0.007}
\renewcommand{\SizeMatInteractionGrade}{0.046}
\renewcommand{\SizeMatInteractionGradeLow}{0.005}
\renewcommand{\SizeMatInteractionGradeHigh}{0.088}
\renewcommand{\SizeMatInteractionGradeP}{0.027}
\renewcommand{\SizeMatInteractionDfwFall}{2.4}
\renewcommand{\SizeMatInteractionDfwFallLow}{0.8}
\renewcommand{\SizeMatInteractionDfwFallHigh}{4.0}
\renewcommand{\SizeMatInteractionDfwP}{0.007}
\renewcommand{\SizeMatInteractionGradeSigned}{+0.046}
\renewcommand{\SizeMatSlopeLowScoreDfw}{3.3}
\renewcommand{\SizeMatSlopeLowScoreDfwLow}{0.2}
\renewcommand{\SizeMatSlopeLowScoreDfwHigh}{6.4}
\renewcommand{\SizeMatSlopeLowScoreDfwP}{0.038}
\renewcommand{\SizeMatSlopeLowScoreGradeLower}{0.089}
\renewcommand{\SizeMatSlopeLowScoreGradeLow}{-0.183}
\renewcommand{\SizeMatSlopeLowScoreGradeHigh}{+0.005}
\renewcommand{\SizeMatSlopeLowScoreGradeP}{0.063}
\renewcommand{\SizeMatSlopeHighScoreDfw}{-1.5}
\renewcommand{\SizeMatSlopeHighScoreDfwLow}{-3.7}
\renewcommand{\SizeMatSlopeHighScoreDfwHigh}{+0.8}
\renewcommand{\SizeMatSlopeHighScoreGrade}{+0.003}
\renewcommand{\SizeMatSlopeHighScoreGradeLow}{-0.072}
\renewcommand{\SizeMatSlopeHighScoreGradeHigh}{+0.079}
\renewcommand{\SizePooledGrade}{-0.002}
\renewcommand{\SizePooledGradeLow}{-0.026}
\renewcommand{\SizePooledGradeHigh}{+0.022}
\renewcommand{\SizeAisPrimaryGrade}{+0.131}
\renewcommand{\SizeAisSizeMedian}{25}
\renewcommand{\SizeAisSizeNinetieth}{50}
\renewcommand{\SizeMatGradeFirstAttempts}{24{,}666}
\renewcommand{\SizeFewestGradeFirstAttempts}{425}
\renewcommand{\SizeMusPrimaryGrade}{-0.114}
\renewcommand{\SizeSameStudentGrade}{-0.008}
\renewcommand{\SizeSameStudentGradeLow}{-0.031}
\renewcommand{\SizeSameStudentGradeHigh}{+0.015}
\renewcommand{\SizeHisDfwPerCohort}{-65}
\renewcommand{\SizeHisDfwPerCohortLow}{-127}
\renewcommand{\SizeHisDfwPerCohortHigh}{-3}
\renewcommand{\SizeMatDfwPerCohort}{-53}
\renewcommand{\SizeMatDfwPerCohortLow}{-178}
\renewcommand{\SizeMatDfwPerCohortHigh}{+71}
\renewcommand{\SizeCheDfwPerCohort}{+46}
\renewcommand{\SizeCheDfwPerCohortLow}{-35}
\renewcommand{\SizeCheDfwPerCohortHigh}{+127}
\renewcommand{\SizeAisDfwPerCohort}{+43}
\renewcommand{\SizeAisDfwPerCohortLow}{+17}
\renewcommand{\SizeAisDfwPerCohortHigh}{+69}
\renewcommand{\SizeMusDfwPerCohort}{-9}
\renewcommand{\SizeMusDfwPerCohortLow}{-29}
\renewcommand{\SizeMusDfwPerCohortHigh}{+12}
\renewcommand{\SizeMatDfwPerCohortFewer}{53}
\renewcommand{\SizeMatDfwPerCohortLowFewer}{178}
\renewcommand{\SizeMatDfwPerCohortHighMore}{71}
\renewcommand{\SizeMatAffectedPerCohort}{3{,}708}
\renewcommand{\SizeMatScenarioDfw}{-1.4}
\renewcommand{\SizeMatScenarioDfwLow}{-4.8}
\renewcommand{\SizeMatScenarioDfwHigh}{+1.9}
\renewcommand{\SizeMatSectionsPerYear}{246}
\renewcommand{\SizeMatSectionsAddedPerYear}{111}
\renewcommand{\SizeMatSectionsAddedPct}{45}
\renewcommand{\SizeLowThirdDfwFewer}{4.6}
\renewcommand{\SizeLowThirdDfwLow}{-8.9}
\renewcommand{\SizeLowThirdDfwHigh}{-0.4}
\renewcommand{\SizeLowThirdDfwP}{0.034}
\renewcommand{\SizeLowThirdGrade}{0.12}
\renewcommand{\SizeLowThirdGradeLow}{-0.01}
\renewcommand{\SizeLowThirdGradeHigh}{+0.25}
\renewcommand{\SizeLowThirdGradeP}{0.082}
\renewcommand{\SizeLowThirdDfwInvertedLow}{-8.9}
\renewcommand{\SizeLowThirdDfwInvertedHigh}{-0.2}
\renewcommand{\SizeLowThirdDfwClusterLow}{-8.2}
\renewcommand{\SizeLowThirdDfwClusterHigh}{-1.0}
\renewcommand{\SizeMidThirdDfwFewer}{1.6}
\renewcommand{\SizeMidThirdDfwLow}{-5.1}
\renewcommand{\SizeMidThirdDfwHigh}{+2.0}
\renewcommand{\SizeHighThirdDfwMore}{2.0}
\renewcommand{\SizeHighThirdDfwLow}{-1.3}
\renewcommand{\SizeHighThirdDfwHigh}{+5.2}
\renewcommand{\SizeAffectedMedianSize}{104}
\renewcommand{\SizeAffectedMedianSizeAfterSplit}{37}
\renewcommand{\SizeAffectedShareWithSmallSection}{85}
\renewcommand{\SizeAllSectionsPerYear}{1{,}086}
\renewcommand{\SizeAllSectionsAddedPerYear}{1{,}086}
\renewcommand{\SizeAllDfwPerCohort}{+12}
\renewcommand{\SizeAllDfwPerCohortLow}{-297}
\renewcommand{\SizeAllDfwPerCohortHigh}{+321}
\renewcommand{\SizeNextOriginFirstAttempts}{19{,}271}
\renewcommand{\SizeNextContinuers}{7{,}968}
\renewcommand{\SizeNextBothGrades}{6{,}510}
\renewcommand{\SizeNextDifferentInstructor}{5{,}934}
\renewcommand{\SizeNextGrade}{+0.027}
\renewcommand{\SizeNextGradeLow}{-0.077}
\renewcommand{\SizeNextGradeHigh}{+0.131}
\renewcommand{\SizeNextGradeDifferentInstructor}{+0.026}
\renewcommand{\SizeNextGradeDifferentInstructorLow}{-0.083}
\renewcommand{\SizeNextGradeDifferentInstructorHigh}{+0.135}
\renewcommand{\SizeContinuationPerDoubling}{-2.95}
\renewcommand{\SizeContinuationPerDoublingLow}{-4.9}
\renewcommand{\SizeContinuationPerDoublingHigh}{-1.0}
\renewcommand{\SizeBandSmallLow}{11}
\renewcommand{\SizeBandSmallHigh}{30}
\renewcommand{\SizeBandSmallGrade}{+0.055}
\renewcommand{\SizeBandReferenceLow}{31}
\renewcommand{\SizeBandReferenceHigh}{45}
\renewcommand{\SizeBandMiddleLow}{46}
\renewcommand{\SizeBandMiddleHigh}{75}
\renewcommand{\SizeBandMiddleGrade}{+0.007}
\renewcommand{\SizeBandLargeLow}{76}
\renewcommand{\SizeBandLargeHigh}{120}
\renewcommand{\SizeBandLargeGrade}{-0.058}
\renewcommand{\SizeBandLargestGrade}{-0.085}
\renewcommand{\SizeContrastShuffles}{200}
\renewcommand{\SizeContrastWildRejectMin}{3.5}
\renewcommand{\SizeContrastWildRejectMax}{7.5}
\renewcommand{\SizeContrastClusterRejectMin}{4.5}
\renewcommand{\SizeContrastClusterRejectMax}{9.5}
\renewcommand{\SizeDryShuffles}{200}
\renewcommand{\SizeDrySubjects}{5}
\renewcommand{\SizeDryClusterRejectMin}{5.5}
\renewcommand{\SizeDryClusterRejectMax}{19.0}
\renewcommand{\SizeDryWildRejectMin}{2.0}
\renewcommand{\SizeDryWildRejectMax}{6.5}
\renewcommand{\SizeCurveMatGpaSmallest}{2.4}
\renewcommand{\SizeCurveMatGpaReference}{1.8}
\renewcommand{\SizeCurveReferenceSize}{200}
\renewcommand{\SizeCurveMinStudents}{10}
\renewcommand{\SizeSubjectsModeledCapitalWord}{Twenty}
\renewcommand{\SizeSubjectsOutsideWord}{five}
\renewcommand{\SizeNestedModelsCapitalWord}{Four}
\renewcommand{\SizeSubjectsPassingCapitalWord}{One}

\renewcommand{\XwalkWeightContain}{0.55}
\renewcommand{\XwalkWeightJaccard}{0.20}
\renewcommand{\XwalkWeightTrigram}{0.25}
\renewcommand{\XwalkAutoScore}{0.72}
\renewcommand{\XwalkAutoMargin}{0.08}
\renewcommand{\XwalkMinCandidateScore}{0.45}
\renewcommand{\XwalkFirstRunScore}{0.80}
\renewcommand{\XwalkRecoveryScore}{0.90}
\renewcommand{\XwalkNameRowsAll}{1{,}663}
\renewcommand{\XwalkReviewRows}{39}
\renewcommand{\XwalkNameRows}{1{,}583}
\renewcommand{\XwalkStageAutoPct}{88.3}
\renewcommand{\XwalkStageAdjudicatedPct}{89.7}
\renewcommand{\XwalkStageVintagePct}{89.8}
\renewcommand{\XwalkCoveragePct}{90.7}
\renewcommand{\XwalkUsableStudents}{54{,}461}
\renewcommand{\XwalkPrivateNames}{118}
\renewcommand{\XwalkPrivateNamedStudents}{2{,}261}
\renewcommand{\XwalkPrivateNamedPct}{3.8}
\renewcommand{\XwalkRecoveredInDirectoryPct}{96}
\renewcommand{\XwalkMatchedSchools}{1{,}192}
\renewcommand{\XwalkMatchedDistricts}{749}
\renewcommand{\XwalkSchoolsInTablePct}{94.7}
\renewcommand{\XwalkTableCycleSchools}{19{,}961}
\renewcommand{\XwalkTableCycleEqualPct}{100.0}
\renewcommand{\XwalkTableOtherCyclesLow}{2.0}
\renewcommand{\XwalkTableOtherCyclesHigh}{7.7}
\renewcommand{\XwalkMatched}{49{,}099}
\renewcommand{\XwalkOutsideTexas}{6{,}945}
\renewcommand{\XwalkUnresolved}{2{,}749}
\renewcommand{\XwalkPrivate}{2{,}058}
\renewcommand{\XwalkNoSchool}{859}
\renewcommand{\XwalkRejected}{244}
\renewcommand{\XwalkUnresolvedDualPct}{17.1}
\renewcommand{\XwalkMatchedDualPct}{12.6}
\renewcommand{\XwalkDualZ}{6.09}
\renewcommand{\XwalkDualZClustered}{1.40}
\renewcommand{\XwalkUnresolvedApPct}{2.07}
\renewcommand{\XwalkMatchedApPct}{2.87}
\renewcommand{\XwalkApZ}{-2.82}
\renewcommand{\XwalkApZClustered}{-1.51}
\renewcommand{\XwalkUnresolvedPellPct}{48.9}
\renewcommand{\XwalkMatchedPellPct}{46.6}
\renewcommand{\XwalkPellZ}{2.34}
\renewcommand{\XwalkPellZClustered}{0.72}
\renewcommand{\XwalkUnresolvedFirstCoursePct}{32.5}
\renewcommand{\XwalkMatchedFirstCoursePct}{30.6}
\renewcommand{\XwalkFirstCourseZ}{1.71}
\renewcommand{\XwalkFirstCourseZClustered}{1.02}
\renewcommand{\XwalkUnresolvedScore}{549}
\renewcommand{\XwalkMatchedScore}{555}
\renewcommand{\XwalkScoreZ}{-3.46}
\renewcommand{\XwalkScoreZClustered}{-1.05}
\renewcommand{\XwalkMaxZClustered}{1.5}
\renewcommand{\XwalkUnresolvedSchools}{197}
\renewcommand{\XwalkMatchedFirstGenPct}{47.7}
\renewcommand{\XwalkOutsideTexasTransferPct}{71.0}
\renewcommand{\XwalkOutsideTexasApPct}{1.0}
\renewcommand{\XwalkOutsideTexasDualPct}{3.0}
\renewcommand{\XwalkNoSchoolTransferPct}{100.0}
\renewcommand{\XwalkNoSchoolApPct}{0.0}
\renewcommand{\XwalkNoSchoolDualPct}{0.7}
\renewcommand{\XwalkAllTransferPct}{42.8}
\renewcommand{\XwalkAllApPct}{2.5}
\renewcommand{\XwalkAllDualPct}{11.5}
\renewcommand{\XwalkPrivatePellPct}{25.7}
\renewcommand{\XwalkPrivateFirstGenPct}{27.8}
\renewcommand{\XwalkProvisionStudents}{48{,}168}
\renewcommand{\XwalkProvisionSchools}{1{,}106}
\renewcommand{\XwalkNoApBandApPct}{0.6}
\renewcommand{\XwalkNoApBandDualPct}{30.0}
\renewcommand{\XwalkTopBandApPct}{5.1}
\renewcommand{\XwalkTopBandDualPct}{10.3}
\renewcommand{\XwalkTopBandEdgePct}{5}
\renewcommand{\XwalkNoApBandMedianEnrollment}{266}
\renewcommand{\XwalkTopBandMedianEnrollment}{1{,}598}
\renewcommand{\XwalkPovertyStudents}{53{,}367}
\renewcommand{\XwalkPovertyDistricts}{682}
\renewcommand{\XwalkPovertyMedianStudents}{16.3}
\renewcommand{\XwalkPovertyMedianChildren}{21.9}
\renewcommand{\XwalkPovertyMedianGap}{5.6}
\renewcommand{\XwalkLargestBinLow}{15}
\renewcommand{\XwalkLargestBinHigh}{17.5}
\renewcommand{\XwalkLargestBinPct}{36}
\renewcommand{\XwalkNorthsideStudents}{9{,}573}
\renewcommand{\XwalkNorthsidePovertyPct}{16.3}
\renewcommand{\XwalkNorthEastStudents}{6{,}371}
\renewcommand{\XwalkNorthEastPovertyPct}{15.8}
\renewcommand{\XwalkDistanceNinetieth}{21.2}
\renewcommand{\XwalkDistanceNinetyFifth}{208.5}
\renewcommand{\XwalkFarMiles}{100}
\renewcommand{\XwalkFarLocalHomePct}{63.6}
\renewcommand{\XwalkPrimaryMiles}{10}
\renewcommand{\XwalkStrictMiles}{5}
\renewcommand{\XwalkWithinPrimaryPct}{80.0}
\renewcommand{\XwalkPointWithinPrimaryPct}{83.5}
\renewcommand{\XwalkZipZeroPct}{44.4}
\renewcommand{\XwalkPointZeroPct}{0.3}
\renewcommand{\XwalkMisreadZipStudents}{76}
\renewcommand{\XwalkBestTractSharePct}{22}
\renewcommand{\XwalkTractsPerZip}{9.9}
\renewcommand{\XwalkBestTractShareAllZips}{0.742}
\renewcommand{\XwalkContextCoveragePct}{97.7}

}{}
\maketitle

\begin{abstract}
\noindent
The San Antonio metropolitan area has one of the highest poverty rates among large U.S. metropolitan areas. Within Bexar County, where the city is located, child poverty varies widely across school districts.
We link entrants of the city's public research university, a broad-access Hispanic-Serving Institution, to their sending schools and districts, and examine their preparation scores and first-year outcomes.
Across the nation's school districts, enrollment in widely taken core courses (Algebra~II, Geometry, Biology) shows little relation to child poverty, while enrollment in Calculus and Advanced Placement mathematics and science is far lower where child poverty is higher, nationally and within Bexar County.
Among linked entrants, mathematics preparation differs with the school and district.
Entrants with stronger mathematics preparation earn higher first-year grades in nearly every subject.
Most entrants who do not return after the first year had unsuccessful courses, and mathematics is the largest single source of those outcomes.
Among students who change major, mathematics is one of the subjects with grades furthest below their own averages.
Within the same course and term, larger mathematics sections go with more unsuccessful outcomes for entrants with lower mathematics preparation, which identifies a hypothesis for testing smaller sections.
Recorded between Fall 2014 and Fall 2019, these outcomes offer a reference point for change after the pandemic, test-optional admissions, and generative artificial intelligence.
The analysis is reproducible; adapting it elsewhere requires local definitions and linkage validation.
All results are associations among the students of one institution; establishing causes would require a designed study.
\end{abstract}

\medskip
\noindent\textbf{Keywords:} gateway mathematics; educational equity;
child poverty; record linkage; Hispanic-Serving Institution;
reproducible research

\medskip
\noindent\textbf{Mathematics Subject Classification:} 62P25 (primary);
62H30; 62J05; 97B40

\section{Introduction}\label{sec:intro}

San Antonio, Texas, is the seventh-largest city in the United States
by the July 2025 population estimate \citep{CensusVintage2025}, and
its metropolitan area's poverty rate, 14.2\% in 2022, is among the
highest of the twenty-five largest metropolitan areas, second by point
estimate and statistically indistinguishable from Houston (14.3\%) and
Detroit (13.8\%) \citep{CensusPoverty2022}. The city also occupies a
specific place in the history of American school finance. Texas funds
public schools substantially through local property taxes; the
constitutional challenge to that arrangement, \textit{San Antonio
Independent School District v.\ Rodriguez}, originated in the city's
Edgewood district. The Supreme Court held that education is not a
fundamental right under the federal Constitution and upheld the
finance system under rational-basis review \citep{Rodriguez1973}.
Fifty years later, the school districts of Bexar County still span
a broad range of child poverty, from Alamo Heights
(\CalcAlamoPovertyPct{}\%) to Edgewood (\CalcEdgewoodPovertyPct{}\%) (Section~\ref{sec:context:calculus}). Higher
educational attainment is strongly associated with higher lifetime
earnings and lower unemployment
\citep{cheah2021college,BLSEducationPays}. Students' later earnings
and measured income mobility also differ across the colleges they
attend, and those differences reflect both the students each
institution enrolls and the education it provides
\citep{chetty2020mobility}. The equity question posed by this
geography is therefore concrete: how the schools and districts that
send students to the university relate to the mathematics preparation
those students arrive with, and how that preparation relates to their
first-year performance.

We study this question at The University of Texas at San Antonio (UT
San Antonio), the city's public research university. Three facts
describe the institution today, after the period of the records
analyzed here. UT San Antonio holds the Carnegie R1 research
designation in the 2025 classification \citep{Carnegie2025}, and it
merged with UT Health San Antonio in September 2025
\citep{UTSAMerger2025}. Its access is relatively broad: it accepted
87\% of first-time undergraduate applicants in Fall 2024
\citep{UTSA2024Admissions}. It is a federally designated
Hispanic-Serving Institution whose Fall 2024 enrollment was 60\%
Hispanic and 43\% first-generation \citep{UTSA2024FastFacts}. Its
broad access and its Hispanic-Serving designation make the
institution a relevant setting for examining how recorded school and
district context relates to preparation and first-year performance.

We ask three questions. First, among students who enroll at this
university, how are family-resource markers, sending-school course
enrollment, and district child poverty associated with mathematics
preparation at entry? Second, how are mathematics and verbal
preparation associated with grades in the courses those students take
during their first year? Third, within the institution, how do
first-year course outcomes relate to persistence and to changes of
major, and how does section size relate to grades? National
district-level patterns of
course enrollment provide the context for the first question.

The relationships motivating these questions have substantial
precedents. National evidence already shows that high-poverty schools
offer fewer advanced courses, while basic course offerings are more
widespread \citep{gao2018offerings}. Student records from Florida and
Texas and the CRDC show related unevenness, by poverty and by race,
in who takes advanced courses
\citep{conger2009explaining,klopfenstein2004advanced,xu2021acceleration},
and rigorous course-taking is associated with high-school graduation
and enrollment in four-year colleges \citep{long2012effects}. Linked
studies at the University
of California, San Diego and at Texas public universities have
related family background and sending-high-school characteristics to
college achievement \citep{betts1999determinants,fletcher2010highschool,
black2015highschool}. One of them includes this university. With
records of four Texas public universities, UT San Antonio among them,
for entrants of the 1990s and early 2000s,
\citet{fletcher2010highschool} find that comparing students who
attended the same high school removes most of the Black-White and
Hispanic-White gaps in college grades. Related evidence comes from a
national survey of freshmen and from the graduates of one large
school district \citep{wolniak2010academic,allensworth2020high}. A
statewide study across fifteen public
universities examined how high-school grades, test scores, and
curriculum rigor, measured both directly and relative to the sending
school, relate to college outcomes \citep{bastedo2023contextualized}.
These studies establish that the school context of college entrants
matters. Our contribution is to measure particular relationships
within that context. Our entrants enrolled more than a decade after
those of \citet{fletcher2010highschool}, and we describe the sending
context with measured quantities, district child poverty and recorded
course enrollment at the school, where that study uses an indicator
for each high school.

Course-level analysis also has precedents. Earlier research compared
the predictive contributions of mathematics and verbal scores across
college courses and documented differences in course selection and
grading \citep{ramist1994coursegrades}. Research on admissions-test
validity further shows why predictive associations must be interpreted
in light of selection and background characteristics
\citep{rothstein2004predictions}. We extend this work by examining
first attempts in \LinkSubjects{} subjects under a common specification, comparing
students within the same course and term, and evaluating the
additional predictive contribution of district poverty and measured
school course enrollment when whole sending schools are held out.
The record lacks high-school grades and individual high-school
transcripts, so this comparison does not establish the best available
admissions or placement model.

The third question has precedents as well. Students leave college as
they learn about their own academic performance
\citep{stinebrickner2012learning}. Low grades predict changes
of major \citep{astornefigari2019changes}, and students move away
from science as they learn about their own performance in it
\citep{stinebrickner2014major}. University studies report lower
grades in larger classes \citep{kokkelenberg2008effects}. In
mathematics that
relation has been found to be stronger for lower-ability students
\citep{depaola2013class}, and negative effects of class size have
been reported for Hispanic students at a Hispanic-Serving Institution
\citep{matta2015class}. We describe these relations subject by
subject at one institution. We report how often entrants with an
unsuccessful first-year course are observed enrolled at least one
year after entry, how much each introductory subject contributes to
the unsuccessful outcomes of the entrants who are not, and how much
the relation between a first-year grade and a recorded change of
major depends on which entrants remain enrolled. Students who stopped
out and returned before the observation window closed count as
observed later.

Earlier studies have examined these relations one or two at a time: the
course enrollment of high-poverty schools, the school context of
college achievement, the reach of entry scores across college
courses, grades and persistence, grades and changes of major, and
class size. To our knowledge, this is the first study to examine all
of them together, in depth, for the students of one large
institution. The institutional analyses use one source extract, which
holds the records of the \PopActiveStudents{} students enrolled between Fall
2014 and Fall 2019. Each analysis uses a separately defined sample of
those students: linked entrants for the models of preparation and
grades, a fixed reference cohort for progression, persistence, and
major change, and all students active in the window for the other
institutional descriptions (Table~\ref{tab:populations}).
Working inside one institution is what makes this depth possible: the
record holds every course attempt of every student, and every number
in the paper is regenerated by its programs. The same choice sets the
scope of the claims. The institutional statistics are associations
among the students who applied to, were admitted by, and chose this
institution; statements about Texas or the nation would need other
data. Appendix~\ref{sec:results:position} places the institution on
the national completion landscape so that other institutions can
position themselves relative to these findings.

The first motivation for this study is geographic. Students arrive
at the university from schools and districts that differ in course
enrollment and in child poverty, so a study of preparation at entry
has to measure the schools and districts that send students. In the
2020--21
Civil Rights Data Collection (CRDC), 39\% of the roughly 31{,}100
public middle schools offering grade 7 or 8 did not offer Algebra~I in
that collection cycle \citep{CRDC2021}. In Texas, House Bill 5 of 2013 replaced the
default curriculum with the Foundation High School Program, under
which Algebra~II is required for the STEM endorsement and for the
Distinguished Level of Achievement and is otherwise optional
\citep{TexasHB5_2013,TEAMathFAQ}. National transcript evidence
identifies the highest level of mathematics reached in high school as
a key marker of momentum toward a bachelor's degree
\citep[pp.~30--33]{adelman2006toolbox}. Geographic access to colleges
themselves also varies by place \citep{hillman2016geography}.

The second motivation for this study is temporal: later comparisons
need a reference state measured before the period 2020--2025. Between
2020 and 2025, several
layers of this system changed at once: COVID-19 brought instructional
disruption and emergency grading policies, a large share of admissions
became test-optional, Texas restructured community-college finance
under House Bill 8 \citep{TexasHB8_2023}, and generative artificial
intelligence diffused into routine coursework, with 86\% of surveyed
higher-education students reporting its use in their studies by 2024
\citep{DEC2024StudentSurvey} and 62\% of U.S. students in middle
school and above using it for homework by December 2025
\citep{RAND2026AIHomework}. These changes overlap in time and in
population, so institutional records alone cannot attribute
subsequent movement in outcomes to any one of them. What such records
can support is comparison against a reference state whose outcomes
closed before the changes began. We construct that reference state
here: every measurement of the reference cohort, the first-time
freshmen who entered between Fall 2014 and Fall 2015, closed before
Fall 2020 (Section~\ref{sec:data:cohort}).

The paper proceeds as follows. Section~\ref{sec:data} describes the
institutional database, defines the reference cohort and its
observation window, builds the geography linkage, and defines the
linked entrant models and the section-size models.
Section~\ref{sec:results} opens with the five findings of the paper
and then reports the district course gradients
(Sections~\ref{sec:context:calculus} and~\ref{sec:context:courses}),
a summary of the linked models of preparation and first-year
performance (Section~\ref{sec:results:summary}), and the
institutional results on persistence and major change, relative
subject grading, and section size
(Sections~\ref{sec:baseline:flows} to~\ref{sec:baseline:classsize}).
Section~\ref{sec:discussion} discusses what the findings show
together, their implications and limitations, and a protocol for
post-2020 comparison. Four appendices support the body: the
quantitative methods, which open with the record of rigor and
reproducibility (Appendix~\ref{sec:data:quant}); the additional
results, with the linked models and the section-size models in full
(Appendix~\ref{app:results}); the data inventory, linkage validation,
and governance (Appendix~\ref{app:data}); and the angular
index of the majors in Figure~\ref{figChordShort}
(Appendix~\ref{app:chords}).

\section{Data and Methods}\label{sec:data}

\subsection{The institutional database}\label{sec:data:oir}

The institutional record is an extract of the university's student
information system at course-section grain: one row per student, term,
and course section (\PopExtractRecords{} rows for \PopExtractStudents{}
students), with terms from Fall 2009 through Fall 2019. The
institutional analyses of this paper use the \PopActiveStudents{}
students who hold at least one course record between Fall 2014 and
Fall 2019 (\PopActiveRecords{} rows in those terms); we call them the
students active in the window. A companion per-student table holds
pre-college mathematics credit for \PopActiveTableStudents{} of them.
Coverage of the linkage fields is high: \PopActiveWithNamePct{}\% of
the active students record a sending high-school name and
\PopActiveWithZipPct{}\% a home ZIP code
(Appendix~\ref{app:data:sources}). Every external
dataset joined to this record enters a
provenance-layered PostgreSQL database whose raw layer stores source
files verbatim with SHA-256 hashes; its source registry records
dataset, vintage, path, hash, byte count, acquisition URL, and row
count for each ingested file. The archives acquired for this paper
additionally hold full manifest rows recording acquisition, license,
schema, and lineage detail (Appendix~\ref{app:data:sources}).
Provenance is established by hash and registry entry rather than by
filename or row count. The institutional review board of the study institution approved this
research, and the institutional extract itself is restricted (see the
Ethics approval and Data availability statements).
Appendix~\ref{app:data} holds the full inventory
(Table~\ref{tab:sources}), the source-flaw catalog, and the
inferential conventions.

\subsection{The reference cohort and observation
  window}\label{sec:data:cohort}

The analyses use different populations because they answer different
questions; Table~\ref{tab:populations} in Appendix~\ref{app:data}
lists each with its window and unit. The \emph{reference
cohort} comprises the $N=\InstCohortN$ first-time freshmen who entered
between Fall 2014 and Fall 2015: \InstCohortFirstFallN{} in Fall 2014,
\InstCohortSecondFallN{} in Fall 2015, and \InstCohortSpringSummerN{} in the Spring or
Summer of 2015. A first-time freshman is a student whose first
recorded term bears the registrar's first-time-freshman type code. We
observe each student from the entry term through the same term
\InstHorizonYearsWord{} years later, inclusive. A Fall 2014 entrant is thus
observed through Fall 2018, a Fall 2015 entrant through Fall 2019,
and a Spring or Summer 2015 entrant through the Spring or Summer term
of 2019, so all cohort measurements closed before Fall 2020. For a
fall entrant, the window holds the first \InstHorizonYearsWord{} academic years
and the following fall term. Every reported cohort quantity uses this
inclusive boundary, and the comparison protocol of
Section~\ref{sec:discussion:protocol} binds post-2020 reruns to it.
The strict \InstHorizonYearsWord{}-year variant is reported as a sensitivity
check.

The progression outcome is an enrollment record in a course numbered
strictly above 4000, that is, senior-level coursework, within the
window. The gateway subjects are mathematics (registrar
subject code MAT), writing composition (WRC), and the
required first-year seminar, Academic Inquiry and Scholarship (AIS).
In the cohort tables, exposure to a subject means any course under
its code at any point in the window; a first-year restriction applies
only in the linked entrant models of Section~\ref{sec:data:link}.
The cohort includes \InstCapN{} students flagged for the Coordinated
Admission Program (CAP), who plan to transfer after the first year:
\InstCapSeniorPct{}\% of them reach senior-level coursework at the institution,
against \InstNoCapSeniorPct{}\% of the remaining \InstNoCapN{}. Cohort quantities are
reported for the full cohort; the mathematics odds ratio is \InstNoCapOddsMat{}
among students without the flag.

\paragraph{Window sensitivity and legacy endpoint rules.}
The window sensitivity was recomputed read-only against the source
system. Restricting to the strict \InstHorizonYearsWord{} academic years
lowers the share reaching the senior level from \InstSeniorPct{}\% to \InstSeniorStrictPct{}\%
(\InstSeniorStrictN{} of \InstCohortN{}) and moves the gateway odds ratios from \InstOddsMat{} to
\InstOddsMatStrict{} (MAT), \InstOddsWrc{} to \InstOddsWrcStrict{} (WRC), and \InstOddsAis{} to \InstOddsAisStrict{} (AIS). Boundary
cases are reported so that the estimand is stated as implemented.
\InstExactBoundaryCourseNText{} cohort students touch a course numbered exactly 4000 and are
excluded by the strict inequality, \InstGraduateNumberedN{} touch graduate-numbered courses
(5000 and above) and are counted, and \InstPreEntryRecordsN{} have course records dated
before their entry term, which the queries admit because they impose
no lower bound. The major-change flag compares the first and last
recorded major within the window: \InstMissingMajorN{} cohort students lack one of the
two and the published \InstMajorChangePct{}\% counts them as changes (counting only
confirmed changes gives \InstConfirmedChangePct{}\% on the full cohort and \InstConfirmedChangeCompletePct{}\% among
the \InstMajorCompleteN{} complete cases). The underlying query has no tie-break
within a term; no student of the cohort holds two majors in the
first or the last term of the window, so the flag does not depend on
that order. These are the rules of the cohort tables as first built,
kept so that the published quantities stay comparable, and the
cohort flags should be read as defined by them.
Appendix~\ref{app:results:majorchange} repeats one comparison under
an entry-bounded rule that counts confirmed changes only. Each of
these counts is small against $N=\InstCohortN$; the window-boundary and odds-ratio counts are
regenerated as a machine-readable artifact
(\texttt{cohort\_window\_audit.json}) alongside the tables.

Institutional descriptives may use the \emph{full window} (everyone
enrolled Fall 2014--Fall 2019, entrants and transfers alike); every
figure states which population it describes. The linked models of
preparation use first-time fall entrants of Fall 2014 through Fall
2019. The models of first-year grades and the section-size models use
the entrants of Fall 2014 through Fall 2018, whose first year is
complete in the record (Section~\ref{sec:data:link}). Every
institutional population is selected in scope, per the scope
statement of Section~\ref{sec:intro}.

\subsection{The geography linkage chain}\label{sec:data:linkage}

Contextual measurements reach a student through an audited chain with
two branches. In the district branch, we first match each student's
recorded high-school name to a National Center for Education
Statistics (NCES) school identifier. Candidate schools are compared
within geographic blocks defined by city and ZIP code;
Appendix~\ref{sec:data:quant} describes the matching score. The first
seven digits of the assigned school identifier identify the district
(its local education agency identifier, LEAID). We use the district
identifier to attach the Census Bureau's Small Area Income and
Poverty Estimates (SAIPE) of child poverty, and the school identifier
to attach the course-enrollment measures of the CRDC. These joins on
identifiers are exact. Whether the school was assigned correctly in
the first step is a separate question of validation.
The neighborhood branch resolves the home ZIP to ZIP Code Tabulation
Area (ZCTA) level Opportunity Atlas context
\citep{chetty2018opportunity}, aggregated up from
census tracts with population weights; the resulting variable is
ecological and is flagged as such in the schema. The name-resolution
crosswalk reaches \XwalkCoveragePct{}\% of the \PopActiveTexasName{}
active students who record a Texas high-school name. The per-student
table holds \PopActiveTableStudents{} of the active students, and
their final disposition is \XwalkMatched{} matched,
\XwalkOutsideTexas{} with a high school outside Texas,
\XwalkUnresolved{} unresolved, \XwalkPrivate{} in private schools
outside the NCES universe, \XwalkNoSchool{} with no school recorded,
and \XwalkRejected{} rejected on review (Table~\ref{tab:linkage}; the
per-student table lacks \PopTableShortfall{} of the active students,
Appendix~\ref{app:data:sources}). In the sample, the unresolved
students hold dual-enrollment credit and Pell grants more often than
the matched students. With sending schools as clusters these
differences are within the variation between schools, so the audit
leaves open whether linkage is selective on these markers. The
district poverty of the unresolved students is
unobserved, and the direction of any resulting bias in the
district-linked results is unknown; the selection audits are in
Appendix~\ref{app:data:selection}. Agreement of the assigned
identifiers with the national school universe shows that they are
real schools; the rate at which the correct school was chosen among
similarly named candidates has not been measured against an
adjudicated sample. Two scope rules govern every use
of the chain: a CRDC snapshot is a cycle-specific school report and
is never treated as a student exposure history, and a school point is
never treated as a catchment area.

\subsection{The linked entrant models}\label{sec:data:link}

The linked models ask two questions of one population: which entry
characteristics are associated with mathematics preparation at entry,
and which are associated with first-year performance in each subject.
The analysis plan (variables, outcome rules, models, hypotheses, and
the thresholds for calling a hypothesis supported) was written and
frozen before any model was estimated, and the pipeline was first run
end to end on shuffled outcomes. The prespecification is internal: the
project holds the plan with a SHA-256 record of the plan and the code,
and no external registry timestamps it. Changes made after the freeze
are recorded in an append-only list of deviations, and each is named
where its result is reported. The plan accompanies the code. This
subsection states the population, the measures, and the comparisons;
Appendix~\ref{app:methods:link} gives the full specification.

\paragraph{Population.}
All first-time fall entrants of Fall 2014 through Fall 2019
(\LinkEntrants{} students), the window of the institutional descriptives. The
\emph{linked sample} holds the \LinkSample{} whose sending school resolves
in the crosswalk and whose school and district context is complete;
\LinkSampleScored{} of them have both entry scores. Performance models use the
cohorts through Fall 2018 (\LinkSampleGradeCohorts{} linked entrants), whose first-year
window is complete. CAP entrants are retained, and the flag enters
every model. The plan was first written for entrants of Fall 2009
through Fall 2019 and was first run on those \LinkPlanCohortsWord{} cohorts. The
restriction to Fall 2014 onward was adopted after that run, with its
results known, because institutional conditions in the earlier
cohorts differ. It is recorded as a deviation in the plan (D1), and
nothing else in the plan changed with it. This paper reports only
students enrolled between Fall 2014 and Fall 2019, so the results of
the earlier cohorts are not given. The restricted analysis is
therefore a choice of population made after the plan was frozen, and
the tables and figures that report it say so.

\paragraph{Preparation and performance.}
Preparation at entry is measured by the entry mathematics score: the
institution's recorded highest mathematics section score across the
SAT and the ACT, on the current SAT scale. Within each entry cohort we
subtract the cohort mean and divide by the cohort's standard
deviation, so one unit is one standard deviation of the score (\LinkScoreSD{}
points when the cohorts are pooled). The verbal score is standardized
in the same way. A second measure, the standardized mathematics score
minus the standardized verbal score, describes an entrant's relative
position on the two tests. First-year performance is measured on
first attempts in lower-division courses during the entry fall and
the following spring and summer, in two forms: the grade on the
4-point scale, and a DFW indicator (a D, F, withdrawal, or no-credit
outcome). The record holds no high-school grade average, class rank,
or transcript.

\paragraph{Predictors and models.}
The predictors describe four things: family resources (whether the
student applied for aid, the income rank of applicants, and
first-generation status), demographics (gender and ethnicity), the
child poverty rate of the sending district, and recorded course
enrollment at the sending school (Calculus, AP mathematics, AP
science, and dual enrollment per student). The
\emph{advanced-mathematics index} of a school is the mean of its
standardized Calculus and AP mathematics rates. Three kinds of model
use these predictors. A model with random intercepts for districts
and for schools shows how much of the variation in preparation lies
between districts, between schools, and among entrants from the same
school. For each of the \LinkSubjects{} subjects in which at least \LinkSubjectMinimum{} linked
entrants hold a first attempt, a regression with course-by-term fixed
effects relates grades to the two entry scores and the other
characteristics, so students are compared with other students taking
the same course in the same term. Finally, predictions for sending
schools held out of the fit measure how much each group of predictors
adds to accuracy.

Two hypotheses were fixed in advance. H1: the school's
advanced-mathematics index is positively associated with the
mathematics-minus-verbal measure after family resources. H2: the
mathematics score is positively associated with first-year grades
outside the quantitative subjects after the verbal score. \LinkPlanQuantSubjectsCapitalWord{}
subjects were classed as quantitative before estimation, and
\LinkQuantSubjectsWord{} of them have enough entrants to be modeled: mathematics,
chemistry, economics, statistics, physics, computer science, and
accounting. Every estimate is an association conditional on entry to
the institution.

\paragraph{Section-size models.}
A prespecified addendum to the plan, frozen before any within-course
model was fitted to real grades, asks whether entrants in larger
sections of the same course earn lower grades. Rows are the first
attempts of all first-time fall entrants of Fall 2014 through Fall
2018 in face-to-face lecture sections of more than \SizeMinSection{} students
(\SizeFirstAttempts{} first attempts, \SizeEntrants{} entrants). Section size counts
every student enrolled in the section, whatever the entry cohort. The
models use the base-2 logarithm of section size, so a coefficient is
the difference that goes with a doubling of section size.

\SizeNestedModelsCapitalWord{} nested models tighten the comparison in order: all sections of
the subject; course-by-term fixed effects; the student covariates of
the performance models; and instructor fixed effects, which complete
the primary model. The primary model therefore compares entrants who
take the same course in the same term, with their measured
characteristics held equal and average differences between
instructors removed. It leaves in place any difference specific to an
instructor in one course or one term, and it leaves section
assignment as the administrative process it is. A fifth model adds
the interaction of section size with the entry mathematics score, and
a pooled model with student fixed effects compares each student with
the same student in other courses. \SizeSubjectsModeledCapitalWord{} of the \SizeSubjectsConsidered{} subjects have
enough students and enough variation in section size to enter; the
entry rule is in Appendix~\ref{app:methods:classsize}. Two hypotheses
were fixed in advance for mathematics. H3: larger sections go with
lower grades in the primary model. H4: that association is stronger
for entrants with lower entry mathematics scores.

Dry runs on shuffled grades showed that the usual test with
instructor-clustered standard errors rejects too often in this
design. The intervals and $p$-values of the section-size terms
therefore come from a wild cluster bootstrap over instructors
\citep{cameron2008bootstrap}; the same-student model, with \SizeSameStudentInstructors{}
instructors, keeps clustered intervals
(Appendix~\ref{app:methods:classsize}). A scenario then splits every
section above \SizeCap{} students into equal sections of \SizeCap{} or fewer and
applies the primary coefficient of each subject. The value \SizeCap{} is a
scenario threshold: the record holds no enrollment-cap field, and \SizeCap{}
is the \SizeCapCalculusPercentile{}th percentile of the observed sizes of the Calculus~I and II
lecture sections (median \SizeCalculusMedianSize{}, none above \SizeCalculusSizeBound{}). The scenario describes
an intervention only under three conditions that the observational
models do not establish: that section size causes the estimated
differences, that the association is linear in the logarithm of size
over the range of the split, and that added sections would be taught
as existing ones are.

\subsection{Quantitative methods}\label{sec:data:methods}

One estimator is needed to read
Figures~\ref{figCalculusPanels} and~\ref{figCourseGradients}. The
\emph{district course gradient} starts
from the enrollment rate of a course in a district: enrollees divided
by estimated enrollment in the grades of the course. For a
high-school course these are grades 9--12, and for Algebra~I taken in
grades 7--8 they are those two grades. The denominators describe the
size of the school population in the relevant grades and leave
unmeasured which students had met the prerequisites of a course. The
gradient is
the least-squares slope of that rate on the district child poverty
rate, weighted by the same enrollment. Because courses differ widely
in their base rates, each slope is reported on a relative scale: the
difference in the rate that goes with ten percentage points of child
poverty, as a percentage of the national mean of the course.
Intervals are 95\% intervals from a bootstrap that resamples states
(\GradBootstrapResamples{} resamples).

The grade deviation of Figure~\ref{figHeatmap} is defined where it is
used (Section~\ref{sec:baseline:barriers}), and the section-size
coefficient is defined with its models
(Section~\ref{sec:data:link}).
Appendix~\ref{sec:data:quant} states every estimator in full,
including the record-linkage score, the layout of the major-change
diagram, the smoothing of the class-size curves, and the
specification and inference of the linked entrant models and the
section-size models. The implementing scripts are named in the Code
Availability statement.

\section{Results}\label{sec:results}

The results support five findings. Each is stated here with the
subsection that reports it.

\begin{enumerate}
\item \textbf{Course enrollment and district poverty}
  (Sections~\ref{sec:context:calculus}
  and~\ref{sec:context:courses}).  Districts with higher child 
  poverty have much lower enrollment rates in Calculus and 
  AP mathematics and science. In contrast, enrollment rates 
  in Algebra II, Geometry, and Biology differ little with 
  district poverty. These patterns appear nationally \citep{gao2018offerings}
  and among Bexar County districts. Rates account for the 
  estimated number of students in the relevant grades. 
  They measure how many students enroll, rather than whether 
  schools offer the courses.
\item \textbf{Preparation and grades across the curriculum}
  (Section~\ref{sec:results:summary}). After accounting for verbal 
  scores and the other included characteristics, higher mathematics 
  scores are associated with higher grades among students taking 
  the same course in the same term. The coefficient is
  larger within the quantitative subjects. Outside them the data do
  not establish that the mathematics score has a larger association
  than the verbal score.
\item \textbf{Context and prediction}
  (Section~\ref{sec:results:summary}). Entrants mathematics 
  scores differ with district poverty and advanced-mathematics 
  enrollment at their sending schools. However, once entry 
  scores are included, adding these contextual measures does 
  little to improve predictions of college grades. This 
  remains true when predictions are tested on students from 
  schools excluded from model training. The finding does 
  not mean that school context is unimportant or explain 
  what caused the differences in preparation.
\item \textbf{Unsuccessful first-year outcomes and persistence}
  (Section~\ref{sec:baseline:flows}). Most entrants with no 
  enrollment record a year or more after entry had received a 
  D, F, withdrawal, or no-credit outcome during their first 
  year. Mathematics accounted for more of these outcomes 
  than any other subject. In every introductory subject 
  examined, entrants with an unsuccessful outcome were less 
  likely to be observed enrolled later. This association 
  was concentrated among entrants with unsuccessful outcomes 
  in several subjects. These comparisons do not establish 
  that the course outcomes caused students to leave.
\item \textbf{Section size} (Section~\ref{sec:baseline:classsize}).
  Earlier university studies examined class-size relationships that
  differ among students and comparisons within courses and
  instructors \citep{bandiera2010classsize,karas2021classsize}. Here,
  within the same course and term, a small average association of
  section size with mathematics grades coexists with an interaction
  with preparation: larger sections go with more DFW outcomes at
  lower entry mathematics scores, the direction that
  \citet{depaola2013class} report for mathematics at an Italian
  university. The finding motivates a designed evaluation at this
  institution and does not establish an effect of reducing section
  size.
\end{enumerate}

Sections~\ref{sec:baseline:flows} and~\ref{sec:baseline:barriers}
also describe changes of major among the students who take
mathematics. The audited record linkage
(Section~\ref{sec:data:linkage}; Appendix~\ref{app:data}) and the
pre-2020 reference quantities (Appendix~\ref{sec:declaration})
support these findings and later replication: they allow populations
and outcomes to be compared under stated definitions, with linkage
failures and selection limits reported alongside the results.

The district analyses use published aggregates, and districts are
their units. Student records enter a model in two places, the linked
entrant models (Appendix~\ref{sec:link}) and the section-size models
(Section~\ref{sec:baseline:classsize}), under the rules of
Section~\ref{sec:data:link}. Every result is an association, and
nothing in this section supports a causal reading.

\subsection{District Calculus enrollment, child poverty, and
  income}\label{sec:context:calculus}

\begin{figure}[tbp]
  \centering
  \includegraphics[width=\textwidth]{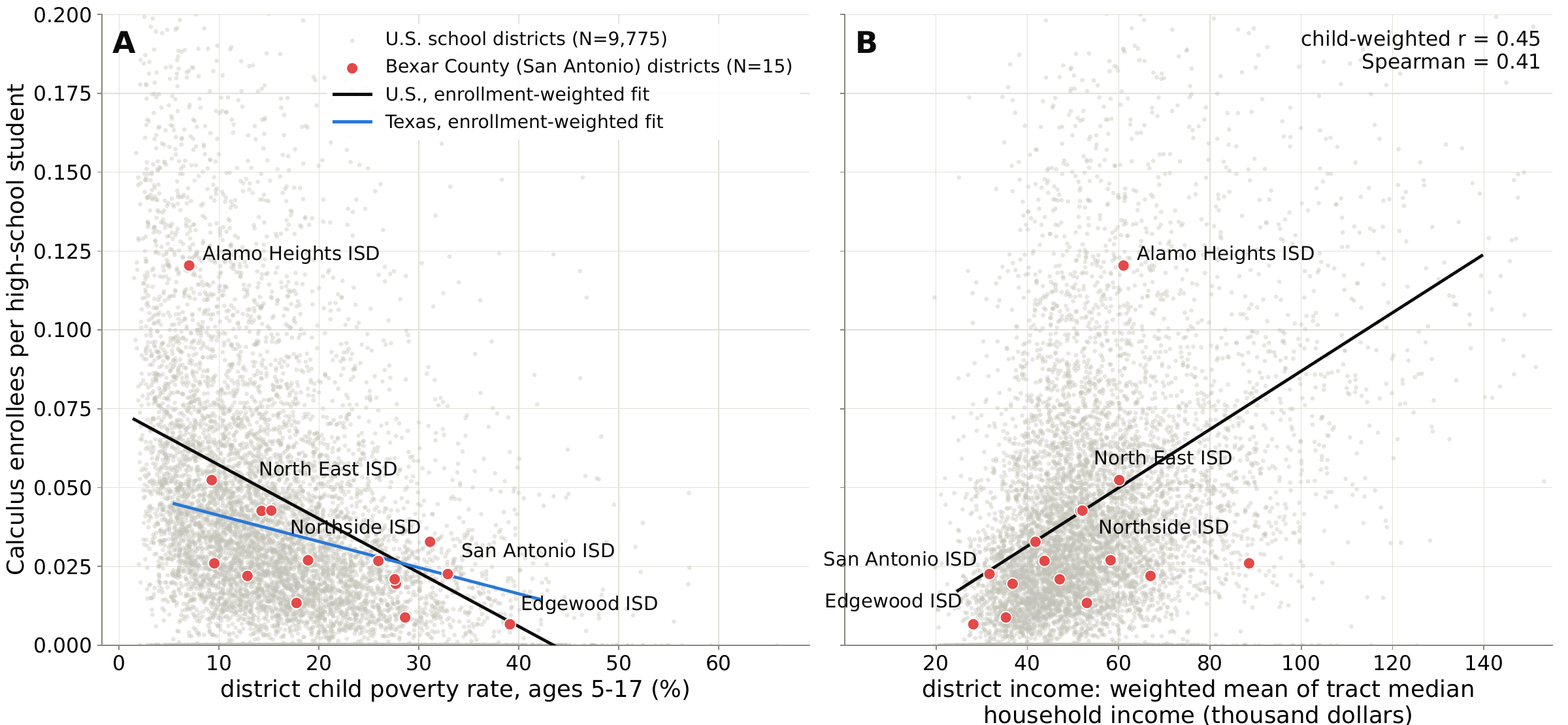}
  \caption{District Calculus enrollment per high-school
    student (CRDC 2017--18) against (A) the district child poverty
    rate (SAIPE 2017) and (B) district income, the weighted mean of
    tract median household incomes (ACS 2012--16, the tract
    covariate distributed with the Opportunity Atlas; tracts
    weighted by childhood-cohort count times land-area share), for
    the 9,790 districts of the
    common frame (matched, at least 200 children aged 5--17,
    a high-school course sequence reported, no missing-coded
    course item). The denominator is estimated high-school
    enrollment: total enrollment of the schools serving grades
    9--12, prorated by grade span. (A) The black line is the U.S.
    enrollment-weighted least-squares fit, slope $-0.171$ per
    unit of the poverty rate (state-clustered standard error
    0.020), which is $-36.9\%$ of the national mean
    (0.046) per ten percentage points of child poverty; the
    blue line is
    the same fit within Texas (860 districts). (B) The
    9,775 districts with an income value; the
    child-weighted correlation is 0.45 (Spearman 0.41).
    Red: the school districts of Bexar County, Texas (15 in the
    frame, landmark districts labeled).
    2,321 districts report no Calculus enrollment, and 112
    above the displayed range are not shown. Districts are the
    units; enrollment counts seats within one cycle, with no
    completion or outcome content.}
  \label{figCalculusPanels}
\end{figure}

Figure~\ref{figCalculusPanels} places the school districts of Bexar
County on the national cross-section of Calculus enrollment, against
district child poverty (panel A) and district household income (panel
B). Calculus closes the accelerated mathematics sequence of high
school, and Section~\ref{sec:context:courses} finds that its
enrollment has one of the largest negative associations with district
child poverty among the course families compared. Enrollment is
measured per high-school student: CRDC Calculus enrollees divided by
the estimated high-school enrollment of the district's schools.

Calculus enrollment is lower in districts with higher child poverty.
In the national district frame, \CalcMeanPerHundred{} of every 100 high-school students
are enrolled in Calculus. A district whose child-poverty rate is ten
percentage points higher has, on average, \CalcSeatsPerTenPoints{} fewer Calculus
enrollees per 100 high-school students, a difference equal to \CalcRelDecline{}\%
of the national mean enrollment rate (enrollment-weighted slope
$\CalcSlope$ per unit of the poverty rate, state-clustered standard error
\CalcSlopeSE{}). This is a cross-sectional comparison of districts, and it
gives no estimate of how enrollment would change in a district whose
poverty rate changed. The unweighted fit agrees ($\CalcSlopeUnweighted$, standard
error \CalcSlopeUnweightedSE{}), and within Texas the slope is shallower ($\CalcTexasSlope$,
heteroskedasticity-robust standard error \CalcTexasSlopeSE{}; \CalcTexasRelDecline{}\% of the Texas
mean per ten percentage points). The Bexar districts span a broad range of child
poverty, from the Alamo Heights Independent School District (ISD; \CalcAlamoPovertyPct{}\%
child poverty) to Edgewood ISD (\CalcEdgewoodPovertyPct{}\%), the two poles of the
school-finance history recounted in Section~\ref{sec:intro}, and the
two sit at opposite ends of the national distribution: \CalcAlamoPerHundred{} Calculus
enrollees per 100 high-school students in Alamo Heights, \CalcAlamoRatio{} times
the national value, against \CalcEdgewoodPerHundred{} in Edgewood, \CalcEdgewoodRatio{} times. The two
largest districts of the county, Northside ISD and North East ISD,
sit just below the national value (\CalcLargestTwoRatio{} times), and San Antonio ISD
sits at half of it. Single-district values in one cycle are noisy
under the disclosure adjustments of the public-use file and serve as
illustration.

The income panel shows the same ordering. Its income measure is a
weighted average of tract median household incomes. The tract values
come from the American Community Survey (ACS) 2012--16 covariate
distributed with the Opportunity Atlas
\citep{chetty2018opportunity}. We combine them over the
tracts of a school district, weighting each tract by the Atlas
childhood-cohort count times the share of its land area in the
district \citep{EDGEGRF2017}. Appendix~\ref{sec:data:quant}
identifies the boundary file and its vintages. We call this weighted
mean of tract medians \emph{district income}; it is distinct from the
median income of the district's households. Across Bexar County the income ordering runs from Edgewood and
San Antonio ISD near \$\CalcIncomeLowNear{} to Alamo Heights ISD above \$\CalcIncomeHighAbove{}
(income and child poverty correlate at $\CalcIncomePovertyCorr$, unweighted, across
districts nationally). Calculus enrollment per high-school student
rises with district income: the child-weighted correlation is \CalcIncomeCorr{}
(Spearman \CalcIncomeSpearman{}).

Of the \CalcDistricts{} districts, \CalcDistrictsZero{} report no Calculus
enrollment in the cycle. The qualifications of the enrollment measure
apply to both figures and are stated at the end of
Section~\ref{sec:context:courses}.

\subsection{The poverty gradient across course
  families}\label{sec:context:courses}

\begin{figure}[tbp]
  \centering
  \includegraphics[width=0.92\textwidth]{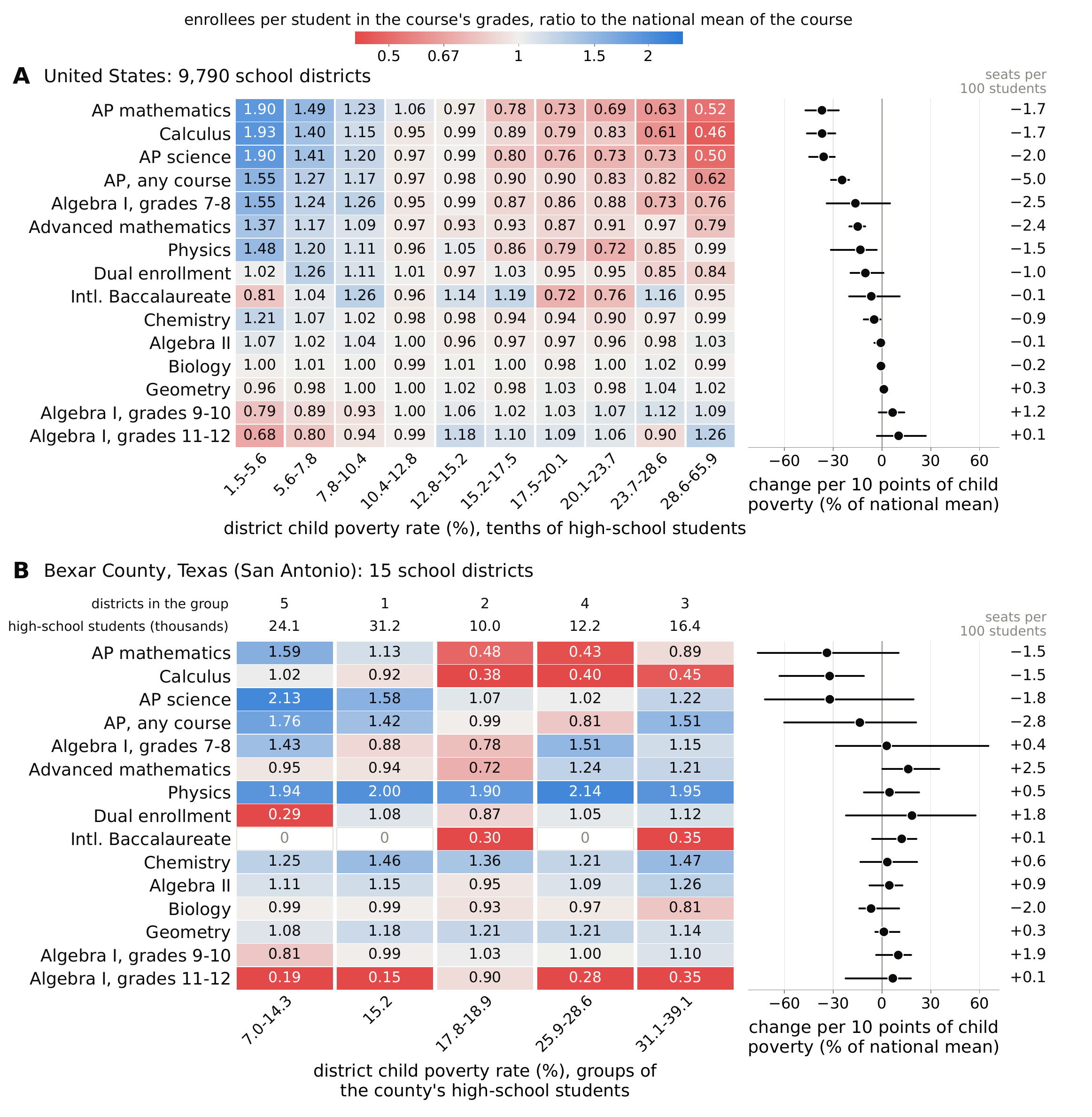}
  \caption{Course enrollment and district child poverty, by
    course family (CRDC 2017--18; SAIPE 2017). Each count is
    divided by estimated enrollment in the grades of the course
    (grades 9--12, or grades 7 and 8 for Algebra~I in grades
    7--8), and each cell is the ratio of that rate to the
    national rate of the same course (1.00 is the national rate;
    blue above, red below; logarithmic color scale). (A) United
    States, the 9,790 districts of
    Figure~\ref{figCalculusPanels}. Left: tenths of high-school
    students ordered by district child poverty. Right: the
    weighted least-squares slope of the rate on child poverty,
    in percent of the national mean per ten percentage points of
    child poverty, with 95\% intervals from a state-cluster
    bootstrap (5,000 resamples); the numbers at the right
    restate each slope in enrollees per 100 students. Rows are
    ordered by this gradient in both panels. (B) The same
    display for the 15 school districts of Bexar County,
    Texas. The districts are ordered by child poverty and merged
    into 5 groups that each hold at least a tenth of the
    county's high-school students.
    The slope is computed over the 15 districts, with 95\%
    intervals from a bootstrap that resamples districts. A
    cell marked 0 reports no enrollment. Districts are the
    units; enrollment counts seats within one cycle, with no
    completion or outcome content.}
  \label{figCourseGradients}
\end{figure}

Figure~\ref{figCourseGradients} extends the comparison to the \GradCourseFamiliesWord{}
course families reported in the CRDC school files. Panel~A uses the
\CalcDistricts{} districts of Figure~\ref{figCalculusPanels}, and panel~B
applies the same measure to the \GradBexarDistrictsWord{} school districts of Bexar
County. Each course is measured against estimated enrollment in its
own grades, and each slope is on the relative scale of
Section~\ref{sec:data:methods}: the difference in enrollment that
goes with ten percentage points of child poverty, in percent of the
national mean of the course (Equation~\eqref{eq:relgradient}). On
this scale three widely taken core courses have small estimated poverty
slopes whose intervals include zero: Biology ($\GradBiologyRel\%$, 95\%
bootstrap interval $\GradBiologyRelLow$ to $\GradBiologyRelHigh$), Algebra~II ($\GradAlgebraTwoRel\%$, $\GradAlgebraTwoRelLow$ to
$\GradAlgebraTwoRelHigh$), and Geometry ($\GradGeometryRel\%$, $\GradGeometryRelLow$ to $\GradGeometryRelHigh$). The result
concerns district averages; it does not establish equal access or
equal enrollment among the schools of a district.

The negative associations are concentrated in the accelerated
sequence. AP mathematics
(\GradApMathRelDecline{}\%), Calculus (\CalcRelDecline{}\%, interval \GradCalculusRelDeclineLow{} to \GradCalculusRelDeclineHigh{}), and AP science
(\GradApScienceRelDecline{}\%) have the largest declines. The profile by tenths shows the same
contrast without a linear fit: for the tenth of high-school students
in the lowest-poverty districts (child poverty below \GradLowestTenthPovertyMax{}\%), Calculus
enrollment per student is \GradCalculusLowestTenthRatio{} times the national value, and for the
highest-poverty tenth (above \GradHighestTenthPovertyMin{}\%) it is \GradCalculusHighestTenthRatio{} times, against \GradBiologyLowestTenthRatio{}
and \GradBiologyHighestTenthRatio{} for Biology. AP
enrollment in any course (\GradApAnyRelDecline{}\%), Algebra~I taken in grades 7--8,
advanced mathematics (\GradAdvMathRelDecline{}\%), and Physics (\GradPhysicsRelDecline{}\%) occupy the middle
of the ordering. For Algebra~I taken in grades 7--8, the point
estimate of the primary cycle is a decline of \GradEarlyAlgebraRelDecline{}\% per student in
those grades, with an interval that includes zero ($\GradEarlyAlgebraRelDeclineLow$ to \GradEarlyAlgebraRelDeclineHigh{});
in the 2015--16 cycle the estimate is \GradEarlyAlgebraFirstCycleRelDecline{}\% and its interval
excludes zero (\GradEarlyAlgebraFirstCycleRelDeclineLow{} to \GradEarlyAlgebraFirstCycleRelDeclineHigh{}). Algebra~I taken in grades 9--10 has a
small positive estimate ($\GradAlgebraOneNineTenRel\%$, interval $\GradAlgebraOneNineTenRelLow$ to $\GradAlgebraOneNineTenRelHigh$), the
sign expected where fewer students complete the course before high
school. The AP
categories overlap with one another and with Calculus, so the upper
rows are correlated measurements of advanced course-taking and their
order is a display order. The pattern is consistent with the
structure of the mathematics sequence: a student who completes
Algebra~I before grade 9 can reach Calculus by grade 12. Enrollment
at the end of that accelerated path is lower in districts with higher
child poverty, the estimate for its start has the same sign with less
precision, and enrollment in the middle of the sequence shows a
near-zero association.

The same contrast appears in the county that contains San Antonio.
Panel~B of Figure~\ref{figCourseGradients} repeats the display for
the \GradBexarDistrictsWord{} school districts of Bexar County. We order the districts
by child poverty and combine them into \GradBexarGroupsWord{} groups, each holding at
least \GradBexarMinSharePct{}\% of the county's estimated high-school enrollment: Alamo
Heights, Fort Sam Houston,
Randolph Field, Lackland, and North East ISDs (child poverty of \GradBexarGroupOnePovertyLow{}\%
to \GradBexarGroupOnePovertyHigh{}\%); Northside ISD (\GradBexarGroupTwoPoverty{}\%); Judson and East Central ISDs
(\GradBexarGroupThreePovertyLow{}\% to \GradBexarGroupThreePovertyHigh{}\%); Southwest, Southside, South San Antonio, and
Harlandale ISDs (\GradBexarGroupFourPovertyLow{}\% to \GradBexarGroupFourPovertyHigh{}\%); and Somerset, San Antonio, and
Edgewood ISDs (\GradBexarGroupFivePovertyLow{}\% to \GradBexarGroupFivePovertyHigh{}\%). Calculus is taken at about the
national rate in the \GradBexarLowerGroupsWord{} lowest-poverty groups (\GradBexarGroupOneCalculusRatio{} and \GradBexarGroupTwoCalculusRatio{} times)
and at less than half of it in the other \GradBexarUpperGroupsWord{} (\GradBexarGroupThreeCalculusRatio{}, \GradBexarGroupFourCalculusRatio{}, and \GradBexarGroupFiveCalculusRatio{}
times). Algebra~II, Geometry, and Biology stay close to the national
rate in every group (between \GradBexarCoreRatioMin{} and \GradBexarCoreRatioMax{} times). The county's
Calculus gradient is a decline of \GradBexarCalculusRelDecline{}\% of the national mean per ten
percentage points of child poverty (95\% interval \GradBexarCalculusRelDeclineLow{}\% to \GradBexarCalculusRelDeclineHigh{}\%, from
a bootstrap that resamples the \GradBexarDistrictsWord{} districts), against \CalcRelDeclineRounded{}\%
nationally. The point estimates for AP mathematics (\GradBexarApMathRelDecline{}\%) and AP
science (\GradBexarApScienceRelDecline{}\%) are also close to the national values, with intervals
that include zero, and the gradients of the three widely taken core
courses are within seven points of zero.

Two qualifications apply to the county panel. \GradBexarDistrictsCapitalWord{} districts give
wide intervals, and among the three steepest courses only the
Calculus interval excludes zero. And the local decline is specific to
Calculus and Advanced Placement: within the county, advanced
mathematics has a positive gradient ($\GradBexarAdvMathRel\%$, interval \GradBexarAdvMathRelLow{} to \GradBexarAdvMathRelHigh{}),
and Algebra~I in grades 7--8 and dual enrollment show no decline.

The contrast holds under the checks reported in
Appendix~\ref{app:results:gradients}. Across the \GradCyclesWord{} CRDC cycles
held (2015--16, 2017--18, and 2021--22), the gradients of the three
widely taken core courses stay within four points of zero, while the
Calculus gradient is \GradCalculusFirstCycleRelDecline{}\%, \CalcRelDecline{}\%, and \GradCalculusLastCycleRelDecline{}\%. State fixed
effects, a broader district frame, and a restriction to schools whose
grade span needs no estimate, each leave the Calculus gradient between
\GradCalculusChecksRelDeclineMin{}\% and \GradCalculusChecksRelDeclineMax{}\% and the widely taken core courses within three points
of zero. Within Texas, where House Bill 5 made Algebra~II optional
outside the endorsements that require it \citep{TexasHB5_2013},
Algebra~II enrollment per high-school student shows no poverty
gradient ($\GradTexasAlgebraTwoRel\%$).

The relative scale divides each gradient by the course's own mean,
so a course with a small base can look steep while involving few
students. Figure~\ref{figCourseGradients} therefore also prints each
gradient in seats: the difference in enrollees per 100
students in the course's grades that goes with ten percentage points
of child poverty. In
seats the largest declines are in AP enrollment in any course (\GradApAnySeats{}
per 100 high-school students; state-clustered interval \GradApAnySeatsLow{} to \GradApAnySeatsHigh{}),
Algebra~I in grades 7--8 (\GradEarlyAlgebraSeats{} per 100 students in grades 7 and 8,
with an interval that includes zero, $\GradEarlyAlgebraSeatsLow$ to \GradEarlyAlgebraSeatsHigh{}), and advanced
mathematics (\GradAdvMathSeats{}; \GradAdvancedMathSeatsLow{} to \GradAdvancedMathSeatsHigh{}), ahead of
AP science (\GradApScienceSeats{}), Calculus (\CalcSeatsPerTenPoints{}), and AP mathematics (\GradApMathSeats{}); the
widely taken core courses change by \GradCoreSeatsMax{} or less. Measured in students,
the point estimates for early Algebra~I and for advanced mathematics
are at least as large as the estimate for Calculus.

District income sorts the courses the same way. Per high-school
student, the child-weighted correlation of income with enrollment is
between $\GradIncomeCoreCorrMin$ and $\GradIncomeCoreCorrMax$ for the widely taken core courses, against
\GradIncomeAdvMathCorr{} for advanced
mathematics, \CalcIncomeCorr{} for Calculus, and \GradIncomeApCorr{} for AP mathematics and for
AP science (\CalcDistrictsWithIncome{} districts). Income geography tracks the advanced
end of the curriculum and says little about the widely taken core
courses.

Four qualifications apply. Enrollment counts seats within one cycle,
so the comparison is silent on the grade at which students take a
course, on completion, and on outcomes. District aggregates can
conceal sorting among the schools of one district. The high-school
denominator is an estimate built from total school enrollment and
grade span, checked above against schools that need no estimate. And
Algebra~I in grades 7--8 is measured on the districts of the frame
that serve those grades; elementary districts that feed separate
secondary districts sit outside the frame.

\subsection{Preparation at entry and first-year
  grades}\label{sec:results:summary}

The two figures above describe school districts. The remaining
results describe students at the institution. This subsection states
what the linked entrant models of Section~\ref{sec:data:link} find;
Appendix~\ref{sec:link} reports them in full.

\emph{Mathematics preparation at entry differs with the sending
school and district.} Among the \LinkSample{} linked entrants of Fall 2014
through Fall 2019, \LinkVarBetweenPct{}\% of the variation in the entry mathematics
score lies between sending schools and districts, and \LinkVarWithinPctRound{}\% lies among
entrants from the same school. Entrants from districts with higher
child poverty arrive with lower scores, by \LinkPovertyPoints{} points per standard
deviation of district child poverty. Entrants from schools with more
enrollment in advanced mathematics arrive with higher scores, by \LinkIndexPoints{}
points per standard deviation of the school's advanced-mathematics
index. One standard deviation of the score is \LinkScoreSD{} points. H1 concerns
the mathematics score relative to the verbal score, and it is
supported with a small coefficient ($\LinkHOne$ standard deviations per
standard deviation of the index; interval \LinkHOneLow{} to \LinkHOneHigh{}).
That relative measure varies almost entirely among entrants from the
same school (\LinkGapVarWithinPct{}\% of its variation). District
poverty and school context are therefore associated with the level of
both scores and only weakly with the difference between them.

\emph{Entrants with stronger mathematics preparation earn higher
first-year grades in nearly every subject.} Among students taking the
same course in the same term, and with the verbal score held equal,
the mathematics score has a positive coefficient in all \LinkSubjects{}
subjects, and the coefficient passes false-discovery control in
\LinkSubjectsPassing{} of them. One standard deviation of the score goes with \LinkPooledMathOtherTwo{} higher
grade points outside the quantitative subjects (interval \LinkPooledMathOtherLow{} to
\LinkPooledMathOtherHigh{}), which supports H2, and with \LinkPooledMathQuantTwo{} higher grade points within
them (\LinkPooledMathQuantLow{} to \LinkPooledMathQuantHigh{}). In mathematics courses themselves it goes with
\LinkMatMathGrade{} higher grade points and a DFW probability \LinkMatMathDfwPoints{} percentage points
lower. Outside the quantitative subjects the verbal score has a
similar coefficient (\LinkPooledVerbalOtherTwo{} grade points), and
the difference between the two scores has an interval that includes
zero. Within the quantitative subjects the verbal coefficient is the
smaller (\LinkPooledVerbalQuantTwo{}).

\emph{District poverty and school course enrollment add little to the
prediction of grades once the entry scores are included.} When whole
sending schools are held out, the full set of predictors reduces the
squared error of predicted grades by a median \LinkSkillMedianPct{}\% across subjects.
The entry scores and the route of pre-college credit account for most
of that reduction (a median \LinkPrepAdded{} of the \LinkSkillMedian{}). District poverty
changes it by a median $\LinkPovertyAddedSigned$ and the school variables by
$\LinkSchoolAdded$. This result concerns prediction with the variables at
hand. District poverty keeps a negative coefficient in several
subjects, and the analyses do not estimate how much of the
association between poverty and grades is caused or mediated by
school course provision or by preparation.

These results describe entrants of Fall 2014 through Fall 2019, a
restriction adopted after the plan's first run on \LinkPlanCohortsWord{} entry
cohorts, with its results known (Section~\ref{sec:data:link}).

\subsection{First-year mathematics, persistence, and major
  change}\label{sec:baseline:flows}

The remaining results describe the institution, where the available
evidence locates academic risk in first-year mathematics.
The exposure is nearly universal. In the reference cohort
(Section~\ref{sec:data:cohort}), \InstMatTakersPct{}\% of the
entrants took at least one mathematics course within the observation
window, \InstSeniorPct{}\% reached senior-level coursework, and
\InstMajorChangePct{}\% have a recorded change of major
(Appendix~\ref{sec:baseline:cohort}). An internal
institutional report, available with institutional approval, puts
the share of new students whose degree plans include a quantitative
requirement at roughly three-quarters and the share of service
courses among the roughly 12{,}000 seats the mathematics department
serves annually at about 95\% \citep{gutierrez2019}. The
results below use two populations: the reference cohort, and the
\emph{full window} of everyone enrolled between Fall 2014 and Fall
2019.

The national record shows the same concentration. Nationally, 93\%
of the mathematics-course enrollments in the mathematics departments
of four-year colleges and universities are in courses at the
precollege, introductory, or calculus level, a level that includes
linear algebra and differential equations
\citep[Table~S.2]{CBMS2015}. National studies describe a related
pattern in developmental mathematics. Remedial coursetaking is common
among students who begin at public two-year and four-year institutions
\citep{chen2016remedial}, and among community-college students
referred to a developmental mathematics sequence a minority complete
it \citep{bailey2010referral}. In one randomized trial, students
assessed as needing remedial algebra who were assigned to
college-level statistics with a weekly workshop passed at a higher
rate than students assigned to the remedial course
\citep{logue2016mainstreaming}.

We begin with who remains. The comparison uses the \PersistPopulationN{}
entrants of the reference cohort without the Coordinated Admission
Program flag; the \PersistCapN{} entrants with the flag transfer after the
first year by the design of that program, and \PersistCapLeftPct{}\% of them
hold no later record. We call an entrant \emph{observed enrolled
later} when the record holds a course in a term one year or more
after the entry term, through the end of the observation window. An
entrant who stopped out and returned counts as observed later. Of the
\PersistPopulationN{} entrants, \PersistLeftN{} (\PersistLeftPct{}\%) were not observed enrolled
later. The record cannot tell whether such an entrant left higher
education or transferred to another institution, so these are
departures from this institution. The comparisons of this subsection
are unadjusted and were added after the analysis plans were frozen.

Unsuccessful first-year outcomes are common, and they go with not
being observed later. We count each first attempt of the first year
in an introductory course that ended in a D, F, withdrawal, or
no-credit outcome as one DFW outcome. The entrants hold \PersistOutcomes{}
such outcomes, which is \PersistOutcomeRatePct{}\% of their \PersistFirstAttempts{} first
attempts, and \PersistAnyDfwN{} entrants (\PersistAnyDfwPct{}\% of those with such an
attempt) hold at least one. The outcome falls in the first year, so
it precedes the later enrollment it is compared with. The share of
entrants not observed later rises with the number of DFW outcomes:
\PersistLeftPctWithNone{}\% of the entrants with none, \PersistLeftPctWithOne{}\% with one,
\PersistLeftPctWithTwo{}\% with two, \PersistLeftPctWithThree{}\% with three, and \PersistLeftPctWithFourPlus{}\% with
four or more. Of the \PersistLeftN{} entrants not observed later,
\PersistLeftWithDfwN{} (\PersistLeftWithDfwPct{}\%) held at least one DFW outcome.

Mathematics is the largest single source of these outcomes
(Figure~\ref{figPersistenceDFW} in
Appendix~\ref{app:results:majorchange}). Three percentages describe
its place, and each has its own denominator. First, mathematics
accounts for \PersistMatOutcomeSharePct{}\% of all first-year DFW outcomes in
introductory courses, against \PersistSecondSubjectOutcomeSharePct{}\% for the next subject
(\PersistMatOutcomeShareWithRepeatsPct{}\% when repeated attempts within the year are counted).
Second, \PersistMatOutcomeRatePct{}\% of the \PersistMatFirstAttempts{} first attempts in mathematics
end in a DFW, the highest rate among the subjects compared. Third,
among the entrants not observed later, mathematics accounts for
\PersistLeftMatOutcomeSharePct{}\% of the DFW outcomes, more than any other subject (the
next contributes \PersistLeftSecondSubjectOutcomeSharePct{}\%), and \PersistLeftWithDfwHoldingMatPct{}\% of those
entrants who held a DFW held one in mathematics. These percentages
describe course outcomes. A separate comparison describes students:
\PersistMatLeftWithPct{}\% of the \PersistMatDfwN{} mathematics entrants with at least one DFW
were not observed later, against \PersistMatLeftWithoutPct{}\% of the mathematics
entrants without one, a difference of \PersistMatDiff{} percentage points
(pointwise \InstDfwIntervalLevel{}\% interval \PersistMatDiffLo{} to \PersistMatDiffHi{}).

The mathematics difference is much smaller among entrants with no
DFW in another introductory subject: \PersistMatAloneLeftWithPct{}\% against
\PersistMatAloneLeftWithoutPct{}\%, a difference of \PersistMatAloneDiff{} points whose interval includes
zero ($\PersistMatAloneDiffLo$ to \PersistMatAloneDiffHi{}). Among mathematics entrants who also
hold a DFW in another subject, the rates are \PersistMatWithElsewhereLeftWithPct{}\% and
\PersistMatWithElsewhereLeftWithoutPct{}\%. The record therefore shows a broad pattern of
unsuccessful coursework among the entrants not observed later, and it
leaves open whether mathematics has a part of its own in their
departure. Splitting by the outcomes of other courses is a
description: those outcomes may share their reasons with the
mathematics outcome. A difference of the same sign appears in every
one of the \PersistSubjectsWord{} introductory subjects with enough entrants,
and it is larger than in mathematics in \PersistSubjectsAboveMatWord{} of them
(panel~B of that figure and Table~\ref{tab:persistencedfw}): a DFW in
the first-year seminar or in writing composition goes with not being
observed later more often than one in mathematics. Each of these
intervals is pointwise and lies above zero; the comparisons are
exploratory and are not adjusted for the number of subjects.

Departure has many reasons, and the record holds few of them, so
these are statements of share. What the record supports is this: most
entrants who are not observed later hold DFW outcomes, and
mathematics contributes more of those outcomes than any other
subject. The comparisons do not establish that a course outcome
caused a departure.

Major change connects first-year mathematics to the rest of the
degree for the entrants who remain. We explored how a DFW on a first
attempt in a first-year introductory course relates to a recorded
change of major during the observation window, in the reference
cohort. The two majors compared are the first and the last on
record, so the comparison does not date the change against the
course outcome. Among the \InstDfwMatLaterN{} entrants with such an attempt in
mathematics who were observed enrolled at least one year after entry,
\InstDfwMatLaterChangedWithPct{}\% of the \InstDfwMatLaterWithN{} with a DFW changed major, against \InstDfwMatLaterChangedWithoutPct{}\% of the
\InstDfwMatLaterWithoutN{} without one, a difference of \InstDfwMatLaterDiff{} percentage points (\InstDfwIntervalLevel{}\%
interval \InstDfwMatLaterDiffLo{} to \InstDfwMatLaterDiffHi{}). Among all \InstDfwMatAllN{} entrants with such an
attempt, whatever their later enrollment, the rates are \InstDfwMatAllChangedWithPct{}\% and
\InstDfwMatAllChangedWithoutPct{}\%, a difference of \InstDfwMatAllDiff{} points (\InstDfwMatAllDiffLo{} to \InstDfwMatAllDiffHi{}). The association
therefore depends substantially on whether entrants remain observable
beyond their first year: an entrant who leaves has less opportunity
to record a change of major. Mathematics is the introductory subject
in which the most entrants receive a DFW. Mathematics and physics are
the only subjects whose difference is positive with an interval that
excludes zero in both populations, and physics counts \InstDfwPhyLaterWithN{}
later-observed entrants with a DFW against \InstDfwMatLaterWithN{} in mathematics
(Table~\ref{tab:majorchangedfw} in
Appendix~\ref{app:results:majorchange}). Per student, the difference
among later-observed entrants is larger in computer science and
chemistry (\InstDfwCsLaterDiff{} and \InstDfwCheLaterDiff{} points), which fewer entrants take, and
in information systems and physics it is close to that of mathematics
(\InstDfwIsLaterDiff{} and \InstDfwPhyLaterDiff{} points). These
comparisons do not establish whether the course outcome
preceded the change of major or whether either caused the other. They
are the reason the next two figures follow the students who took
mathematics and changed major.

\begin{figure}[tbp]
  \centering
  \includegraphics[width=\textwidth]{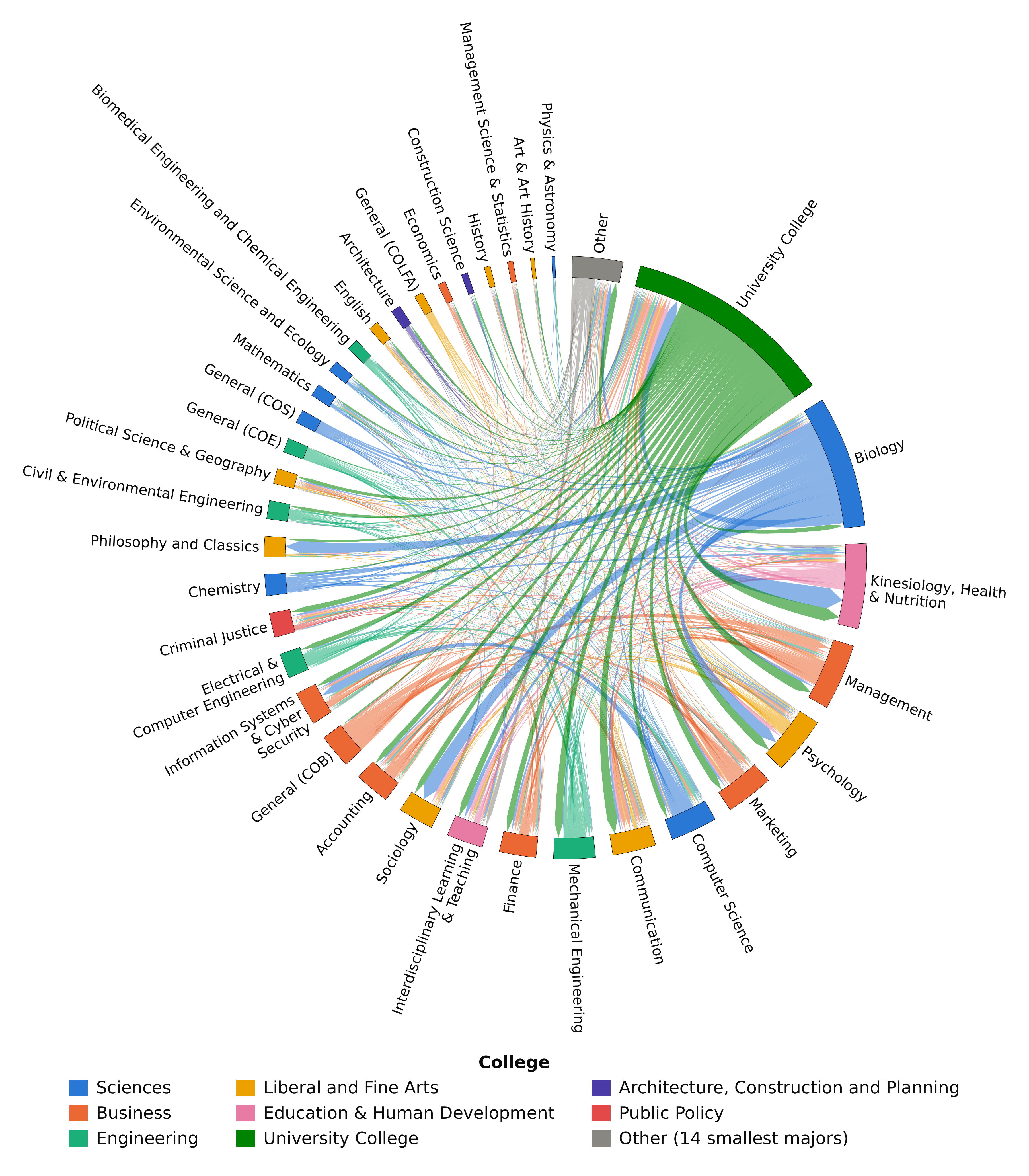}
  \caption{Major-change flows among students who took MAT courses
    and changed major (Fall 2014--Fall 2019). The figure draws the
    \InstChordDrawnFlows{} flows of at least \InstChordMinFlowWord{}
    students, which hold \InstChordDrawnStudents{} of the
    \InstChordStudents{} students; smaller flows are not drawn. Each
    sector is one major; arc length is
    proportional to the number of students it donated plus received. 
    Ribbons point from donor to acceptor and inherit the donor's
    color; color denotes college. A major named General is the
    college-wide undeclared or general-studies major of the college
    in parentheses. The \InstChordOtherMajorsWord{} majors with the
    smallest flows are
    collected in the gray sector labeled Other.
    Table~\ref{tab:majorangles} gives the size
    of every sector.}
  \label{figChordShort}
\end{figure}

Figure~\ref{figChordShort} shows the major-change flows among the
students who took MAT courses and changed major within the full Fall
2014--Fall 2019 window (entrants and transfers alike). A ribbon of
the figure is a count of students, so the figure draws only the flows
of at least \InstChordMinFlowWord{} students: \InstChordDrawnFlows{}
flows, which hold \InstChordDrawnStudents{} of the
\InstChordStudents{} students (\InstChordDrawnPct{}\%). The
\InstChordWithheldFlows{} smaller flows are not drawn. Reading the
ring: each sector is one major, sized by the students it donated plus
received in the drawn flows; ribbons run from donor to acceptor in
the donor's color.
Each sector is labeled with the name of its major. Three structural
facts stand out. First, University College, which holds undeclared
and general-studies students, is the dominant donor: the large green
ribbons fan out from it to every college, which is the intended
function of an advising unit. Second, the large receiving majors are
spread across colleges (Biology; Kinesiology, Health \& Nutrition;
Management; Psychology; Marketing; Computer Science), so major change
among MAT-takers is a campus-wide circulation. Third, flows between
named majors concentrate within colleges (ribbons tend to stay near
their own color), with visible cross-college exceptions into Business
and Health-adjacent majors. Table~\ref{tab:majorangles} lists the
majors in alphabetical order with the angular position and size of
each sector (Appendix~\ref{app:chords}).

\subsection{Relative subject grading among
  major-changers}\label{sec:baseline:barriers}

\begin{figure}[tbp]
  \centering
  \includegraphics[width=\textwidth]{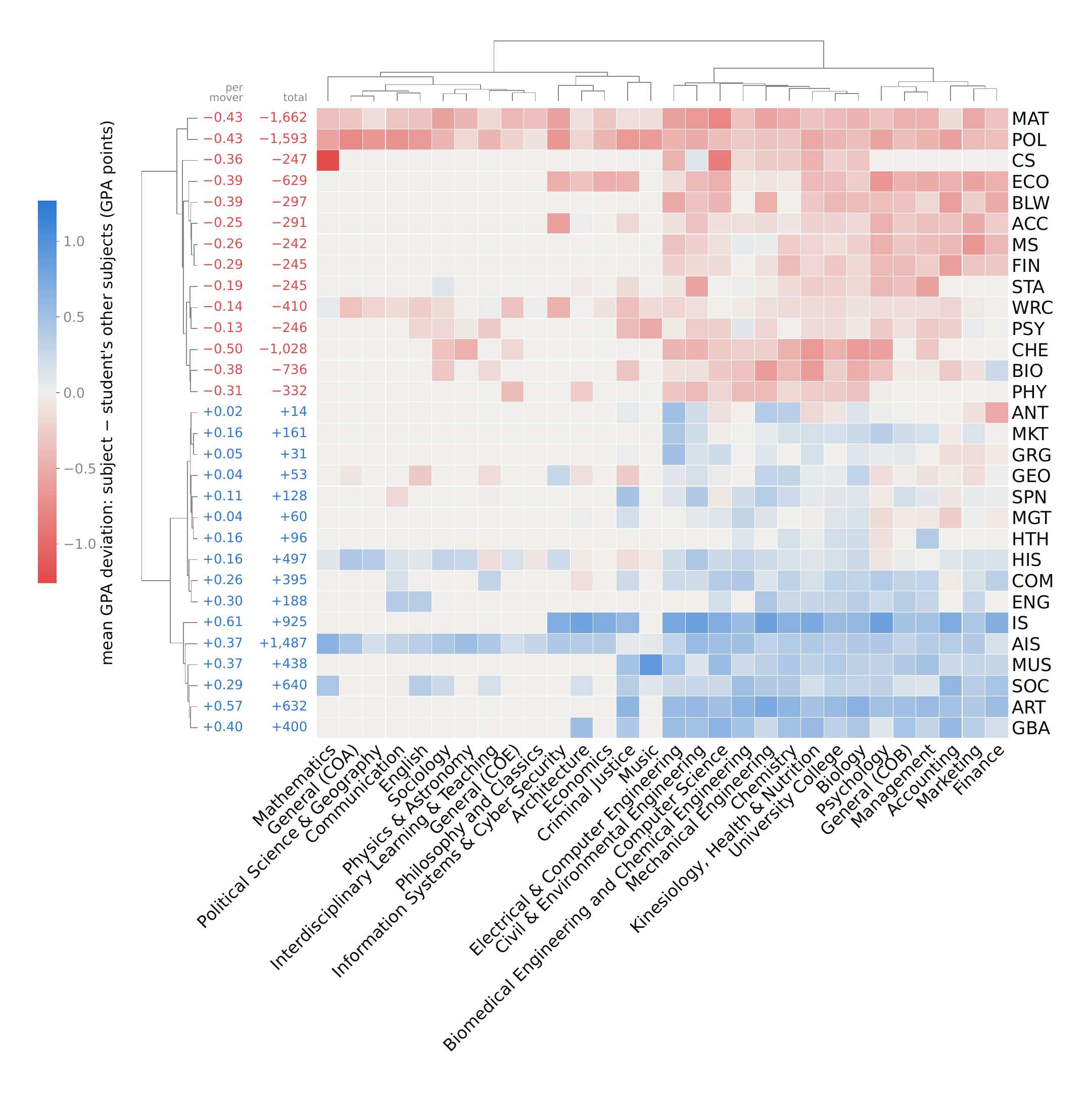}
  \caption{Differences between grades in one subject and grades in
    all other subjects, among students who changed major: first-time
    freshmen of the reference cohort whose first and last major
    within the observation window (Section~\ref{sec:data:cohort})
    differ. Rows are subjects, and columns are the students' initial
    majors. Each cell is the mean, over students whose initial major
    is the column, of their GPA in the row subject minus their mean
    GPA in all other subjects (Equation~\eqref{eq:deviation}). Red:
    lower grades in the subject than in the students' other subjects;
    blue: higher grades. Grades are means over course rows that count
    in the grade average. Cells with fewer than \InstHeatMinCell{}
    students are blank; the display keeps the \InstHeatTopKept{}
    highest-volume subjects and initial majors. The two left columns
    summarize each subject over its observed cells: the mean
    difference (GPA points per mover) and that mean times the number
    of movers observed (GPA points in total across movers). Rows and
    columns are ordered by Ward hierarchical clustering; unobserved
    cells enter the clustering as zero.}
  \label{figHeatmap}
\end{figure}

Figure~\ref{figHeatmap} compares each student's grades in one subject
with that student's average grade in all other subjects. We average
those differences among students of the reference cohort who began in
the same major and have a recorded change of major (\emph{movers},
below). Rows show subjects, and columns show the students' initial
majors. A negative cell means that these students earned lower grades
in the row subject than in their other subjects, and a positive cell
means that they earned higher grades.
Equation~\eqref{eq:deviation} gives the calculation. Two columns at
the left summarize each row over its observed cells. The first
reports the mean difference, in GPA points per mover. The second
multiplies that mean by the number of movers observed, so it also
reflects how many movers took the subject.

The subjects form two blocks. In the block at the top of the figure,
movers earned lower grades than in their other subjects. Per mover,
chemistry shows the largest negative difference ($\InstHeatChePerMover$ GPA points; \InstHeatCheMovers{} movers over \InstHeatCheCells{}
initial majors), followed by mathematics and political science ($\InstHeatMatPolPerMover$
each) and by business law, economics, and biology ($\InstHeatBlwEcoBioPerMoverLow$ to
$\InstHeatBlwEcoBioPerMoverHigh$). In total across movers, mathematics and political science
lead ($\InstHeatMatTotal$ and $\InstHeatPolTotal$ GPA points over \InstHeatMatMovers{} and \InstHeatPolMovers{}
movers): both are observed in all \InstHeatDonorMajorColumns{} initial-major columns, and
their cells are negative in the columns of nearly every initial
major. Chemistry is third ($\InstHeatCheTotal$). In the block at the bottom of the
figure, movers earned higher grades than in their other subjects.
Per mover it is led by information systems
($\InstHeatIsPerMover$) and art ($\InstHeatArtPerMover$); in total it is led by the first-year
seminar AIS ($\InstHeatAisTotal$ over \InstHeatAisMovers{} movers, at $\InstHeatAisPerMover$ per mover),
which nearly every mover takes, ahead of information systems
($\InstHeatIsTotal$). The
column clustering adds a related observation: the contrast is
largest for students who left engineering, science, and business
majors (the right-hand column block from Electrical and Computer
Engineering through Finance), whose negative cells in the upper rows
and positive cells in the lower rows are the strongest in the
matrix. Grades are means over course rows that count in the grade
average, so withdrawals and credit or no-credit rows do not enter.
All of this is descriptive of grades among movers; nothing in the
figure orders cause and effect between difficulty and departure.
The deviations are relative recorded grades. Preparation, course
mix, grading conventions, and selection into changing major can each
contribute to them, so a deficit does not measure how demanding the
instruction of a subject is. Unobserved cells enter the clustering as
zero, so the order of rows and columns is a display order, and the
dendrograms are no evidence of natural academic groupings.

\subsection{Section size and grades}\label{sec:baseline:classsize}

\begin{figure}[tbp]
  \centering
  \includegraphics[width=\textwidth]{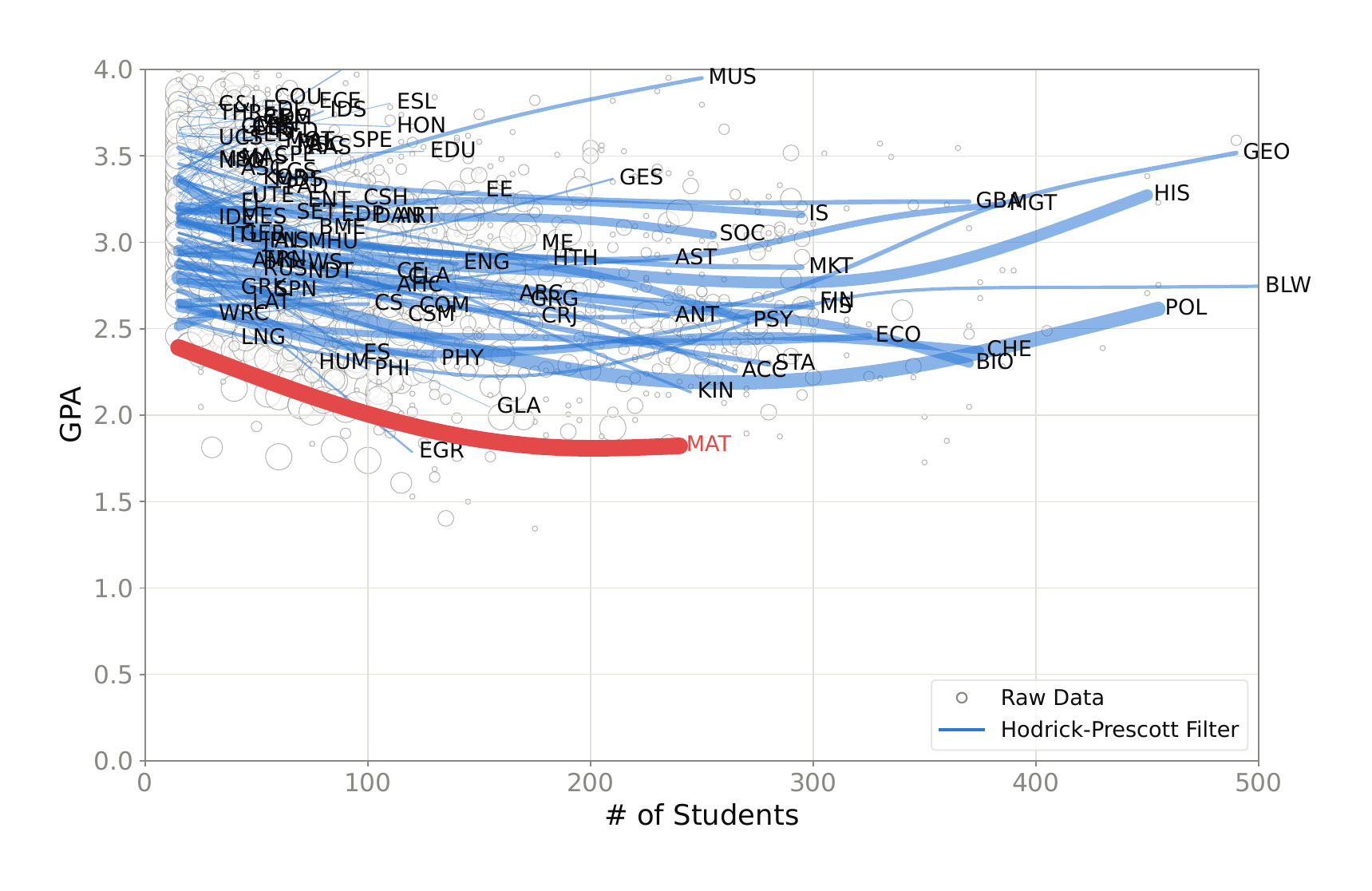}
  \caption{Mean section GPA (course rows that count in the grade
    average) vs.\ class size for the observed
    subjects, Fall 2014--Fall 2019; the displayed grid runs to \SizeCurveGridMax{}
    students. Each curve is one subject: section sizes are binned in
    \SizeCurveBin{}-student increments, each observed size weighted equally within
    its bin, interpolated, and smoothed with a Hodrick--Prescott
    filter ($\lambda=\SizeCurveSmoothing$;
    Appendix~\ref{app:methods:institution}); line
    width is proportional to students served; MAT is drawn in red.
    Each bubble shows the mean section GPA for a class-size group. 
    Its area is proportional to the square of the number of 
    distinct class sizes in that group.
    Only sections with more than \SizeCurveMinStudents{} students enter the
    analysis.}
  \label{figClassSize}
\end{figure}

Figure~\ref{figClassSize} plots mean section GPA against section
size, one smoothed curve per subject, with line width proportional to
students served. Two facts are visible at a glance. Across subjects,
mean section GPA is broadly flat in section size over the observed
range, and subject identity separates the curves far more than size
does: the vertical spread between subjects at any fixed size is
roughly two GPA points. And the MAT curve (red) is the lowest of the
high-volume subjects across the displayed range of section sizes,
falling from
about \SizeCurveMatGpaSmallest{} mean GPA in the smallest sections to about \SizeCurveMatGpaReference{} near \SizeCurveReferenceSize{}
students and flattening beyond; a few low-volume curves dip below it
at particular sizes. The curves compare sections across the courses
of a subject, so they describe grading outcomes and hold no design
contrast.

The prespecified section-size models supply that contrast for
first-time entrants: their primary model compares entrants within one
course and term (Section~\ref{sec:data:link}). This subsection states
their three results for mathematics. Appendix~\ref{app:results:classsize} reports
the models in full, with the estimates of every subject
(Figure~\ref{figClassSizeLadder}), the scenario
(Table~\ref{tab:classsizescenario}), and the sensitivity checks.

\emph{Larger mathematics sections go with more DFW outcomes among
entrants with lower entry mathematics scores.} The association
between section size and outcomes depends on preparation, as H4
proposed, and H4 is supported. Per standard
deviation of the entry mathematics score, the coefficient on section
size rises by \SizeMatInteractionGrade{} grade points (\SizeMatInteractionGradeLow{} to \SizeMatInteractionGradeHigh{}; $p=\SizeMatInteractionGradeP$) and the
DFW coefficient falls by \SizeMatInteractionDfwFall{} percentage points (\SizeMatInteractionDfwFallLow{} to \SizeMatInteractionDfwFallHigh{};
$p=\SizeMatInteractionDfwP$). The size association at a given entry score is a
separate contrast with its own interval. For
an entrant one standard deviation below the mean entry score, a
doubling of section size goes with a DFW probability \SizeMatSlopeLowScoreDfw{} percentage
points higher (\SizeMatSlopeLowScoreDfwLow{} to \SizeMatSlopeLowScoreDfwHigh{}; $p=\SizeMatSlopeLowScoreDfwP$) and a grade \SizeMatSlopeLowScoreGradeLower{} points lower
($\SizeMatSlopeLowScoreGradeLow$ to $\SizeMatSlopeLowScoreGradeHigh$; $p=\SizeMatSlopeLowScoreGradeP$). For an entrant one standard
deviation above the mean, the fitted associations are $\SizeMatSlopeHighScoreDfw$
percentage points ($\SizeMatSlopeHighScoreDfwLow$ to $\SizeMatSlopeHighScoreDfwHigh$) and $\SizeMatSlopeHighScoreGrade$ grade points
($\SizeMatSlopeHighScoreGradeLow$ to $\SizeMatSlopeHighScoreGradeHigh$). The DFW association at the lower entry score
is therefore distinguishable from zero, and the grade association at
that score has the same direction with an interval that reaches zero.

\emph{Averaged over all entrants, the association is small.} Across
all mathematics sections the coefficient is
$\SizeMatAllSectionsGrade$ grade points per doubling of section size ($\SizeMatAllSectionsGradeLow$ to
$\SizeMatAllSectionsGradeHigh$), which reproduces the falling curve. With course-by-term effects the coefficient is
$\SizeMatCourseTermGrade$, so
the aggregate relationship largely reflects differences between
course offerings: the Calculus~I and II lecture sections of the
record hold a median of \SizeCalculusMedianSize{} students and none holds more than \SizeCalculusSizeBound{},
while one in six sections of the courses below Calculus holds more
than \SizeLargeSectionBound{}. In the primary model, which adds the student covariates
and instructor effects, the coefficient is $\SizeMatPrimaryGrade$
($\SizeMatPrimaryGradeLow$ to $\SizeMatPrimaryGradeHigh$; $p=\SizeMatPrimaryGradeP$), with a matching difference of $\SizeMatPrimaryDfw$
percentage points in the DFW rate ($\SizeMatPrimaryDfwLow$ to $\SizeMatPrimaryDfwHigh$). The estimate
has the sign H3 proposed (larger
sections, lower mathematics grades), and its interval includes zero,
so the prespecified criterion for H3 is not met. Across the \SizeSubjectsModeled{} subjects that enter the models, the pooled primary
coefficient is $\SizePooledGrade$ grade points per doubling ($\SizePooledGradeLow$ to
$\SizePooledGradeHigh$), and outside
mathematics the size-by-score interaction has mixed signs.

\emph{The result identifies a hypothesis for testing smaller
sections.} The scenario of Section~\ref{sec:data:link} splits every
mathematics lecture section above \SizeCap{} students. For the third of first attempts with the lowest
entry scores, the projection is \SizeLowThirdDfwFewer{} fewer DFW outcomes per 100
affected first attempts ($\SizeLowThirdDfwLow$ to $\SizeLowThirdDfwHigh$; $p=\SizeLowThirdDfwP$) and \SizeLowThirdGrade{} more grade points
($\SizeLowThirdGradeLow$ to $\SizeLowThirdGradeHigh$; $p=\SizeLowThirdGradeP$). For the middle third it is \SizeMidThirdDfwFewer{} fewer
DFW outcomes ($\SizeMidThirdDfwLow$ to $\SizeMidThirdDfwHigh$), and for the highest third \SizeHighThirdDfwMore{} more
($\SizeHighThirdDfwLow$ to $\SizeHighThirdDfwHigh$). Over
all affected first attempts
the projected change is $\SizeMatScenarioDfw$ percentage points ($\SizeMatScenarioDfwLow$ to $\SizeMatScenarioDfwHigh$),
and the split would add \SizeMatSectionsAddedPerYear{} sections to the \SizeMatSectionsPerYear{} taught per
year. The projected reduction in DFW outcomes is therefore
concentrated among the entrants with lower preparation. The
projection describes an intervention only under the conditions of
Section~\ref{sec:data:link}, which the observational models do not
establish. Sections were assigned administratively, so every estimate
in this subsection is an association, and the test that the result
calls for is a designed one.

\section{Discussion}\label{sec:discussion}

\subsection{What the findings show
  together}\label{sec:discussion:findings}

The five findings of Section~\ref{sec:results} are interrelated. Enrollment at
the advanced end of the high-school mathematics sequence is lower
where child poverty is higher, across the nation and among the
districts of Bexar County. Entrants from poorer districts, and from
schools with less advanced-mathematics enrollment, arrive at this
institution with lower mathematics scores. Entry scores are
associated with first-year grades in nearly every subject, and
unsuccessful first-year outcomes, of which mathematics is the largest
single source, go with not being observed enrolled later. Every link
is an association. The links from entry scores onward describe
students who entered this institution, and the analyses leave open
how much of one link is produced by another.

Three views of the same institutional record, computed under
different designs, describe mathematics as a low-grading, widely
taken subject. Students who changed major earned mathematics grades
below their own averages in other subjects, second to chemistry per
student and largest in total (Figure~\ref{figHeatmap}). In the
section-level record, MAT is the lowest-grading high-volume subject
across the displayed range of section sizes
(Figure~\ref{figClassSize}). In the reference
cohort, taking a MAT course at any point in the window associates
with elevated odds of having no senior-level enrollment in the
window, an
estimate that holds under the cohort restriction that reverses the
naive AIS association (Table~\ref{tab:oddsnaivecohort}). The three
views describe different populations and quantities. They set the
scene in which the linked results operate; the direct estimate of how
preparation relates to performance is the linked analysis.

Persistence gives these views a practical reading. Mathematics is the
course in which a difficult first year is recorded for the most
entrants: it is the largest single source of the unsuccessful
outcomes of the entrants who are not observed later
(Section~\ref{sec:baseline:flows}). That makes it a practical place
to find these students early. The record sets limits on the
statement. Departure has many reasons that the record does not hold,
and the record cannot tell a transfer from leaving higher education.
A DFW goes with not being observed later in every introductory
subject, and among mathematics entrants with no DFW elsewhere the
difference is small, with an interval that includes zero. Among the
entrants who remain, a first-year DFW in mathematics goes with a
recorded change of major more often, by an amount that depends on
which entrants remain observable. Neither comparison sorts cause and
effect.

The intake findings give the phrase ``access institution''
measurable content: sending-district composition, recorded course
enrollment at the sending school, and the type of pre-college credit
entrants hold. The
institution draws from districts less poor than the state's
child-weighted center (Figure~\ref{figSendingPoverty}), dominated by
two large urban districts that sit just below the national fitted
line for Calculus enrollment (Figure~\ref{figCalculusPanels}). The credit type students arrive
with tracks the recorded AP mathematics enrollment of their school:
AP arrival rises and dual-enrollment arrival falls across the
enrollment bands (Figure~\ref{figProvisionBands}), and the schools
with zero recorded AP mathematics enrollment concentrate among the
smallest campuses.
This intake description combines the credit that students hold at
arrival with the context of their sending schools and districts,
which is measured later for many of them: the school profiles use the
2017--18 CRDC snapshot, which postdates entry for many students. The
profiles describe the environments associated with the enrolled
population; they do not establish the opportunities each student
encountered before college.

The methodological record supports conclusions portable to any
institution repeating the exercise. The sign of an association can
depend on the comparison group: the required first-year seminar
reverses from strongly adverse to strongly favorable when the
comparison is restricted to a defined entry cohort with a fixed
horizon (Appendix~\ref{sec:results:design}). The sign can
likewise depend on the denominator frame: \GradFrameMismatchDistrictsWord{} frame-mismatch
districts, of more than ten thousand, reverse the unweighted national
gradient of a course-enrollment rate computed per resident child
(Appendix~\ref{app:data:defects}).
Every material source flaw in the catalog produced plausible output
with no error or warning, and each was found by a check aimed at
something else. The standing rules that follow (cohort restriction,
denominator filters, frame-against-frame tests, uncertainty computed
at the level the exposure varies) are conclusions of this paper in
their own right.

The national position completes the baseline. In the 2019--20 Scorecard
data, the institution's completion rate was \ScoreBaselinePointsBelow{} percentage points
below the value fitted at its Pell share. In the 2024--25 data it was
\ScoreCurrentPointsAbove{} points above the corresponding fitted value, at an essentially
unchanged Pell share (\ScoreBaselinePellShare{} to \ScoreCurrentPellShare{},
Figure~\ref{figScorecard}). These are descriptive comparisons with
the national curve of each data year. The baseline fixes the first
coordinate, and the movement spans the disruption period, which is
the kind of change the reference quantities exist to measure.

\subsection{Implications for equity and
  measurement}\label{sec:discussion:implications}

The equity reading starts from the measured associations of
Section~\ref{sec:results:summary}. Entry preparation may help
identify students who need academic support, through placement
instruments, early-term signals, and section-level monitoring. The
small contribution of the district and school variables to predictive
accuracy does not establish that geographic information is
unnecessary for support or equity decisions: it concerns these
predictors under two evaluation designs, and district poverty keeps a
negative adjusted coefficient in several subjects. The data also do
not identify which school or family processes produced the
differences in preparation that are already present at entry.

The record also points to one thing the institution controls:
section size. The projection of Section~\ref{sec:baseline:classsize}
is specific enough to guide where small sections might go. It is
concentrated in mathematics and among less-prepared entrants, whose
entry scores are also associated with district poverty and school
course enrollment, whereas splitting every
large mathematics section would add \SizeMatSectionsAddedPct{}\% more sections and splitting
every large section in the \SizeSubjectsModeled{} subjects would double them
(Table~\ref{tab:classsizescenario}). This is a hypothesis
with a prespecified observational basis. Sections were
assigned administratively, so the test it calls for is a designed
assignment of less-prepared entrants to small mathematics sections.
The present comparisons do not establish what would happen if the
institution reassigned students or added instructors to split
sections.

The recorded mathematics score is associated with performance beyond
mathematics courses. Outside the quantitative subjects its
association is comparable to that of the verbal score. Both scores
also reflect general preparation, which the record cannot separate
from them (Section~\ref{sec:discussion:limitations}), so these
results do not establish that mathematics support would improve
performance in other subjects.

The comparison across course families locates the district-level
differences at the advanced end of the mathematics sequence
(Section~\ref{sec:context:courses}). Among entrants a related pattern
appears from the student side: enrollment in advanced mathematics at
the sending school is associated with the entry score and with the
level of the first mathematics course.
National transcript evidence places the threshold of momentum toward
a bachelor's degree above Algebra~II
\citep[pp.~30--33]{adelman2006toolbox}, so the courses with the
largest negative association with district poverty are the courses
that evidence, drawn from the high-school classes of 1982 and 1992,
ties most closely to degree completion. Enrollment counts are
silent on course content, timing, and completion. Where transcripts
permit, a measurement program for the gateway
should record the highest mathematics course each entrant completed
in high school.

Admissions policies and community-college finance can change who
enters the university and what credit entrants bring. Test-optional
admissions changes the information available about applicants, and
House Bill 8 restructured the finance of the community colleges that
award dual-enrollment credit \citep{TexasHB8_2023}. These changes may
alter the credit-arrival shares of Figure~\ref{figProvisionBands} as
well as later outcomes. A post-2020 comparison should therefore
describe intake alongside performance, so that an observed difference
is not automatically attributed to experiences after entry.

District, school, and neighborhood measures describe different
populations and should be joined only through audited crosswalks
(Section~\ref{sec:data}). In these data, district income is more
strongly associated with advanced-course enrollment than with
enrollment in the widely taken core courses
(Section~\ref{sec:context:courses}), so an area-income measure cannot
substitute for every kind of course-enrollment measure. The
major-change results describe movement that an enrollment snapshot
would miss: students leave University College for majors in every
college, and the subjects in which students who change major earn
grades above or below their own averages can be identified
(Figures~\ref{figChordShort} and~\ref{figHeatmap}). These results do
not establish whether a particular move benefited the student.

\subsection{Limitations}\label{sec:discussion:limitations}

The study has nine limitations. Some arise from studying one institution and using school, district, and neighborhood measures to describe students’ context. Others reflect missing information about which institutions awarded dual-enrollment credit and when student characteristics were recorded. The accuracy of the school matches also remains unverified, but it can be assessed by auditing the existing materials.

\begin{enumerate}
\item \textbf{Selected scope.} Every student-linked statistic
  describes the students of one institution
  (Section~\ref{sec:intro}). Nothing generalizes to Texas high-school
  graduates or to the nation.
\item \textbf{School-context vintage and linkage accuracy.} The
  sending-school credit profile of
  Appendix~\ref{sec:context:provision} uses the CRDC 2017--18
  snapshot, which postdates many entrants' high-school years. The
  linked models use the cycle nearest each entrant's senior year,
  within \LinkSnapshotMaxLagWord{} year of it. The 2013--14 cycle recorded dual
  enrollment as an indicator only, so for the \LinkDualSubstituteLinked{} linked
  entrants assigned that cycle the dual-enrollment rate is that of
  2015--16, as many as \LinkDualSubstituteMaxLagWord{} years after their senior year. In both uses the variables
  describe school context near entry and leave each student's own
  coursework unmeasured. Between the 2015--16 and 2017--18 cycles the
  per-student rates hold the order of the sending schools imperfectly
  (rank correlations of 0.62 to 0.78), so school-level readings
  inherit measurement error of unknown size. The accuracy of the
  accepted school matches has not been measured against an
  adjudicated sample, and the linked models make that accuracy more
  consequential; an audit of a stratified sample of accepted matches
  is the next step for the linkage.
\item \textbf{Ecological context.} Opportunity Atlas outcome measures
  are doubly ecological (area averages of predicted cohort outcomes)
  at ZCTA level, and the median-income variable of
  Figure~\ref{figCalculusPanels} is an ACS 2012--16 tract covariate
  distributed with the Atlas. %
\item \textbf{Unobserved credit partners.} The records hold no
  partner-institution or delivery-site field for dual-enrollment
  credit, so credit taught on a college campus cannot be
  distinguished from credit taught in the high school, and grading
  standards cannot be held fixed across awarding partners.
\item \textbf{A shared record.} The three designs that agree on the
  mathematics gateway share one institutional record, so they can
  share its faults; the agreement is corroboration within one
  source, and replication at other institutions is the appropriate
  external check.
\item \textbf{Associational, selected, and without a high-school
  record.} The linked models estimate associations conditional on
  entry. Admission and enrollment select the population. Selection
  into enrollment can change the associations between background and
  preparation, so they should not be generalized to all high-school
  students. The record holds no high-school grade
  average, rank, or transcript, so the two entry scores absorb general
  preparation. The extract holds one aid-application flag, one value of
  adjusted gross income (AGI), and one highest score per student, with no aid year and no
  test date, so the timing of these variables relative to entry rests
  on institutional practice that the extract does not document; AGI
  is absent for the \LinkNonApplicantPct{}\% who did not apply for aid, and the primary
  estimates are stable when the aid variables are left out. The
  prespecification is internal, without an external timestamp. Three
  choices were made after the plan was frozen and are recorded as
  deviations: the restriction to entrants of Fall 2014 onward, the
  joint covariance of the pooled intervals, and the training-only
  construction of the prediction targets.
\item \textbf{Persistence at one institution.} The record ends at the
  institution. An entrant with no course record a year or more after
  entry may have left higher education or transferred, so persistence
  here means continued enrollment at this institution. The
  comparisons by first-year course outcome are unadjusted for
  preparation and course load, they were added after the analysis
  plans were frozen, and a withdrawal can be part of a departure
  already under way.
\item \textbf{Section size without a design.} The section-size models
  compare students within a course offering under administrative
  assignment. Enrollment cap, room, and meeting
  time were not analyzed, so no discontinuity design was performed; 
  differences between the small and large sections of a course remain
  possible. The preparation interaction rests on one prespecified
  test ($p=\SizeMatInteractionGradeP$ for grades, $\SizeMatInteractionDfwP$ for DFW), is specific to
  mathematics, and awaits replication. The projection for the lowest
  third of entry scores excludes zero for DFW outcomes and includes
  zero for grades once its interval comes from the bootstrap, and the
  scenario inherits each of these conditions together with its own
  assumptions about causation, linearity, and staffing.
\item \textbf{No outcome tables by group.} No cohort outcome in this
  paper is tabulated by race and ethnicity, Pell status, or
  first-generation status, although the intake composition makes
  those dimensions central at a Hispanic-Serving Institution. The
  linked models include family resources, first-generation status,
  gender, and ethnicity as predictors under a prespecified plan and the
  disclosure-control regime of Appendix~\ref{app:data:governance}
  (aggregate cells of at least \InstDisclosureMinCellWord{} students); outcome tables
  disaggregated by group are the immediate next
  study in this program.
\end{enumerate}

\subsection{A protocol for post-2020
  comparison}\label{sec:discussion:protocol}

The baseline supports a specific comparison protocol: the same
pipelines applied to entrants of Fall 2020 and later, with the same
observation window, comparing each reference quantity with its
baseline counterpart under identical population definitions, designs,
and generated tables. Repeating the analyses can show which measured
quantities changed. If measured intake remains stable while the
progression quantities change, the difference is recorded after
entry; this does not identify its cause. Changes in the
credit-arrival shares describe intake. Changes in the grade
deviations or in the class-size curves may reflect different
students, different courses, or different grading practices. Changes
in the coefficients of the linked models describe different
associations between recorded preparation and first-year grades. The
Scorecard panel adds a national comparison, provided that reporting
units and definitions stay the same across data years. The first
post-period coordinate is already available
(Appendix~\ref{sec:results:position}), and the comparison within the
institution will say which of these quantities moved beneath it.

Three changes complicate comparisons with the pre-2020 baseline.
Emergency grading 
policies alter the meaning of every GPA-based instrument, so
deviation-matrix and class-size comparisons across the 2020 and 2021
terms must treat pass/fail-heavy semesters as their own labeled
regime. Test-optional admissions thins the standardized-score fields
that the baseline's audits used for selection checks, so post-period
audits will need replacement covariates. And the diffusion of
generative artificial intelligence into coursework can move grades
and learning independently. The grade-based measures describe
recorded grades, and a change in recorded grades need not imply a
corresponding change in learning. Each hazard changes what
a difference means, and some (the 2025 merger among them) may break
entity comparability outright; labeling populations, vintages, and
regimes is the practice that preserves the interpretation of each
comparison where comparison remains possible.

In summary, we have documented associations that link district child
poverty and course enrollment, mathematics preparation at entry, and
first-year performance at one Hispanic-serving research university
with broad access, and we have provided the pre-2020 reference
quantities, linkage
infrastructure, validation record, and estimand definitions required
for the post-2020 comparisons that the 2020--2025 disruptions now
make necessary.

\section*{Statements and Declarations}

\paragraph{Funding.}
This material is based upon work supported by the National Science
Foundation under Award No.~2518973 \citep{NSF2518973}. Any opinions,
findings, and conclusions or recommendations expressed in this
material are those of the author and do not necessarily reflect the
views of the National Science Foundation.

\paragraph{Competing interests.}
The author declares no competing financial or non-financial
interests.

\paragraph{Ethics approval.}
The analysis of institutional student records was conducted under
institutional review board protocol FY24-25-316, ``Supporting Student
Learning and Success in Early Undergraduate Mathematics Courses via
AI-Enhanced Education.'' The protocol provides for retrospective data
analysis: non-interventional secondary analysis of existing
institutional administrative records for students who attended 
between Fall 2014 and Fall 2019, with no intervention, no contact or
interaction with students, and no new data collection, and with every
exposure and outcome window closing before the Spring 2020
instructional disruption. The records were obtained from the
institution under data request 201908048-IN, approved on August 25,
2019, which authorized storage of the data for future research use.
The disclosure-control regime is described in
Appendix~\ref{app:data:governance}: individual records are queried in
place and never ingested into the analysis database, and only
aggregate cells supported by at least \InstDisclosureMinCellWord{} students are published,
with withheld-cell counts recorded alongside published counts.

\paragraph{Consent to participate.}
The study involved no contact or interaction with students and no new
data collection. It is a secondary analysis of existing administrative
records conducted under the protocol and data request named above.

\paragraph{Data availability.}
The public-source data (SAIPE, CRDC, NCES EDGE and Common Core of
Data (CCD), Opportunity Atlas, College Scorecard, and the Census
relationship files) are
available from their originating agencies at the URLs cited in the
references. An inventory that accompanies the code lists every public
input file with its dataset, vintage, SHA-256 hash, size, and source
address (\DataSourceFiles{} files, each verified against the file on
disk), and a manifest holds the acquisition records of the
\DataManifestArchivesWord{} archives acquired for the study. Each
figure and table is released with the aggregate inputs that a second
manifest names. For most exhibits these inputs hold the values shown;
for the heatmap, the class-size curves, and the cohort table they
hold summary statistics, and redrawing those exhibits requires the
institutional records. The institutional extract is restricted under
the terms of the protocol and data request and cannot be shared; its
results are computationally reproducible by authorized analysts.

\paragraph{Code availability.}
Every figure, table, and computed number of this paper is generated by
a named program in the analysis repository, which accompanies the
paper for review and release. A master script runs the programs in
dependency order; each analysis writes its results to a statistics
file, and the text reads every computed number from those files by
name (Appendix~\ref{app:data:repro}). The analysis plan of the linked entrant
models and its section-size addendum, each frozen before estimation,
are included with the code, together with their hashed records and
the dated list of deviations made after each freeze. Every file named
in the record of the addendum is included in its frozen state. Of the
files named in the record of the original plan, those unchanged since
the freeze are included in that state. The plan, the data-assembly
program, and the run script were amended by the first logged
deviation, and a loader of public course data was extended afterward;
the repository includes their current versions, and the recorded
hashes identify their frozen states.

\paragraph{Author contributions.}
The sole author designed the study, performed the analyses, and wrote
the manuscript.

\paragraph{Use of generative artificial intelligence.}
The author prepared the initial draft drawing on a previously 
authored internal report of approximately 10{,}000 pages.
During manuscript preparation, the author used Visual Studio Code extensions 
for Anthropic's Claude Code (Claude Fable 5.2 Max Ultracode) and OpenAI's Codex 
(GPT-6.1 Sol Extra High).  
The AI tools assisted with revising prose, synthesizing analyses, and 
generating Python code to replace an existing MATLAB pipeline. 
Each system's output was critically reviewed by the other system and by 
the author. Separate review rounds examined the manuscript as a whole, 
individual sections, and software components. Every component underwent 
at least nine rounds of review aimed at identifying errors, unsupported 
claims, and weaknesses in interpretation. 
No individual-level student records were supplied to either AI system. 
Quantitative analyses were executed locally by code that queried databases
stored on an encrypted drive, using a computer requiring authentication at 
startup and for operating-system access. 
The author designed the study, directed software development, and reviewed 
and edited all generated material. The author takes full responsibility for 
the study's design and implementation, the analyses, and the manuscript's content.

\bibliographystyle{unsrtnat}
\bibliography{references}

\appendix
\counterwithin*{figure}{section}
\counterwithin*{table}{section}
\counterwithin*{equation}{section}
\renewcommand{\thefigure}{\thesection\arabic{figure}}
\renewcommand{\thetable}{\thesection\arabic{table}}
\renewcommand{\theequation}{\thesection\arabic{equation}}
\providecommand{\theHfigure}{}
\providecommand{\theHtable}{}
\providecommand{\theHequation}{}
\renewcommand{\theHfigure}{\thesection.\arabic{figure}}
\renewcommand{\theHtable}{\thesection.\arabic{table}}
\renewcommand{\theHequation}{\thesection.\arabic{equation}}
\section{Quantitative Methods}\label{sec:data:quant}

This appendix opens with the record of rigor and reproducibility of
the study. It then states each estimator and its respective figure or
table, if any. Each analysis is computed on its own population and
observation window, listed in Table~\ref{tab:populations}. The
inferential conventions that govern the estimators are in
Appendix~\ref{app:data:inference}, and the implementing scripts are
named in the Code Availability statement. The associations between
gateway subjects and progression use the reference entry cohort with
its fixed window, and Appendix~\ref{sec:results:design} reports the
naive/cohort comparison that motivates defining an entry population
before computing an association.

\subsection{Rigor and reproducibility}\label{app:data:repro}

\paragraph{The repository.}
Every program of the study is in one repository, which accompanies
the paper and is public at
\mbox{\url{https://github.com/biomathematicus/gateway}}. Its root holds a description of the repository, a master
script, an inventory of the public input files, and the records of
the prespecified analyses. One folder holds the programs, and one
holds the aggregate outputs that this manuscript reads. The master
script runs the programs in dependency order: the loading of the
public data, the district analyses, the matching of high-school names
to schools, the institutional baseline, the linked and section-size
models, and the generation of the numbers of this manuscript. The
figures and tables are exported by the same programs that compute the
statistics.

\paragraph{No computed quantity is typed in the manuscript.}
Each analysis writes its results to a statistics file beside its
figure. A macro module names every quantity that the text prints and
reads its value from that file, or from the constant of the program
that sets it. For each analysis the pipeline then writes two \LaTeX{}
files. The first declares every quantity with
\verb|\newcommand| and gives all of them the same
initial value, a red warning that reads \WarningRunPipeline. The
second assigns each computed value with
\verb|\renewcommand|. For the national mean of
Calculus enrollment (Section~\ref{sec:context:calculus}), the two
files hold these lines:
\begin{quote}\small\noindent
\verb|\newcommand{\CalcMeanPerHundred}{\WarningRunPipeline}|\\
\verb|\renewcommand{\CalcMeanPerHundred}{|\texttt{\CalcMeanPerHundred}\verb|}|
\end{quote}
The manuscript loads the declarations in its preamble and the values
after the document begins. Its source says
\verb|\CalcMeanPerHundred| where the reader sees
\CalcMeanPerHundred{}. A quantity whose value no program supplied
therefore prints as the red warning wherever it is used, in the text,
in a table, or in a caption, so a missing or incomplete run of the
pipeline is conspicuous. The \MacroModulesWord{} macro modules name
\MacroTotal{} quantities. A check program lists every number still
typed in the sources and compares it with the list of literals that
stay text, each with its reason: calendar labels, names that hold
digits, units, numerals of formulas, statistics quoted from cited
sources, and statistical conventions.

\paragraph{Inputs and released outputs.}
The context database rebuilds from source files with recorded hashes.
An inventory in the repository lists every public input file with its
hash (\DataSourceFiles{} files, each verified against the file on
disk), and a manifest names the program and the released inputs of
each figure and table. The released outputs are an explicit list of
aggregate files, each scanned under the disclosure rule of
Appendix~\ref{app:data:governance}.

\paragraph{Restricted records.}
The institutional extract is restricted, so its results are
computationally reproducible by authorized analysts, and the
repository holds the aggregate outputs that the manuscript reads. The
older institutional-baseline programs read the extract through a
local query cache. Each cache file is stored with the hash of the
query that produced it, and a program reads the source again whenever
its query differs from the recorded one. For the submitted run these
programs were executed against an empty cache, and every table,
statistics file, and backing data file they write was identical to
the earlier output. The hash identifies the query; a change in the
source tables themselves calls for the refresh option documented with
the code. The loaders apply the standing data rules
(Appendix~\ref{app:data:defects}) so that the flaws documented there
cannot silently re-enter through future analysis code.

\paragraph{Prespecified analyses.}
The linked entrant models and the section-size models were specified
before estimation, and the pipeline was first run end to end on
shuffled outcomes (Section~\ref{sec:data:link}). The plan, its
addendum, their hashed records, and the dated list of deviations are
in the repository, and the Code availability statement says which
frozen files it holds.

\subsection{Record linkage}\label{app:methods:linkage}

\paragraph{Record-linkage scoring.}
High-school names are resolved by a containment-first score. Let $A$
and $B$ be the token sets of the two names after school-type tokens
are stripped. With containment
$c(A,B) = |A \cap B| / \min(|A|,|B|)$, Jaccard similarity
$J(A,B) = |A \cap B| / |A \cup B|$, and character-trigram similarity
$t(A,B)$, the match score is
\begin{equation}
  s(A,B) \;=\; \XwalkWeightContain{}\,c(A,B) \;+\; \XwalkWeightJaccard{}\,J(A,B) \;+\; \XwalkWeightTrigram{}\,t(A,B),
  \label{eq:linkscore}
\end{equation}
computed within hierarchical blocks on city, ZIP, normalized city
key, and state. Auto-acceptance requires $s \geq \XwalkAutoScore$; within the
state-level blocks it additionally requires a margin of at least
\XwalkAutoMargin{} over the runner-up, while within city and ZIP blocks the block
itself supplies the disambiguation and the margin requirement is
waived (Appendix~\ref{app:data:linkage}).

\subsection{District course gradients}\label{app:methods:district}

\paragraph{Weighted least squares and diagnostics.}
The district course gradient is the least-squares slope of enrollees
per estimated student in the course's grades on the child poverty
rate, weighted by that enrollment; standard errors are clustered at
the state level, and a fit within one state uses
heteroskedasticity-robust standard errors
(Figure~\ref{figCalculusPanels}). The frame holds the matched
districts with at least \GradMinChildren{} children aged 5--17 (the pre-treatment
denominator rule of Appendix~\ref{app:data:defects}) that report
Algebra~II enrollment, the marker used here for a district operating
a high-school course sequence; a neutral frame that admits every
district reporting any core high-school mathematics or science
course is a robustness variant.

\paragraph{Relative gradients across course families.}
The weighted estimator is applied, on that frame, to each of
\GradCourseFamiliesWord{} course families reported in the CRDC school
files: Algebra~I in grades 7--8, 9--10,
and 11--12; Geometry; Algebra~II; advanced mathematics; Calculus;
Biology; Chemistry; Physics; AP enrollment in any course, in
mathematics, and in science; International Baccalaureate; and dual
enrollment. Within a district that serves high-school grades, the
CRDC skip code on a course item marks a school reporting zero classes
in the subject and enters the district sum as zero. Every other
reserve code is missing data, and a district holding one in any course
family leaves the common frame (\GradIncompleteDistricts{} of the \GradDistrictsBeforeExclusion{} districts in
2017--18), so all \GradCourseFamiliesWord{} families are compared on the same \CalcDistricts{}
districts (Appendix~\ref{app:data:defects}). Raw
slopes scale with the base rate of each course, so the reported
quantity for course $c$ is the relative gradient
\begin{equation}
  g_c \;=\; 100 \times \frac{0.10\,\hat\beta_c}{\bar y_c},
  \label{eq:relgradient}
\end{equation}
where $\hat\beta_c$ is the weighted slope of the enrollment rate
(enrollees over the denominator) on the child poverty rate (a
proportion) and $\bar y_c$ is the pooled national rate, total
enrollees over the total denominator of the frame; $g_c$ is the
change in the rate per ten percentage points of child poverty, in
percent of the national mean of the course. The
denominator is estimated enrollment in the grades of the course. For
a high-school course that is estimated enrollment in grades 9--12, the
denominator of Figure~\ref{figCalculusPanels}: the CRDC total
enrollment of each
school serving any of grades 9--12, prorated by the share of its
K--12 grades that are high-school grades. For Algebra~I in grades
7--8 it is enrollment in grades 7 and 8, estimated the same way. The
two Algebra~I counts reported for grades 9--10 and for grades 11--12
are each divided by the full grades 9--12 estimate. These
denominators describe the size of the school population in the
relevant grades; they do not identify which students had met the
prerequisites of a course, so a rate is enrollment per estimated
student and leaves individual eligibility unmeasured. A
per-capita variant, reported in the text for comparison, divides by
resident children aged 5--17 (SAIPE). Intervals are 95\%
percentile intervals from a state-cluster bootstrap with \GradBootstrapResamples{}
resamples; each resample draws states with
replacement and re-estimates both the slope and the national mean,
and differences between courses and the rank of a course are computed
within resamples. The same estimator is applied to estimated
high-school enrollment per resident child, which explains the
per-capita variant. Robustness
variants repeat the computation on the 2015--16 and 2021--22 cycles,
on a neutral frame that admits every district reporting enrollment in
any core high-school mathematics or science course, with the
incomplete districts kept, and with state
fixed effects; within-state slopes use heteroskedasticity-robust
standard errors, since one state supplies one cluster
(Figure~\ref{figCourseGradients}).

\paragraph{The county panel.}
Panel~B of Figure~\ref{figCourseGradients} repeats the profile and
the gradient for the \GradBexarDistrictsWord{} districts of Bexar County in the frame,
a district belonging to the county when more than half of its schools
are located there. Every rate is divided by the national rate of the
course, as in panel~A, so 1.00 is the national rate in both panels. A
county holds too few districts for tenths: the districts are taken in
order of child poverty and merged until a group holds at least a
tenth of the county's estimated high-school enrollment, and a short
tail joins the last group. The gradient is the enrollment-weighted
slope of the rate on child poverty over the \GradBexarDistrictsWord{} districts,
divided by the national mean of the course. Its 95\% interval is a
percentile interval from a bootstrap that resamples the districts,
the level at which child poverty varies (\GradBootstrapResamples{} resamples). \GradBexarDistrictsCapitalWord{}
districts are few clusters, so these intervals are approximate and
wide.

\paragraph{Geographic aggregation.}
Area-level context variables are aggregated from smaller to larger
geographic units through hashed relationship files, and never in the
other direction. Opportunity Atlas tract values
aggregate to ZCTAs as population-weighted means over the Census 2010
ZCTA--tract relationship \citep{CensusZCTA2010}, and to school
districts as
child-count-weighted means over the land-area shares of the NCES
Education Demographic and Geographic Estimates (EDGE) 2017 geographic
relationship file \citep{EDGEGRF2017} (TIGER 2017 school-district
boundaries, which
represent school year 2015--16, on 2010 tract definitions); every
aggregated variable stays
ecological and is flagged as such in the schema. Neither file is ever
used to assign a student to a tract or a district
(Figure~\ref{figCalculusPanels}).

\paragraph{Rank persistence and agreement.}
Provision stability across CRDC cycles is measured by the Spearman
rank correlation, between cycles, of each school's per-student
enrollment rate in a course. Cohen's $\kappa$ is used where a binary
agreement statistic is needed, and it exposed a degenerate offering
indicator in one cycle (Appendix~\ref{app:data:defects}).

\paragraph{Weighted distributions.}
The sending-district poverty analysis uses weighted histograms and
weighted medians: the same district-level poverty rates weighted once
by entrants sent and once by children held, so the two curves differ
only in the weighting (Figure~\ref{figSendingPoverty}).

\subsection{Institutional descriptives}\label{app:methods:institution}

\paragraph{Flow matrices and circular layout.}
Major-change flows are square donor$\times$acceptor count matrices
over the union of majors. The \InstChordOtherMajorsWord{} majors with
the smallest total flow are collected in one sector labeled Other; a
flow between two of them becomes a flow inside that sector. A ribbon
of the diagram states a count of students through its width, so a
flow of fewer than \InstChordMinFlowWord{} students is not drawn
(\InstChordWithheldFlows{} flows, which hold
\InstChordWithheldStudents{} students in all). Each sector's arc
length is proportional to the students it donated plus received in
the drawn flows, and sectors are ordered by that size, so no ribbon,
no arc, and no row of the index table rests on fewer than
\InstChordMinFlowWord{} students. Sectors
are labeled with the name of the major, because a major can hold
several registrar subject codes and no single code stands for it. The
ring is rotated rigidly so that the longest name sits at \InstChordAnchorDeg{} degrees,
measured counter-clockwise from the 3 o'clock direction, and the
angular index of Table~\ref{tab:majorangles} applies to the chord
diagram (Figure~\ref{figChordShort}).

\paragraph{Leave-one-out deviation and hierarchical clustering.}
For subject $s$ and initial major $m$ (the donor major of the flow
diagrams), let $C_m$ be the set of
major-changers whose first major is $m$. The deviation matrix entry
is
\begin{equation}
  D_{s,m} \;=\; \frac{1}{|C_{s,m}|} \sum_{i \in C_{s,m}}
  \bigl( G_{i,s} - \bar G_{i,-s} \bigr),
  \label{eq:deviation}
\end{equation}
where $G_{i,s}$ is student $i$'s grade-point average (GPA) in subject
$s$ over course rows that count in the registrar's grade average
(withdrawals and credit or no-credit rows do not enter), $\bar G_{i,-s}$ is the same student's mean GPA in all other
subjects, and $C_{s,m} \subseteq C_m$ is the subset observed in
subject $s$; cells with fewer than \InstHeatMinCell{} students are blank. Each
subject is summarized over its observed cells on two scales: the
student-weighted mean deviation (GPA points per mover) and that mean
times the movers observed (the total across movers). Rows and
columns are ordered by Ward hierarchical clustering on Euclidean
distance, with dendrograms displayed at the margins
(Figure~\ref{figHeatmap}).

\paragraph{Binning and Hodrick--Prescott smoothing.}
Section-level mean GPA is binned in \InstSectionSizeBin{}-student class-size increments,
interpolated on a common grid, and smoothed with the
Hodrick--Prescott filter, the minimizer of
\begin{equation}
  \sum_t (y_t - \tau_t)^2 \;+\;
  \lambda \sum_t \bigl( \tau_{t+1} - 2\tau_t + \tau_{t-1} \bigr)^2,
  \label{eq:hp}
\end{equation}
solved as a pentadiagonal linear system with $\lambda = 10^4$. As
$\lambda \to \infty$ the filter converges to a straight line,
equivalent to a linear regression, so the chosen value states how
much curvature the display retains (Figure~\ref{figClassSize}).

\paragraph{Introductory-course outcomes and major change.}
For each subject, the entrants of the reference cohort who hold a
first attempt in a course numbered below \InstIntroBelow{} during the first year
are split by whether any such first attempt ended in a D, F,
withdrawal, or no-credit outcome (DFW), with the grade codes of the
linked models. The share that changed major (first and last recorded
majors within the window differ) is compared between the two groups:
the difference with a Wald interval, and the odds ratio with an
interval by the log (Woolf) method. Two populations are reported side
by side: the entrants observed enrolled at least one year after
entry, defined by an enrollment record in a term one year or more
after the entry term, so that an entrant who stopped out and returned
qualifies; and all entrants with such an attempt
(Table~\ref{tab:majorchangedfw}).

\begin{figure}[tbp]
  \centering
  \includegraphics[width=\textwidth]{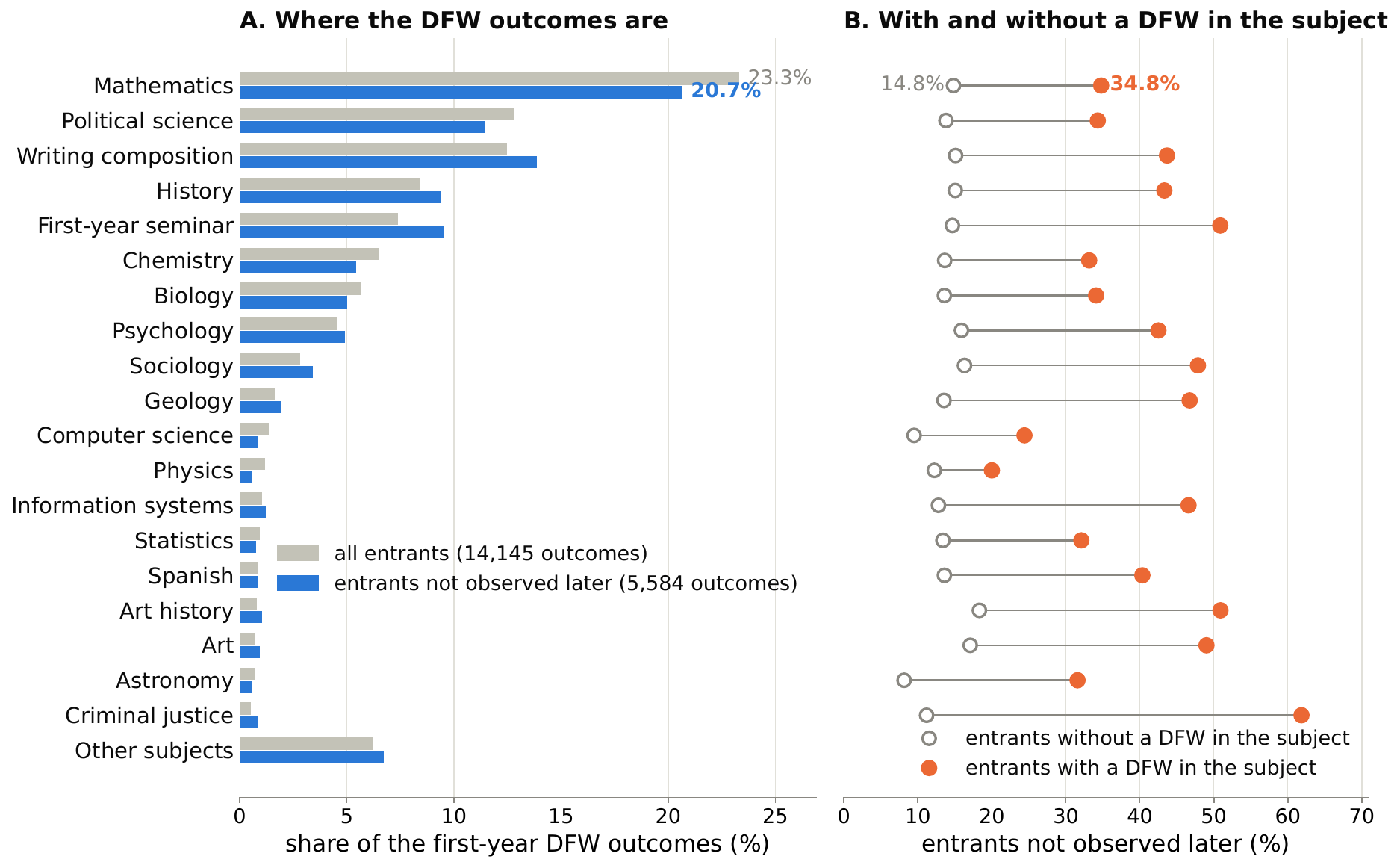}
  \caption{First-year DFW outcomes and persistence, by subject, among
    the \PersistPopulationN{} entrants of the reference cohort without the
    Coordinated Admission Program flag. A DFW outcome is a first
    attempt of the first year in an introductory course that ended in
    a D, F, withdrawal, or no-credit outcome. Not observed later: no
    course record in a term one year or more after the entry term.
    (A)~Share of the DFW outcomes that fall in each subject, among
    all entrants (gray, \PersistOutcomes{} outcomes) and among the entrants not
    observed later (blue, \PersistLeftOutcomes{} outcomes held by \PersistLeftWithDfwN{}
    entrants). (B)~Share of entrants not observed later among those
    with a DFW in the subject (orange) and those who took the subject
    without one (open circle). Subjects with at least \PersistMinTakers{}
    entrants, ordered by their share of all DFW outcomes; the other
    subjects are pooled in panel~A. The comparisons are unadjusted.}
  \label{figPersistenceDFW}
\end{figure}

\paragraph{Introductory-course outcomes and persistence.}
The same split of the entrants of each subject is compared on
persistence. The outcome is the complement of the rule above: an
entrant is not observed later when the record holds no course in a
term one year or more after the entry term, through the end of the
observation window. The course outcome falls in the first year, so it
precedes this outcome. The comparison leaves out the entrants flagged
for the Coordinated Admission Program, who transfer after the first
year by the design of that program. Mathematics is examined further
in two ways: by the kind of outcome (a D or F in a completed course,
or a withdrawal or no-credit outcome only, since a withdrawal can be
part of a departure already under way), and by whether the entrant
also holds a DFW in another introductory subject
(Table~\ref{tab:persistencedfw}). The DFW outcomes are also counted
course by course: each first attempt of the first year in an
introductory course that ended in a DFW is one outcome, so an entrant
can hold several in one subject. The share of these outcomes that
falls in each subject is computed among all entrants and among the
entrants not observed later, and the share of entrants not observed
later is computed by the number of DFW outcomes an entrant holds
(Figure~\ref{figPersistenceDFW}).

\paragraph{Cohort-restricted odds ratios.}
For each subject, a $2\times2$ table of exposure (took the subject
within the observation window) against outcome (senior-level
enrollment within the same window, without regard to the grade
earned) over the reference entry cohort, with Fisher's exact test,
95\% confidence intervals by the log (Woolf) method, and
Benjamini--Hochberg control at $q=\InstFdrQ$ where families of subjects
are tested. The naive counterpart computes the same table over all
students in the window and is reported only as the design comparison
of Appendix~\ref{sec:results:design}
(Tables~\ref{tab:cohortbaseline} and~\ref{tab:oddsnaivecohort}).

\paragraph{Two-sample tests.}
Where displayed quantities are mean GPAs, panels are compared with
Welch's $t$ (unequal variances), because grade distributions pile up
at 4.0 and rank tests lose power against real mean gaps there; where
a rank-sum convention is reproduced from the prior operational study
(per-donor GPA ratios), the Mann--Whitney test is used, with
Benjamini--Hochberg adjustment in both cases.

\paragraph{Locally weighted regression and centroids.}
The national completion expectation is a lowess fit (fraction \ScoreLowessFraction{}) of
completion on Pell share (the share of undergraduates receiving
federal need-based Pell Grants, a standard low-income marker) across
predominantly-bachelor's institutions, evaluated at the institution's
Pell share to give a
Pell-conditional descriptive residual; the residual has no
uncertainty band and no covariate adjustment. Two centroids summarize
each panel: the enrollment-weighted institutional centroid (mean
position weighted by current undergraduate enrollment, while the
completion coordinate remains a first-time full-time cohort outcome)
and the unweighted mean across institutions, computed in raw
coordinates and plotted on the display scale
(Figure~\ref{figScorecard}).

\subsection{Linked entrant models: specification}\label{app:methods:link}

Section~\ref{sec:data:link} states the population, the measures, and
the hypotheses of the linked entrant models. This subsection gives
the full specification of the predictors, the outcomes, and the
models.

\paragraph{Predictors.}
Family resources: whether the student applied for aid, the adjusted
gross income (AGI) of applicants as a percentile rank within entry
cohort, and first-generation status; the \LinkNonApplicantPct{}\% of all first-time
fall entrants who did
not apply have no recorded income and form their own category. The
extract holds one aid-application flag and one AGI value per student
and no field identifying the aid year, so income is a family-resource
marker of unverified timing; the rank transformation changes its
units and leaves its timing as recorded.
Appendix~\ref{sec:link} repeats the primary estimates without the aid
variables. Demographics: gender and ethnicity. District: the district
child poverty rate, averaged over the four SAIPE vintages before
entry. School: Calculus, AP mathematics, AP science, and
dual-enrollment enrollees per student and total enrollment, from the
CRDC cycle nearest the senior year (2013--14, 2015--16, or 2017--18,
within \LinkSnapshotMaxLagWord{} year of it). There is one exception. The 2013--14 cycle
recorded dual enrollment as an indicator only, so for the entrants
assigned that cycle the dual-enrollment rate is the rate of 2015--16,
as many as \LinkDualSubstituteMaxLagWord{} years after their senior year. The
\emph{advanced-mathematics index} is the mean of the standardized
Calculus and AP mathematics rates. These school variables describe
recorded course enrollment at the sending school near the entrant's
senior year; they are distinct from the courses a school offers and
from the coursework an individual entrant completed. Preparation: the
mathematics and verbal entry scores, and the route of any pre-college
mathematics credit (none; dual enrollment only; examination, AP or
the College-Level Examination Program, CLEP). Each score is the
institution's recorded highest section score across the SAT and the
ACT, with the ACT and the pre-2016 SAT expressed on the current SAT
scale by the institution; the extract holds the converted values and
no test date or concordance version. Within each entry cohort we
subtract the cohort mean from each mathematics score and divide by
that cohort's standard deviation, and we standardize verbal scores
separately in the same way, so one unit is one cohort-specific
standard deviation (\LinkScoreSD{} points for mathematics when the cohorts are
pooled). Other continuous predictors are standardized over the linked
sample. The record holds no high-school grade average, class rank, or
transcript.

\paragraph{Outcomes.}
Preparation at entry is measured in two ways: the standardized
mathematics score, and the standardized mathematics score minus the
standardized verbal score, which describes an entrant's relative
position on the two tests. This second measure is a difference of two
separately standardized scores and differs from the raw difference in
section points. First-year performance is measured on first attempts
in lower-division courses during the entry fall and the following
spring and summer: the grade (quality points on rows that count in
the grade average) and a DFW indicator (a D, F, withdrawal, or
no-credit outcome). Performance is therefore conditional on the
courses an entrant took. A subject enters when at least \LinkSubjectMinimum{} linked
entrants hold a first attempt in it; \LinkSubjects{} subjects qualify, \LinkQuantSubjectsWord{} of
them classed as quantitative before estimation (MAT, CHE, ECO, STA,
PHY, CS, ACC). The grade models use the first attempts with a graded
outcome, so the number of students in a grade model is smaller than
the count that admits the subject and falls just below \LinkSubjectMinimum{} in \LinkSubjectsGradeBelowMinimumWord{}
subject.

\paragraph{Models.}
A nested random-intercept model (district, school within district),
fitted by restricted maximum likelihood, partitions the variance of
each preparation outcome into a part between districts, a part between
schools of the same district, and a part among entrants from the same
school. Blocks of predictors then enter in order, and the change in
each component is reported. We separate each aid applicant's income
rank into the mean rank among entrants from the same sending school
and the student's deviation from that mean. The first coefficient
compares students from schools with different mean income ranks; the
second compares entrants from the same sending school, who may have
attended in different years. Both use the same income-rank scale.
This is the between and within decomposition of a covariate
\citep{mundlak1978pooling}. The difference between the two
coefficients describes a contextual association and does not
identify an effect of attending a particular school. Each
subject model is a least-squares regression with course-by-term fixed
effects, so students are compared with other students taking the same
course in the same term;
standard errors are clustered on sending school, and
Benjamini--Hochberg control of the false-discovery rate at $q=\LinkFdrLevel$
is applied across subjects \citep{benjamini1995fdr}.
Coefficients are pooled across subjects by random effects
\citep{dersimonian1986meta}. The same
students and sending schools contribute to several subject models, so
the pooled intervals use the joint school-clustered covariance of the
subject coefficients: each row's influence on a subject's coefficient
is summed within sending schools, and the cross-products of those sums
across subjects give the covariance between subject estimates. The
plan specified pooling under independence; the joint covariance is a
correction made after the freeze (deviation D2).

Out-of-sample skill is measured for a linear model, ridge regression,
and a random forest in two prediction tasks. In the first, \LinkFoldsWord{} folds
hold out whole sending schools, and the target is a student's grade
relative to the mean grade of the same course and term. That mean is
computed from the training folds only, so no held-out outcome enters
the target or the prediction; a held-out row whose course and term
have no training row takes the training mean of its course (\LinkHeldOutFallbackPct{}\% of
rows). This task describes how well entry characteristics order
students within a course offering whose average outcome is known. It
is retrospective, because the average of a course and term is
unavailable before the term ends. In the second task the models are
fitted on the entrants of Fall 2014 and Fall 2015 and predict the
grades of the entrants of Fall 2016 through Fall 2018, with the mean
grade of the same course in the earlier cohorts as the baseline (the
subject mean for the \LinkLaterFallbackPct{}\% of rows in a course the earlier cohorts did
not take). Every outcome baseline of this task is estimated from the
training cohorts. The task evaluates prediction across cohorts. The
extract does not establish when every recorded predictor became
available: the score fields have no test date, and the aid fields
have no aid year. In both tasks skill is the proportional reduction in
squared error against the baseline, and the contribution of a block of
predictors is the skill it adds when it enters in order and the skill
lost when it is dropped. The plan computed the course-by-term means
from all rows before the folds were formed; the training-only
construction is a correction made after the freeze (deviation D2), and
Appendix~\ref{sec:link} reports both. The variability of a block's
contribution is assessed by resampling sending schools, with the same
draw applied to every subject and the fitted predictions kept fixed.
The resulting intervals describe variability in the evaluation sample
given those predictions; they do not include the variability from
fitting and tuning new models.

\subsection{Section-size models: entry rule, inference, and checks}\label{app:methods:classsize}

Section~\ref{sec:data:link} states the comparisons and the hypotheses
of the section-size models. This subsection gives the entry rule, the
inference, and the checks.

\paragraph{Entry rule.}
A subject enters when at least \SizeEntryMinStudents{} students hold eligible rows and
section size varies after course-by-term and instructor effects are
removed (standard deviation of the logarithm at least \SizeEntryMinSdLogSize{}); \SizeSubjectsModeled{} of
the \SizeSubjectsConsidered{} subjects qualify, and the \SizeSubjectsOutsideWord{} left out hold their sections to
one size or are thin.

\paragraph{Inference.}
Dry runs of the pipeline, on grades shuffled within each course and
term, showed that the $t$ test with instructor-clustered standard
errors rejects too often in this design (\SizeDryClusterRejectMin{}\% to \SizeDryClusterRejectMax{}\% of the time at a
nominal 5\%, across the \SizeDrySubjects{} subjects of the calibration
run on file, with \SizeDryShuffles{} shuffles each). The $p$-value of
every section-size term therefore
comes from a wild cluster bootstrap of the $t$ statistic over
instructors, with the null imposed and \SizeBootstrapDraws{} draws, a method of the
family studied by \citet{cameron2008bootstrap}; under the same
shuffles it rejected \SizeDryWildRejectMin{}\% to \SizeDryWildRejectMax{}\% of the time. The 95\% interval of a
term is symmetric: the estimate plus or minus the 95th percentile of
the absolute bootstrap $t$ statistic times the instructor-clustered
standard error. For the quantities emphasized in the text we also
report the interval obtained by inverting the bootstrap test over a
range of null values, and the two agree closely. The slope of section
size at a given entry score, and the scenario for a third of the
entry-score distribution, are linear contrasts of the size
coefficient and the interaction. Each contrast is tested as one
coefficient of the reparametrized model with the same bootstrap, so
its interval includes the covariance of the two terms. The plan
specified clustered-$t$ intervals for these contrasts; the bootstrap
intervals replace them as a correction made after the freeze. Under
shuffled grades the bootstrap test of these contrasts rejected \SizeContrastWildRejectMin{}\%
to \SizeContrastWildRejectMax{}\% of the time at a nominal 5\% (\SizeContrastShuffles{} shuffles), against \SizeContrastClusterRejectMin{}\%
to \SizeContrastClusterRejectMax{}\% for the clustered $t$ test. The student fixed-effects model keeps clustered
intervals over its \SizeSameStudentInstructors{} instructors. Across subjects,
Benjamini--Hochberg control at $q=0.05$ is applied to the bootstrap
$p$-values. Pooled coefficients and totals use intervals that
include the dependence between subject estimates produced by shared
instructors and shared students, through a covariance matrix built
in three steps. Each row's influence on its subject's coefficient is
summed within instructors and within students, which gives a
two-way clustered covariance between subjects
\citep{cameron2011multiway}. That matrix is converted to a
correlation matrix and projected to the nearest positive
semidefinite matrix by setting negative eigenvalues to zero. The
correlations are then scaled by the bootstrap standard error of each
subject, the half-width of its interval divided by 1.96. The result
is therefore a hybrid: cross-subject correlations from the two-way
clustering, marginal uncertainty from the wild bootstrap. The
estimated cross-subject correlations are close to zero, so the
totals change little. The pooled intervals of the linked entrant
models use a separate construction, the joint school-clustered
covariance described above.

\paragraph{Checks and quantities seen before the freeze.}
Because a higher grade in a smaller section can reflect easier
grading, the grade in the next mathematics course, given the grade in
the first, serves as a limited sensitivity check: it uses only the
entrants who continue to another mathematics course, and it
conditions on the first grade. Two quantities survive the shuffle and were
seen before the freeze: the across-course coefficient, and the rate
at which entrants take a later mathematics course, which is reported
as descriptive. Instructor identifiers are replaced by a surrogate
inside the query, and no result is indexed by section or instructor.

\section{Additional Analysis Results}\label{app:results}

This appendix reports the analyses that support the results of
Section~\ref{sec:results}. Each context analysis follows a standalone
protocol: one source (or one audited set of aggregate sources) per
figure, the source population named in the caption, units held at the
source's own level, and no row-level student record in any plotted
dataset. Two context figures (Figures~\ref{figProvisionBands}
and~\ref{figSendingPoverty}) weight school-level or district-level
aggregates by entrant counts, and their estimands remain
entrant-conditioned in that sense. In the context figures shared
geography never creates a person-level link. Student records enter a
model in the linked entrant models of Appendix~\ref{sec:link} and in
the section-size models, under the rules of
Section~\ref{sec:data:link}. Nothing in this appendix supports a
causal reading.

\subsection{District course gradients: robustness and the
  per-capita measure}\label{app:results:gradients}

This subsection supports Section~\ref{sec:context:courses}.

Six checks support the contrast. Across the universe cycles held,
the gradients of the three widely taken core courses stay within four
points of zero (between $\GradCoreCyclesRelMin\%$ and $\GradCoreCyclesRelMax\%$ in 2015--16, 2017--18,
and 2021--22), while the Calculus gradient is \GradCalculusFirstCycleRelDecline{}\%, \CalcRelDecline{}\%, and
\GradCalculusLastCycleRelDecline{}\%. Keeping the districts with incomplete totals
(\GradIncompleteDistrictsWord{} in 2017--18) moves none
of these gradients by more than half a point. A neutral frame, which
admits every district reporting any core high-school mathematics or
science course, gives \GradCalculusNeutralFrameRelDecline{}\% for Calculus and between $\GradCoreNeutralFrameRelMin\%$ and
$\GradCoreNeutralFrameRelMax\%$ for the widely taken core courses. State fixed effects leave
the Calculus gradient at \GradCalculusStateFeRelDecline{}\% and the widely taken core courses within
three points of zero. A denominator check restricted to the
\GradExactSpanSchools{} schools whose grade span is exactly 9--12, where enrollment
needs no proration, gives \GradCalculusExactSpanRelDecline{}\% for Calculus (interval \GradCalculusExactSpanRelDeclineLow{} to \GradCalculusExactSpanRelDeclineHigh{})
and values within two points of zero for the widely taken core courses.
And within Texas (\CalcTexasDistricts{} districts) AP mathematics (\GradTexasApMathRelDecline{}\%), Calculus
(\CalcTexasRelDecline{}\%), and Algebra~I in grades 7--8 (\GradTexasEarlyAlgebraRelDecline{}\%) decline steeply.
Texas also supplies a policy check. House Bill 5 made Algebra~II
optional outside the endorsements that require it, 2017--18 is the
first cycle in which all four high-school grades entered under the
Foundation High School Program \citep{TexasHB5_2013}, and Algebra~II
enrollment per high-school student shows no poverty gradient there
($\GradTexasAlgebraTwoRel\%$, heteroskedasticity-robust $z=\GradTexasAlgebraTwoZ$). Texas supplies one
cluster, so these are within-state descriptive slopes.

A per-capita measure tells a different story for a reason unrelated
to courses. Divided by resident children aged 5--17, estimated
high-school enrollment itself declines by \GradHsEnrollmentPerChildRelDecline{}\% per ten percentage
points of child poverty (interval \GradHsEnrollmentPerChildRelDeclineLow{} to \GradHsEnrollmentPerChildRelDeclineHigh{}), and that decline attaches to
every course: per resident child the widely taken core courses appear
to fall at the rate of high-school enrollment (\GradCorePerChildRelDeclineMin{}\% to \GradCorePerChildRelDeclineMax{}\%), and
Calculus and AP mathematics and science reach \GradAdvancedPerChildRelDeclineMin{}\% to \GradAdvancedPerChildRelDeclineMax{}\%. Measuring each course on the students in its own grades removes
the enrollment geography and leaves the course.

\subsection{Sending-school provision and arriving
  credit}\label{sec:context:provision}

\begin{figure}[tbp]
  \centering
  \includegraphics[width=\textwidth]{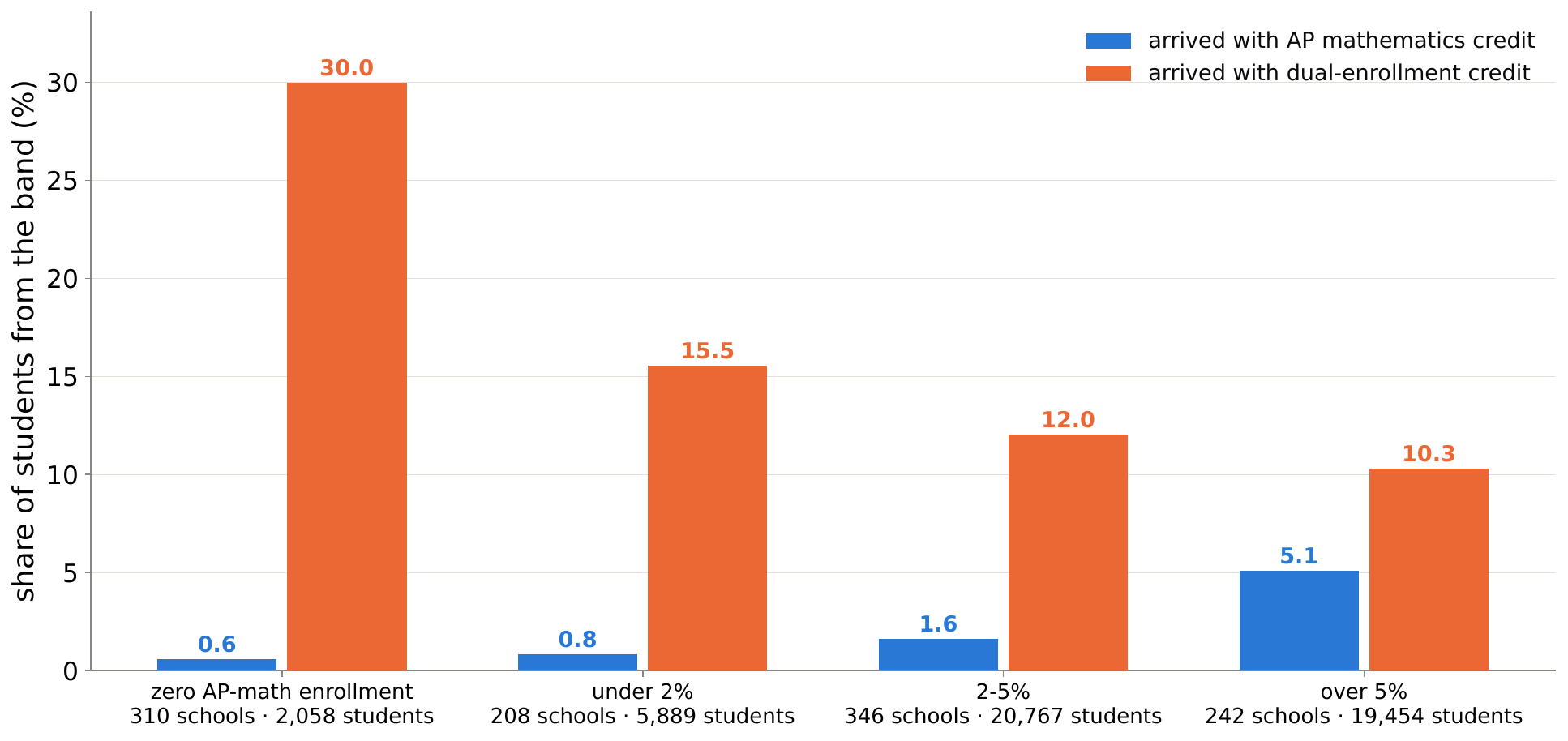}
  \caption{Credit arrival by sending-school AP-mathematics
    provision, among students active between Fall 2014 and Fall
    2019 whose high school is resolved. Sending schools are
    banded by the share of their own students enrolled in AP
    mathematics (CRDC 2017--18); bars show the share of the
    students from each band who arrived with AP credit (blue)
    and with dual-enrollment credit (orange); school and
    student counts under each band (1,106 schools,
    48,168 students in total). AP arrival rises from
    0.6\% to 5.1\% across the bands while
    dual-enrollment arrival falls from 30.0\% to
    10.3\%: the two arrival shares move in opposite
    directions across the bands. Provision is zero recorded
    \emph{enrollment} in the cycle, described at the school
    level; the exhibit profiles the sending-school context and
    is never treated as a student exposure history. Every cell
    holds at least ten students.}
  \label{figProvisionBands}
\end{figure}

Figure~\ref{figProvisionBands} profiles the AP mathematics provision
of the sending high schools, for the \XwalkProvisionStudents{} active
students whose high school resolves to a school of the 2017--18 CRDC
(\XwalkProvisionSchools{} schools).
Reading left to right, students from schools with zero recorded
AP-mathematics enrollment arrive with AP credit at
\XwalkNoApBandApPct{}\% and with dual-enrollment credit at
\XwalkNoApBandDualPct{}\%; students from schools where more than
\XwalkTopBandEdgePct{}\% of students take AP mathematics arrive at
\XwalkTopBandApPct{}\% and \XwalkTopBandDualPct{}\%. The two arrival
shares move in opposite directions across the provision bands, and
the zero-provision schools are small campuses (median enrollment
\XwalkNoApBandMedianEnrollment{}, against
\XwalkTopBandMedianEnrollment{} in the highest band). The figure is an
intake profile built from school-band aggregates, without adjustment
for school size or student characteristics; the linked models of
Appendix~\ref{sec:link} relate entry preparation to sending-school
context with adjustment. Provision here is a 2017--18
snapshot applied to students whose high-school years mostly precede
it. Between the 2015--16 and 2017--18 cycles the sending schools
hold their order imperfectly on the per-student rates used here (rank
correlations of 0.78 for AP science, 0.76 for AP mathematics, and
0.62 for Calculus and for dual enrollment); the snapshot is
therefore later school context, and it approximates each entrant's
own high-school offerings only to the extent that provision held
still.

\subsection{Child poverty of the sending
  districts}\label{sec:context:poverty}

\begin{figure}[tbp]
  \centering
  \includegraphics[width=\textwidth]{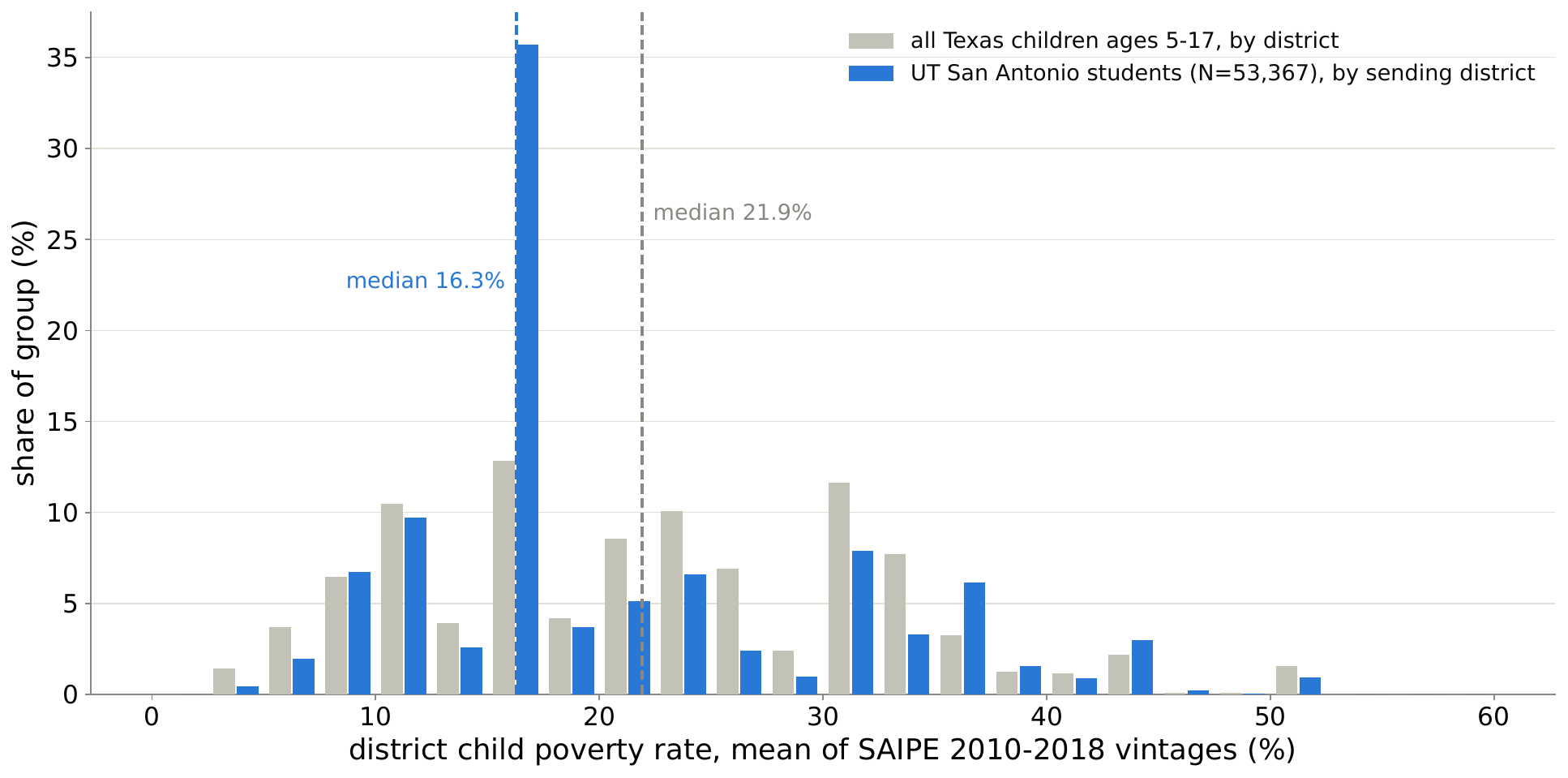}
  \caption{Child poverty of the sending districts. Gray: the
    distribution of district child poverty (mean SAIPE rate,
    2010--2018 vintages) over all Texas children aged 5--17,
    weighting each district by the children it holds. Blue: the
    same distribution over the 53,367 students active
    between Fall 2014 and Fall 2019 whose sending high school
    resolves to a Texas public school district
    (682 districts), weighting each district by the
    students it sent. Weights are aggregate counts; no
    student-level record enters the exhibit. Dashed lines mark
    the weighted medians.
    The exhibit reaches 88.9\% of
    the 60,045 students with a Texas high-school name. The
    district poverty of the other students is unobserved, so
    the blue distribution describes the matched students only.}
  \label{figSendingPoverty}
\end{figure}

Figure~\ref{figSendingPoverty} describes the distribution of child
poverty across the districts that send students to the institution,
for the \XwalkPovertyStudents{} active students whose high school
resolves to a Texas public school district (\XwalkPovertyDistricts{}
districts).
Weighted by the students each district sent, the median
sending-district child poverty rate is \XwalkPovertyMedianStudents{}\%;
weighted by the children each district holds, the Texas median is
\XwalkPovertyMedianChildren{}\%. The institution's intake therefore
sits about \XwalkPovertyMedianGap{} points below the state's
child-weighted center, and the single
\XwalkLargestBinLow{}--\XwalkLargestBinHigh{}\% bin holds
\XwalkLargestBinPct{}\% of these students because the two dominant
sending districts, Northside ISD (\XwalkNorthsideStudents{} students,
\XwalkNorthsidePovertyPct{}\% child poverty on the 2010--2018 mean)
and North East ISD (\XwalkNorthEastStudents{} students,
\XwalkNorthEastPovertyPct{}\%), both sit in it (the top sending
districts are tabulated in the figure's companion file). In the
sample the unresolved students receive Pell grants more often than
the matched students, a difference within the variation between
sending schools. Their district poverty is
unobserved, so the direction of any difference between the matched
distribution and the distribution over all students is unknown
(Appendix~\ref{app:data:selection}). Both
distributions are aggregates of district-level SAIPE rates; no
student-level income enters the analysis.

\subsection{From sending context to preparation and first-year
  performance}\label{sec:link}

The context analyses describe districts and schools. This subsection
turns to the students. Among the \LinkSample{} linked entrants
of Fall 2014 through Fall 2019, it estimates how family, school, and
district characteristics are associated with mathematics preparation
at entry, and how preparation is associated with first-year
performance in each of \LinkSubjects{} subjects. One standard deviation of the
entry mathematics score is \LinkScoreSD{} points; every estimate is conditional on
entry to the institution. The restriction to these cohorts was adopted
after the plan's first run on \LinkPlanCohortsWord{} entry cohorts
(Section~\ref{sec:data:link}); the last paragraphs of the subsection
describe what the restriction changes in the design.

\begin{figure}[tbp]
  \centering
  \includegraphics[width=\textwidth]{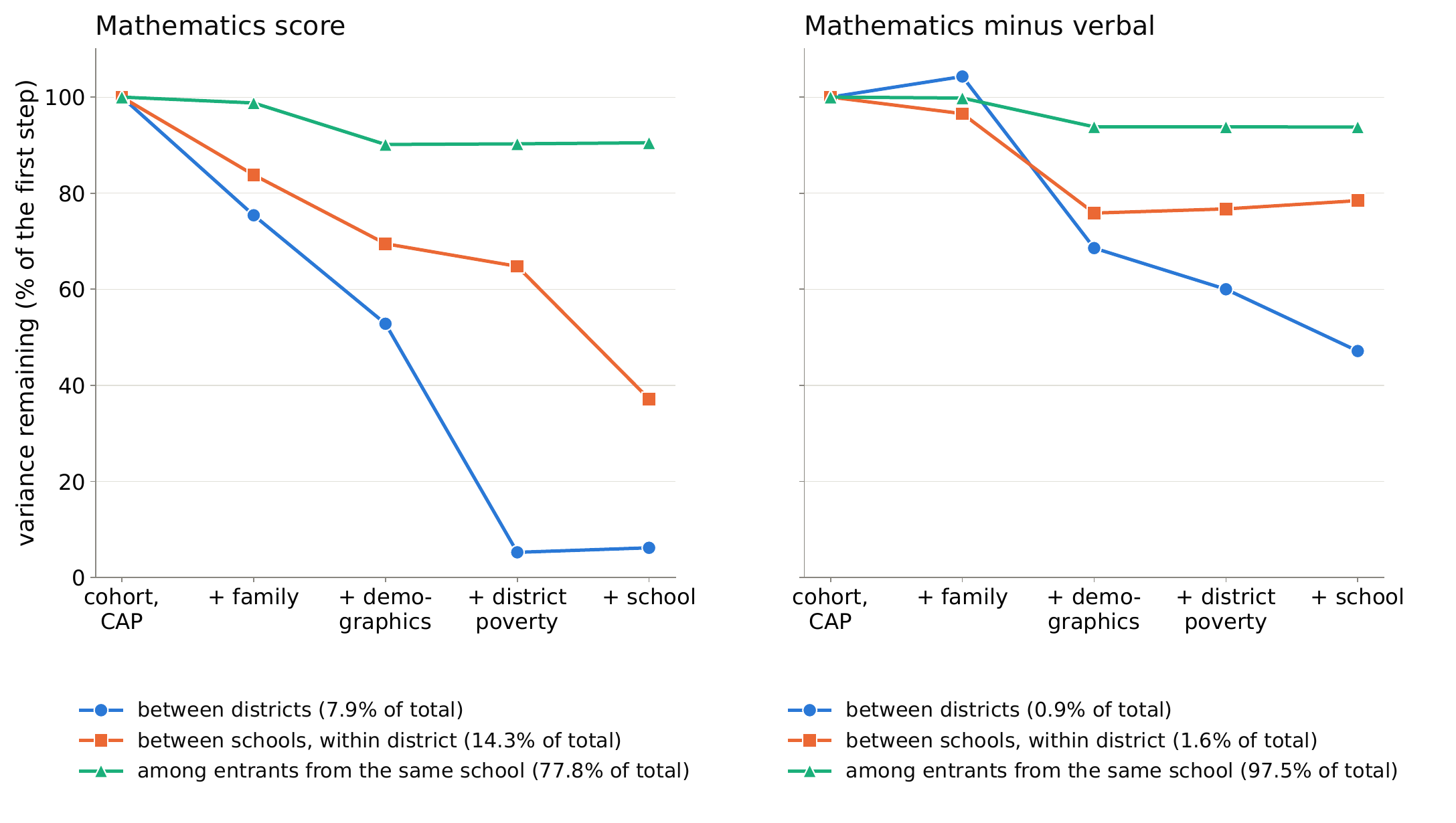}
  \caption{Where the variation in entry preparation sits, and
    which predictor blocks account for it, among the
    23,478 linked first-time fall entrants of Fall
    2014 to Fall 2019 (882 sending schools in
    529 districts). Each line follows one
    variance component of a nested random-intercept model as blocks
    enter in order: family resources (aid application, AGI rank,
    first-generation status), demographics, district child poverty,
    and the school block (advanced-mathematics index, AP science
    and dual-enrollment rates, enrollment). Left: the mathematics
    score, standardized within entry cohort. Right: the mathematics
    score minus the verbal score. Legend percentages give each
    component's share of total variance at the first step. All
    quantities are conditional on entry to the institution.}
  \label{figPreparationVariance}
\end{figure}

\paragraph{Where preparation varies.}
After accounting for entry cohort and CAP status, the random-intercept
model attributes \LinkVarDistrictPct{}\% of the remaining variance of the mathematics
score to differences between districts and \LinkVarSchoolPct{}\% to differences
between schools of the same district, \LinkVarBetweenOnePct{}\% in all. The other \LinkVarWithinPct{}\%
is variation among entrants from the same school
(Figure~\ref{figPreparationVariance}). This partition describes where
scores vary and does not identify the causes of the differences. When
family resources, demographics, and district child poverty enter the
model, the between-district component falls by \LinkVarDistrictReducedPct{}\% and the
between-school component by \LinkVarSchoolReducedPct{}\%; adding the school block brings the
reduction in the between-school component to \LinkVarSchoolReducedWithSchoolPct{}\%. In the full model,
one standard deviation of the school's
advanced-mathematics index goes with $\LinkIndexPointsSigned$ score points and one
standard deviation of district child poverty with $\LinkPovertyPointsSigned$; one
standard deviation of AGI rank goes with $\LinkIncomeRankPoints$ points, and
first-generation entrants score \LinkFirstGenPoints{} points lower than entrants who
are not first-generation. Without the aid variables, whose timing the
extract does not record, the two context coefficients are $\LinkNoAidIndexPoints$ and
$\LinkNoAidPovertyPoints$ score points.

\paragraph{Income between schools and within schools.}
The association of income with the mathematics score is substantially
larger between sending schools than within them: \LinkIncomeBetween{} standard
deviations of the score per standard deviation of income rank between
schools, against \LinkIncomeWithin{} among entrants from the same school, a
difference of \LinkIncomeBetweenMinusWithin{} ($z=\LinkIncomeBetweenMinusWithinZ$). The difference is consistent with
socioeconomic sorting across schools. The comparison does not separate
the influence of schools from that of families, neighborhoods, or
selection into the university.

\paragraph{Mathematics relative to verbal.}
For the standardized mathematics score minus the standardized verbal
score, \LinkGapVarWithinPct{}\% of the variation lies among entrants from the same
school; \LinkGapVarBetweenPct{}\% lies between schools and districts.
H1, that the school's advanced-mathematics index is positively
associated with this measure after family resources, is supported by
its pre-set rule with a small coefficient: $\LinkHOne$ standard
deviations per standard deviation of the index (interval \LinkHOneLow{} to
\LinkHOneHigh{}), or \LinkHOnePoints{} points on the scale of the mathematics score when one
standard deviation is taken as \LinkScoreSD{} points. The estimate is positive in
all \LinkSensitivityRunsWord{} sensitivity runs, and its interval reaches zero in \LinkSensitivityReachingZeroWord{} of
them (entrants through Fall 2015, and CAP entrants taken alone); it is
unchanged without the aid variables ($\LinkNoAidHOne$, \LinkNoAidHOneLow{} to \LinkNoAidHOneHigh{}). Income
rank shows no association with this measure among entrants from the
same school ($\LinkGapIncomeWithin$). The largest coefficients are those of
international students ($\LinkGapInternational$ standard deviations relative to
Hispanic students, with a standard error of \LinkGapInternationalSE{}), Asian
students ($\LinkGapAsian$), and men ($\LinkGapMale$ relative to women). With sending schools held out, the
predictors explain \LinkMathHeldOutPct{}\% of the variation in the mathematics score and
\LinkGapHeldOutPct{}\% of the variation in the difference between the two scores.
Among these entrants, district poverty and school context are
associated with the level of both scores and only weakly with the
difference between them.

\begin{figure}[p]
  \centering
  \includegraphics[width=0.80\textwidth]{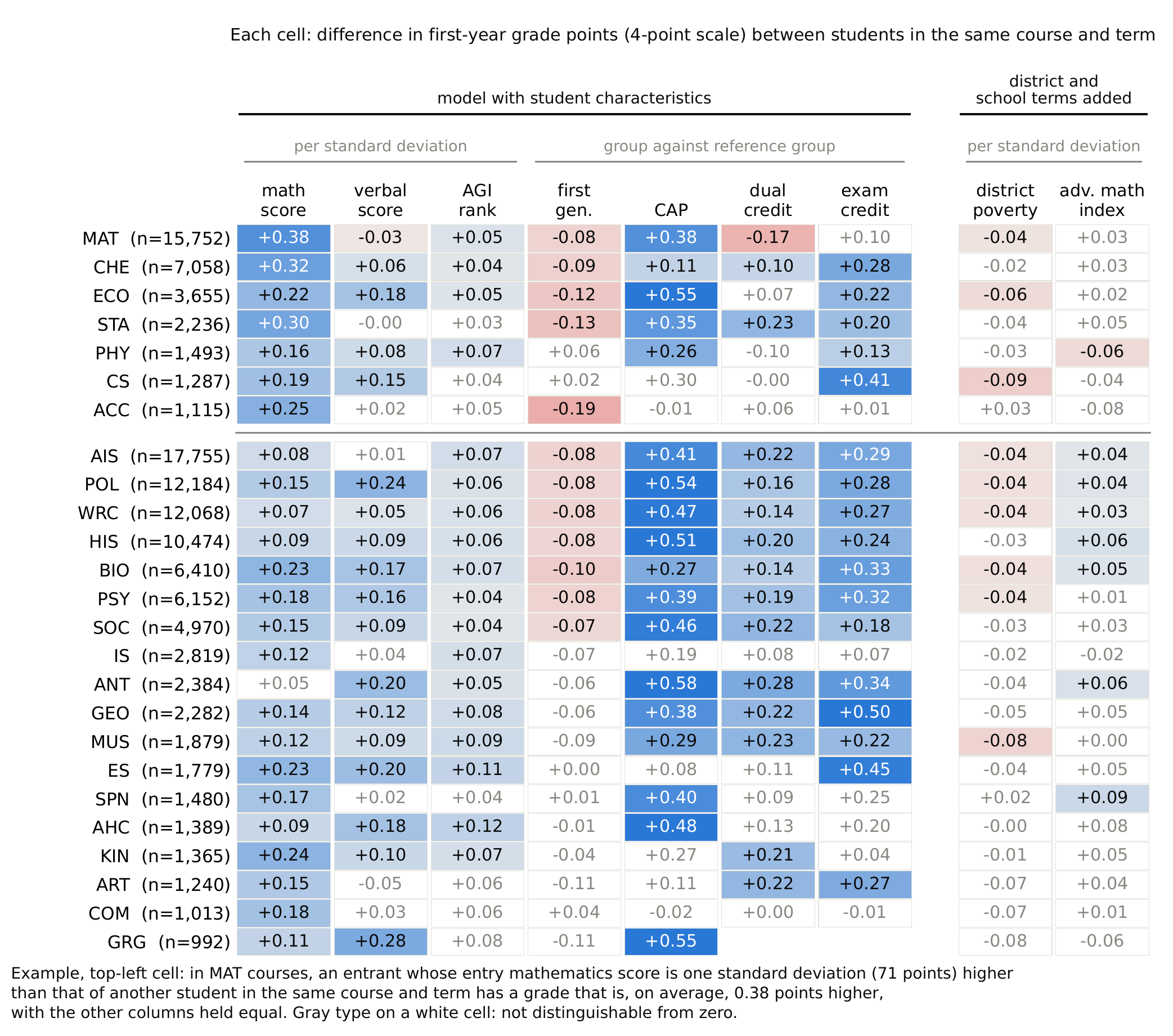}
  \caption{Associations between entry characteristics and
    first-year grades in the 25 subjects in which at
    least 1,000 linked entrants of Fall 2014 to Fall 2018 hold a
    first attempt. Each subject has two regressions with
    course-by-term fixed effects, so students are compared within
    the same course and term: the first seven columns come from a
    model with student characteristics, and the last two from the
    same model with district and school terms added. Each cell is
    the difference in grade points (4-point scale) that goes with
    the column's characteristic, the other characteristics of its
    model held equal; the figure prints a worked example for the
    top-left cell (0.38 grade points is about a third of
    a letter grade). Columns marked per standard deviation give
    the difference that goes with one standard deviation of the
    characteristic. Columns marked group against reference
    compare first-generation entrants with entrants who are not
    first-generation, CAP entrants with other entrants, and
    entrants holding each type of pre-college mathematics credit
    with entrants holding none. A shaded cell (blue positive, red
    negative) has an interval that excludes zero: score columns
    under false-discovery control at $q=0.05$ across subjects,
    other columns at unadjusted $p<0.05$. The 7 quantitative
    subjects sit above the rule. The count $n$ is the number of
    linked entrants with a graded first attempt and can fall
    below the 1,000 that admit a subject (the smallest is
    992). Standard errors are clustered on sending school;
    every cell is an association conditional on entry.
    Subject codes: MAT, mathematics; CHE, chemistry; ECO, economics; STA, statistics; PHY, physics; CS, computer science; ACC, accounting; AIS, the first-year seminar; POL, political science; WRC, writing composition; HIS, history; BIO, biology; PSY, psychology; SOC, sociology; IS, information systems; ANT, anthropology; GEO, geology; MUS, music; ES, environmental science; SPN, Spanish; AHC, art history and criticism; KIN, kinesiology; ART, art; COM, communication; GRG, geography.}
  \label{figSubjectPredictors}
\end{figure}

\paragraph{Preparation and first-year grades.}
Table~\ref{tab:subjectpooled} and Figure~\ref{figSubjectPredictors}
report the subject models, which compare students taking the same
course in the same term. Each row of the figure is one subject. The
first seven columns come from the model with student characteristics,
and the last two from the same model with district and school terms
added. Each cell is the difference in grade points that goes with the
column's characteristic, the other characteristics of its model held
equal. The top-left cell, for example, says that in mathematics
courses an entrant whose entry mathematics score is one standard
deviation (\LinkScoreSD{} points) higher than that of another student in the same
course and term has, on average, a grade \LinkMatMathGrade{} points higher. H2 is
supported: after
the verbal score is accounted for, the mathematics score has a
positive coefficient under false-discovery control in \LinkOtherSubjectsPassing{} of
the \LinkOtherSubjects{} non-quantitative subjects and in all \LinkQuantSubjects{} quantitative subjects.
Outside the quantitative subjects one standard deviation of the
mathematics score is associated with \LinkPooledMathOther{} higher grade points
(interval \LinkPooledMathOtherLow{} to \LinkPooledMathOtherHigh{}), against \LinkPooledVerbalOther{} for the verbal score, and
the difference between the two has an interval that includes zero
($\LinkPooledDiffOther$, $\LinkPooledDiffOtherLow$ to $\LinkPooledDiffOtherHigh$). Within the quantitative subjects
the mathematics coefficient is \LinkPooledMathQuant{} grade points (\LinkPooledMathQuantLow{} to \LinkPooledMathQuantHigh{})
and the verbal coefficient \LinkPooledVerbalQuant{}. These pooled values give every
subject a similar say. Weighted by first attempts, the mathematics
coefficient is \LinkWeightedMathOther{} grade points outside the quantitative subjects
and \LinkWeightedMathQuant{} within them, where mathematics is both the largest subject
and the one with the largest coefficient; the \LinkSubjectsPassing{} subjects in which
the mathematics coefficient passes hold \LinkPassingAttemptsPct{}\% of the first attempts.
In mathematics courses themselves the
coefficient is \LinkMatMathGrade{} grade points and $\LinkMatMathDfwPointsSigned$ percentage points of DFW
probability, and the verbal coefficient is slightly negative
($\LinkMatVerbalGrade$). The verbal coefficient exceeds the mathematics coefficient
in \LinkVerbalAboveMathSubjectsWord{} subjects, and the difference passes false-discovery
control in \LinkVerbalAboveMathPassingWord{} of them, political science and anthropology.
Coefficients estimated on all
entrants match the linked sample, a logit model agrees in sign with
the linear DFW model in every subject, and without the aid variables
the pooled mathematics coefficient outside the quantitative subjects
is \LinkNoAidPooledMathOther{}, with \LinkNoAidSubjectsPassing{} of \LinkSubjects{} subjects passing as before.

The same students and sending schools contribute to several subject
models. The intervals above use the joint covariance of the subject
coefficients (Section~\ref{sec:data:link}); the estimated correlation
between the mathematics coefficients of two subjects averages \LinkSubjectCorrMean{} and
reaches \LinkSubjectCorrMax{}. We also average, across the subjects outside the
quantitative group, the difference between the mathematics
coefficient and the verbal coefficient, weighting each subject by its
number of first attempts. With the covariance included, the interval
for this average includes zero ($\LinkWeightedDiffOtherLow$ to $\LinkWeightedDiffOtherHigh$). Under the
independence the plan assumed, the same interval excluded zero
(\LinkWeightedDiffOtherLowIndep{} to \LinkWeightedDiffOtherHighIndep{}), and the pooled interval of the
mathematics coefficient outside the quantitative subjects was
\LinkPooledMathOtherLowIndep{} to \LinkPooledMathOtherHighIndep{}.
The \LinkSubjectsPassing{} positive subject results are therefore correlated observations
on overlapping students, and the data give no basis for ranking
mathematics above verbal preparation outside the quantitative
subjects.

\paragraph{Predictive contribution of district and school variables.}
In the first prediction task, sending schools are held out and the
target is a student's grade relative to the mean of the same course
and term, computed from the training folds. The full set of predictors
reduces squared error by a median \LinkSkillMedianPct{}\% across subjects (range \LinkSkillMinPct{}\% to
\LinkSkillMaxPct{}\%; \LinkSkillWeightedPct{}\% when subjects are weighted by first attempts). The
preparation block (scores and credit route) contributes a median \LinkPrepAdded{}
to this proportional reduction in squared error (interval \LinkPrepAddedLow{} to
\LinkPrepAddedHigh{}, from resampling sending schools with the fitted predictions
kept fixed); family resources contribute \LinkFamilyAdded{} and demographics \LinkDemoAdded{}.
District poverty contributes a median \LinkPovertyAdded{} (interval $\LinkPovertyAddedLow$ to
$\LinkPovertyAddedHigh$), between $\LinkPovertyAddedMin$ and $\LinkPovertyAddedMax$ across subjects and
positive in \LinkPovertyAddedPositive{} of the \LinkSubjects{}. The school block contributes a median
$\LinkSchoolAdded$ (interval $\LinkSchoolAddedLow$ to \LinkSchoolAddedHigh{}), between $\LinkSchoolAddedMin$ and
$\LinkSchoolAddedMax$ across subjects and positive in \LinkSchoolAddedPositive{}. The evaluation the plan
specified, with course-by-term means computed from all rows, gave
nearly the same values (\LinkPlanSkillMedianPct{}\% for the full set, \LinkPlanPrepAdded{} for
preparation, \LinkPlanPovertyAdded{} for district poverty, $\LinkPlanSchoolAdded$ for the school
block). In the second task, models fitted on the entrants of Fall
2014 and Fall 2015 predict the grades of the three later cohorts
against the earlier mean grade of each course. Ridge regression
reduces squared error by a median \LinkLaterRidgeMedianPct{}\% (range \LinkLaterRidgeMinPct{}\% to \LinkLaterRidgeMaxPct{}\%), and the
unpenalized linear model, which overfits in the smaller subjects, by
\LinkLaterLinearMedianPct{}\%; district poverty contributes a median \LinkLaterPovertyAdded{} to the linear
model and the school block $\LinkLaterSchoolAdded$. For the DFW outcome the predicted
probability separates outcomes with a median area under the curve of
\LinkAucHeldOut{} in the first task and \LinkAucLater{} in the second, against \LinkAucHeldOutBaseline{} and \LinkAucLaterBaseline{}
for the baseline rate alone.

Adding district poverty and the school course-enrollment variables
therefore produced little change in median prediction accuracy beyond
the entry scores, the credit route, and the other student
characteristics, and the changes varied in sign across subjects. This
result is specific to the available predictors and to these two
evaluation designs. It does not establish that school context is
irrelevant to performance: the adjusted coefficient of district
poverty is negative with an interval excluding zero in several
subjects (Figure~\ref{figSubjectPredictors}), and a coefficient and an
increment in predictive accuracy measure different things. A random
forest did not improve on the linear grade model by more than \LinkForestGainHeldOut{} in
the first task, and it did not improve on ridge regression by more
than \LinkForestGainLater{} in the second. We found no larger improvement with this
alternative model, a comparison that does not rule out interactions
or curvature the forest also missed.

\paragraph{Other predictors.}
Within the same course and term, CAP entrants have grades \LinkCapGradeMin{} to \LinkCapGradeMax{}
points higher than other entrants in \LinkCapSubjects{} of the \LinkSubjects{} subjects.
Examination credit (AP or CLEP)
goes with higher grades in \LinkExamCreditSubjects{} subjects. Dual-enrollment credit goes
with higher grades outside mathematics and with lower grades in
mathematics courses ($\LinkMatDualCredit$), where the examination-credit
coefficient is positive ($\LinkMatExamCredit$) with an interval that includes
zero. First-generation entrants have grades \LinkFirstGenGradeMin{} to \LinkFirstGenGradeMax{} points lower
in the \LinkFirstGenSubjects{} subjects whose interval excludes zero (unadjusted for
multiplicity).

\paragraph{The restriction of cohorts.}
Section~\ref{sec:data:link} states how the restriction to entrants
of Fall 2014 onward was adopted, after the plan's first run on
\LinkPlanCohortsWord{} cohorts and with its results known.
In the design, the restriction leaves \LinkSubjects{} subjects above the
\LinkSubjectMinimum{}-student minimum where the \LinkPlanCohortsWord{} cohorts left 31, and it
places the school snapshot of every entrant within \LinkSnapshotMaxLagWord{} year of the senior year,
against as many as \LinkSnapshotMaxLagEarlierWord{} in the earlier cohorts. One variable is an
exception: for the \LinkDualSubstituteLinked{} linked entrants assigned the 2013--14 cycle,
the dual-enrollment rate is that of 2015--16, as many as
\LinkDualSubstituteMaxLagWord{} years after the senior year.

\paragraph{Scope.}
The limits of scope are those of
Section~\ref{sec:discussion:limitations}: selection into enrollment,
and entry scores that absorb general preparation. Subject models
measure first-attempt grades and DFW outcomes given
the course taken; the course in which an entrant starts is itself
related to preparation, and the
advanced-mathematics index goes with a higher entry mathematics course
($\LinkIndexEntryLevel$ levels on a \LinkEntryLevelsWord{}-level scale per standard deviation). The
analyses relate school context to preparation and preparation to
grades; they do not estimate how much of the association between
poverty and grades is caused or mediated by school course provision
or by preparation.

\begin{table}[tbp]
\centering\footnotesize
\setlength{\tabcolsep}{4pt}
\caption{Coefficients of the two entry scores in the
  first-year subject models, per standard deviation of each score,
  with 95\% intervals. For each outcome and subject class the
  first row is the prespecified random-effects pooling across
  subjects, which gives every subject a similar say; the second
  averages the subject coefficients with weights equal to the
  first attempts in each subject, a descriptive summary on the
  scale of the entrants served.
  Grades are on the 4-point scale; DFW is the probability of a D,
  F, withdrawal, or no-credit outcome, in percentage points.
  The same students and sending schools contribute to several
  subject models, so the intervals use the joint school-clustered
  covariance of the subject coefficients, a correction made after
  the plan was frozen (the plan treated subjects as independent).
  Subjects are quantitative (MAT, CHE, ECO, STA, PHY, CS, ACC) or
  other, as classed before estimation. Population: linked
  entrants of Fall 2014 to Fall
  2018, a restriction adopted after
  the prespecified eleven-cohort run.}
\label{tab:subjectpooled}
\begin{tabular}{@{}llll@{}}
\toprule
Outcome, subject class (subjects) & Mathematics & Verbal & Mathematics minus verbal \\
\midrule
Grade points, other (18) & $+0.140$ {\scriptsize($+0.109$, $+0.171$)} & $+0.112$ {\scriptsize($+0.069$, $+0.156$)} & $+0.027$ {\scriptsize($-0.015$, $+0.068$)} \\
\quad weighted by first attempts & $+0.125$ {\scriptsize($+0.108$, $+0.142$)} & $+0.107$ {\scriptsize($+0.093$, $+0.122$)} & $+0.018$ {\scriptsize($-0.008$, $+0.043$)} \\
Grade points, quantitative (7) & $+0.263$ {\scriptsize($+0.197$, $+0.328$)} & $+0.065$ {\scriptsize($+0.000$, $+0.129$)} & $+0.197$ {\scriptsize($+0.073$, $+0.320$)} \\
\quad weighted by first attempts & $+0.316$ {\scriptsize($+0.291$, $+0.340$)} & $+0.032$ {\scriptsize($+0.012$, $+0.053$)} & $+0.283$ {\scriptsize($+0.244$, $+0.323$)} \\
DFW, points, other (18) & $-2.2$ {\scriptsize($-3.0$, $-1.5$)} & $-1.6$ {\scriptsize($-2.9$, $-0.3$)} & $-0.7$ {\scriptsize($-1.8$, $+0.5$)} \\
\quad weighted by first attempts & $-2.0$ {\scriptsize($-2.5$, $-1.6$)} & $-1.5$ {\scriptsize($-1.9$, $-1.0$)} & $-0.6$ {\scriptsize($-1.3$, $+0.2$)} \\
DFW, points, quantitative (7) & $-5.1$ {\scriptsize($-7.8$, $-2.4$)} & $-0.9$ {\scriptsize($-2.7$, $+0.9$)} & $-4.2$ {\scriptsize($-8.3$, $+0.0$)} \\
\quad weighted by first attempts & $-7.7$ {\scriptsize($-8.4$, $-6.9$)} & $+0.2$ {\scriptsize($-0.3$, $+0.8$)} & $-7.9$ {\scriptsize($-9.0$, $-6.8$)} \\
\bottomrule
\end{tabular}
\end{table}

\subsection{National position of the
  institution}\label{sec:results:position}

\begin{figure}[tbp]
  \centering
  \includegraphics[width=\textwidth]{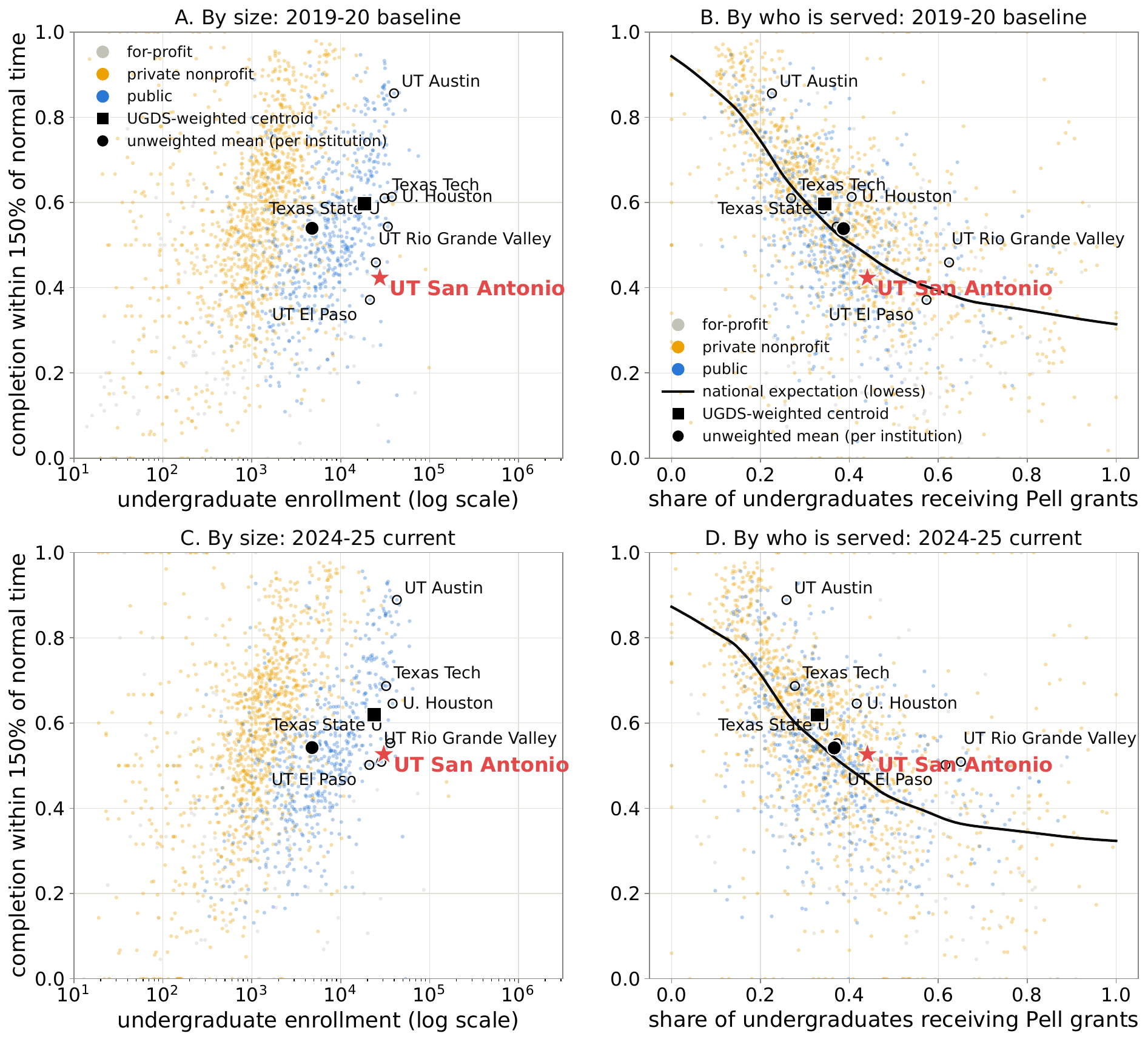}
  \caption{National position of UT San Antonio among the
    predominantly-bachelor's Title IV institutions in the College
    Scorecard, at \ScoreVintagesWord{} explicit vintages. Top row (A, B): the 2019--20
    data year (\ScoreBaselineInstitutions{} institutions), the pre-disruption baseline
    vintage, whose completion cohorts closed before Fall 2019. Bottom
    row (C, D): the 2024--25 data year (\ScoreCurrentInstitutions{} institutions), the most
    recent completion data in the June 2026 release; this row reports
    the legacy pre-merger reporting unit \citep{UTSAMerger2025}. Left
    column: completion of first-time, full-time entrants within
    150\% of normal time vs.\ undergraduate enrollment (log scale).
    Right column: completion vs.\ share of Pell-grant recipients,
    with the national lowess expectation; the gap between UT San
    Antonio and the curve is a Pell-conditional descriptive residual.
    Public institutions blue, private nonprofit yellow, for-profit
    gray; Texas public peers circled. The black square is the
    enrollment-weighted institutional centroid (institutions weighted
    by current undergraduate enrollment; the completion coordinate
    remains a first-time full-time cohort outcome); the black circle
    is the unweighted mean across institutions in raw coordinates,
    one vote per institution regardless of size, plotted here on the
    log axis.}
  \label{figScorecard}
\end{figure}

As of the most recent data year with completion outcomes (2024--25),
UT San Antonio reports \ScoreCurrentUndergraduates{} undergraduates, first-time full-time
completion within 150\% of normal time of \ScoreCurrentCompletion{}, and a Pell share of
\ScoreCurrentPellShare{}; a locally weighted fit of completion on Pell share across the
Scorecard's predominantly-bachelor's institutions puts the
conditional expectation at that Pell share at \ScoreCurrentExpectedCompletion{}, a descriptive
residual of $\ScoreCurrentResidual$ \citep{Scorecard2026}. In the pre-disruption
vintage (2019--20 data year), the same coordinates were \ScoreBaselineUndergraduates{}
undergraduates, completion \ScoreBaselineCompletion{} against an expectation of \ScoreBaselineExpectedCompletion{}, a
residual of $\ScoreBaselineResidual$. One scope note applies: UT San Antonio and UT
Health San Antonio merged effective September 1, 2025, so the
2024--25 row describes the legacy reporting unit rather than the
merged institution \citep{UTSAMerger2025}. The residual measures
position relative to the national expectation; attribution of the
movement between the two vintages is a question reserved for later
comparison studies.

\subsection{Comparison-group definition: a design
  comparison}\label{sec:results:design}

Comparisons that mix entry pathways measure those pathways rather
than the course, and the required first-year seminar demonstrates the
failure. The institution's stored procedures compute, for any
subject, the odds ratio between taking the subject and having no
recorded enrollment in a course numbered above 4000 (the senior
level) among all students in a window. For the
required seminar AIS, that naive design reports an adverse odds ratio
of \InstNaiveOddsAis{} (Table~\ref{tab:oddsnaivecohort}): takers of the required
seminar appear to fare worse. The number is an artifact of the
comparison group. Nearly every first-time freshman takes AIS (\InstAisTakersPct{}\%
of the reference cohort), so the ``unexposed'' group in the naive
design consists almost entirely of transfer entrants, who bring
credit earned elsewhere and reach the senior level quickly, and of
late entrants whose observation window is censored. Restricting both
groups to the reference entry cohort with a fixed window reverses the
association to strongly favorable (\InstOddsAis{}), because within the cohort
the students who skip the required seminar are those who depart
immediately; the small unexposed remainder is itself highly selected,
so even the cohort-restricted AIS estimate describes that selection
as much as the seminar. The mathematics and writing gateways behave
differently under the same restriction: MAT holds (\InstNaiveOddsMat{} naive, \InstOddsMat{}
cohort) and WRC moderates while staying adverse (\InstNaiveOddsWrc{} to \InstOddsWrc{}). The
restriction removes the entry-pathway artifact; it does not adjust
for preparation, major, attempted load, or any other covariate, and
exposure and outcome are measured over the same window, so these
remain contemporaneous marginal associations. The example motivates
the rule followed in the institutional analyses of this paper: each
association is computed on a defined entry population with a stated
window (Table~\ref{tab:populations}), and any
statistic whose comparison group mixes entry pathways is treated as a
measurement of those pathways.

\begin{table}[tbp]
  \centering
  \small
  \caption{Association between core-subject exposure and having no enrollment in a course numbered above 4000 within the window, under two designs. The naive design (columns 2--4; stored procedure over the full Fall 2014--Fall 2019 window) compares takers of each subject with every other student in the window, a group that mixes transfer entrants holding credit earned elsewhere with right-censored late entrants. The cohort design (columns 5--7) restricts both groups to first-time freshmen entering Fall 2014--Fall 2015 with the fixed observation window of Section~\ref{sec:data:cohort}. Odds ratios above one indicate higher odds of having no such enrollment among takers; $p$ from Fisher's exact test; 95\% confidence intervals by the log (Woolf) method. Generated by \texttt{cohort\_tables.py}.}
  \label{tab:oddsnaivecohort}
  \begin{tabular}{@{}lrrl@{\qquad}rrl@{}}
    \toprule
    & \multicolumn{3}{l}{Naive (all students in window)} & \multicolumn{3}{l@{}}{Cohort-restricted (entry cohort)} \\
    \cmidrule(r){2-4}\cmidrule(l){5-7}
    Subject & OR & 95\% CI & $p$ & OR & 95\% CI & $p$ \\
    \midrule
    MAT & 1.47 & [1.43, 1.51] & $<10^{-4}$ & 1.46 & [1.27, 1.67] & $<10^{-4}$ \\
    WRC & 1.80 & [1.74, 1.85] & $<10^{-4}$ & 1.39 & [1.28, 1.52] & $<10^{-4}$ \\
    AIS & 1.81 & [1.76, 1.87] & $<10^{-4}$ & 0.23 & [0.19, 0.28] & $<10^{-4}$ \\
    \bottomrule
  \end{tabular}
\end{table}

\subsection{Introductory-course outcomes, persistence, and major
  change}\label{app:results:majorchange}

This subsection supports Section~\ref{sec:baseline:flows}.

\paragraph{Persistence.}
Table~\ref{tab:persistencedfw} compares, subject by subject, the
share of entrants not observed later between those who received a D,
F, withdrawal, or no-credit outcome (DFW) on a first attempt in an
introductory course during the first year and those who took such a
course without one. It uses the \PersistPopulationN{} entrants of the reference
cohort without the Coordinated Admission Program flag. The
differences run from \PersistDiffMin{} to \PersistDiffMax{} percentage points across the
\PersistSubjectsWord{} subjects, and every interval lies above zero. Among the
\PersistAnyN{} entrants with a first-year attempt in any introductory course,
\PersistAnyDfwN{} (\PersistAnyDfwPct{}\%) hold a DFW in at least one, and \PersistAnyLeftWithPct{}\% of them
are not observed later, against \PersistAnyLeftWithoutPct{}\% of the others.

Counted course by course, \PersistMatOutcomes{} of the \PersistOutcomes{} DFW
outcomes are in mathematics. The
entrants not observed later hold \PersistLeftOutcomes{} of the outcomes,
\PersistLeftMatOutcomes{} of them in mathematics; among the entrants
observed later the mathematics share is \PersistStayMatOutcomeSharePct{}\%, and
\PersistStayWithDfwPct{}\% of them hold at least one DFW outcome.
\PersistFourPlusN{} entrants hold four or more DFW outcomes.

In mathematics, the \PersistMatFailN{} entrants with a D or F in a completed
course are not observed later at \PersistMatFailLeftWithPct{}\% (a difference of
\PersistMatFailDiff{} points from the entrants without a DFW, interval \PersistMatFailDiffLo{} to
\PersistMatFailDiffHi{}), and the \PersistMatWithdrewN{} with a withdrawal or no-credit outcome
only at \PersistMatWithdrewLeftPct{}\%; a withdrawal can be part of a departure already
under way. Of the entrants with a mathematics DFW, \PersistMatGoneAfterTermPct{}\% hold
no record after the term of that course. Of the same entrants,
\PersistMatBothN{} also hold a DFW in another introductory subject and
\PersistMatOnlyN{} do not. In the first group the difference from the entrants
without a mathematics DFW is \PersistMatWithElsewhereDiff{} points (\PersistMatWithElsewhereDiffLo{} to
\PersistMatWithElsewhereDiffHi{}), and in the second it is \PersistMatAloneDiff{} points
($\PersistMatAloneDiffLo$ to \PersistMatAloneDiffHi{}). With the Coordinated Admission Program
entrants included, the mathematics difference is \PersistMatWholeCohortDiff{} points
(\PersistMatWholeCohortDiffLo{} to \PersistMatWholeCohortDiffHi{}), because those entrants leave after the
first year whatever their grades.

Three limits apply. The comparisons are unadjusted for preparation,
first major, and course load. The record does not distinguish leaving
higher education from transferring to another institution. And an
entrant can hold a DFW in several subjects, so the rows are
overlapping descriptions and cannot be added.

\begin{table}[tbp]
\centering\footnotesize
\setlength{\tabcolsep}{4pt}
\caption{First-year outcomes in introductory courses and
  persistence, by subject, in the reference cohort. For each
  subject, the entrants holding a first attempt in a course
  numbered below 2000 during the first year are split by
  whether any such first attempt ended in a D, F, withdrawal, or
  no-credit outcome (DFW). Not observed later: no course record in
  a term one year or more after the entry term, through the end of
  the observation window (Section~\ref{sec:data:cohort}); an
  entrant who stopped out and returned counts as observed, and a
  transfer to another institution cannot be told from leaving
  higher education. The table uses the 8,837 entrants of the
  cohort (10,027) without the Coordinated Admission Program
  flag; the 1,190 entrants with the flag transfer after the
  first year by the design of that program. Parentheses hold
  pointwise 95\%
  intervals: a Wald interval for a difference, in percentage
  points, and the log (Woolf) method for the odds ratio. Subjects
  with at least 400 entrants holding such an attempt,
  ordered by the number with a DFW. An entrant can appear in
  several rows. The comparisons are unadjusted and exploratory:
  they were computed after the analysis plans were frozen, and
  the intervals are not adjusted for the number of subjects.}
\label{tab:persistencedfw}
\begin{tabular}{@{}lrrrrcc@{}}
\toprule
 & & & \multicolumn{2}{c}{Not observed later (\%)} & Difference & \\
\cmidrule(lr){4-5}
Subject & Entrants & With DFW & with DFW & without & (points) & Odds ratio \\
\midrule
Mathematics & 7,782 & 3,164 & 34.8 & 14.8 & $+20.0$ $(18.0,\,21.9)$ & 3.07 $(2.75,\,3.42)$ \\
Political science & 5,050 & 1,673 & 34.3 & 13.8 & $+20.5$ $(18.0,\,23.1)$ & 3.26 $(2.83,\,3.75)$ \\
Writing composition & 6,036 & 1,665 & 43.7 & 15.1 & $+28.6$ $(26.0,\,31.2)$ & 4.36 $(3.84,\,4.95)$ \\
History & 4,829 & 1,076 & 43.3 & 15.1 & $+28.3$ $(25.1,\,31.4)$ & 4.31 $(3.71,\,5.01)$ \\
First-year seminar & 7,995 & 1,046 & 50.9 & 14.7 & $+36.2$ $(33.1,\,39.3)$ & 6.02 $(5.25,\,6.92)$ \\
Chemistry & 3,554 & 890 & 33.1 & 13.6 & $+19.5$ $(16.2,\,22.9)$ & 3.14 $(2.63,\,3.76)$ \\
Biology & 2,667 & 760 & 34.1 & 13.6 & $+20.5$ $(16.8,\,24.2)$ & 3.29 $(2.70,\,4.01)$ \\
Psychology & 2,585 & 647 & 42.5 & 15.9 & $+26.6$ $(22.5,\,30.8)$ & 3.91 $(3.21,\,4.77)$ \\
Sociology & 2,125 & 395 & 47.8 & 16.3 & $+31.5$ $(26.3,\,36.8)$ & 4.71 $(3.72,\,5.96)$ \\
Geology & 1,094 & 229 & 46.7 & 13.5 & $+33.2$ $(26.3,\,40.1)$ & 5.61 $(4.05,\,7.76)$ \\
Computer science & 600 & 168 & 24.4 & 9.5 & $+14.9$ $(7.9,\,22.0)$ & 3.08 $(1.91,\,4.96)$ \\
Information systems & 1,359 & 146 & 46.6 & 12.8 & $+33.8$ $(25.5,\,42.1)$ & 5.95 $(4.13,\,8.58)$ \\
Physics & 554 & 145 & 20.0 & 12.2 & $+7.8$ $(0.5,\,15.0)$ & 1.79 $(1.09,\,2.97)$ \\
Statistics & 881 & 134 & 32.1 & 13.4 & $+18.7$ $(10.4,\,27.0)$ & 3.06 $(2.01,\,4.65)$ \\
Spanish & 627 & 119 & 40.3 & 13.6 & $+26.8$ $(17.4,\,36.1)$ & 4.30 $(2.75,\,6.72)$ \\
Art history & 527 & 112 & 50.9 & 18.3 & $+32.6$ $(22.6,\,42.6)$ & 4.62 $(2.96,\,7.22)$ \\
Art & 797 & 100 & 49.0 & 17.1 & $+31.9$ $(21.7,\,42.1)$ & 4.67 $(3.01,\,7.24)$ \\
Astronomy & 402 & 95 & 31.6 & 8.1 & $+23.4$ $(13.6,\,33.3)$ & 5.21 $(2.87,\,9.44)$ \\
Criminal justice & 407 & 76 & 61.8 & 11.2 & $+50.7$ $(39.2,\,62.1)$ & 12.88 $(7.24,\,22.89)$ \\
\bottomrule
\end{tabular}
\end{table}

\paragraph{Major change.}
Table~\ref{tab:majorchangedfw} compares, subject by subject, the
share of entrants who changed major between those who received a D,
F, withdrawal, or no-credit outcome (DFW) on a first attempt in an
introductory course during the first year and those who took such a
course without one. It reports two populations, because the
comparison depends on which entrants remain observable.

The first population is the \InstLaterObservedN{} entrants of the reference cohort
observed enrolled at least one year after entry: those with an
enrollment record in a term one year or more after the entry term. An
entrant who stopped out and returned qualifies, so the rule is weaker
than enrollment at the one-year anniversary. In this population,
\InstLaterObservedChangePct{}\% changed major. The second population is all entrants, whatever
their later enrollment. The \InstNotObservedN{} entrants with no record after the
first year had less opportunity to record a change of major, and
\InstNotObservedChangePct{}\% of them did. Remaining observable can itself depend on grades,
academic plans, transfer under the Coordinated Admission Program, and
departure, so the first population is a selected one and its
comparison describes later-observed entrants only.

For mathematics the difference is \InstDfwMatLaterDiff{} percentage points among
later-observed entrants (interval \InstDfwMatLaterDiffLo{} to \InstDfwMatLaterDiffHi{}; odds ratio \InstDfwMatLaterOdds{}, \InstDfwMatLaterOddsLo{}
to \InstDfwMatLaterOddsHi{}) and \InstDfwMatAllDiff{} points among all \InstDfwMatAllN{} entrants with a first-year
attempt (\InstDfwMatAllDiffLo{} to \InstDfwMatAllDiffHi{}; odds ratio \InstDfwMatAllOdds{}, \InstDfwMatAllOddsLo{} to \InstDfwMatAllOddsHi{}). For chemistry
the two differences are \InstDfwCheLaterDiff{} points (\InstDfwCheLaterDiffLo{} to \InstDfwCheLaterDiffHi{}) and \InstDfwCheAllDiff{} points
($\InstDfwCheAllDiffLo$ to \InstDfwCheAllDiffHi{}). In writing composition, history, and the first-year
seminar the difference among all entrants is negative ($\InstDfwWrcAllDiff$,
$\InstDfwHisAllDiff$, and $\InstDfwAisAllDiff$ points). Mathematics and physics are the only subjects whose difference is
positive with an interval above zero in both populations. Entrants
whose first recorded major is University College hold no declared
major and change it when they declare one. Without them the
mathematics difference among later-observed entrants is \InstDfwMatDeclaredDiff{}
percentage points among \InstDfwMatDeclaredN{} entrants (odds ratio \InstDfwMatDeclaredOdds{}, \InstDfwMatDeclaredOddsLo{} to
\InstDfwMatDeclaredOddsHi{}).

The flag for a change of major follows the rules of the cohort tables
(Section~\ref{sec:data:cohort}): records before the entry term are
admitted (\InstPreEntryRecordsN{} students hold one), and a missing first or last major
counts as a change (\InstMissingMajorN{} students). No student of the cohort holds two
majors in the first or the last term of the window, so the order of
records within a term does not affect the flag. Under an
entry-bounded rule that keeps only confirmed changes, \InstBoundedConfirmedChangePct{}\% of the
\InstBoundedConfirmedN{} students with both majors recorded changed major
(\InstConfirmedChangeCompletePct{}\% when records before the entry term are admitted, as in
Section~\ref{sec:data:cohort}), and the
mathematics differences are \InstDfwMatBoundedLaterDiff{} points (\InstDfwMatBoundedLaterDiffLo{} to \InstDfwMatBoundedLaterDiffHi{}) among
later-observed entrants and \InstDfwMatBoundedAllDiff{} points (\InstDfwMatBoundedAllDiffLo{} to \InstDfwMatBoundedAllDiffHi{}) among all
entrants.

Two limits of the persistence comparison apply here as well: the
comparisons are unadjusted, and the rows overlap. Two more are
specific to major change. The change of major
is measured from the first to the last recorded major, so its timing
relative to the course outcome is not established. And \InstDfwSubjectsWord{} subjects are compared without an adjustment
for multiplicity. The analysis is descriptive and was computed after
the analysis plans were frozen.

\begin{table}[tbp]
\centering\footnotesize
\setlength{\tabcolsep}{3pt}
\caption{First-year outcomes in introductory courses and later
  major status, by subject, in the reference cohort. For each
  subject, the entrants holding a first attempt in a course
  numbered below 2000 during the first year are split by whether
  any such first attempt ended in a D, F, withdrawal, or no-credit
  outcome (DFW). Changed major: the first and last recorded majors
  within the observation window differ
  (Section~\ref{sec:data:cohort}). Columns 2 to 7 use the 7,250
  entrants observed enrolled at least one year after entry (a
  record in a term one year or more after the entry term; an
  entrant who stopped out and returned qualifies). The last column
  gives the same difference among all entrants with such an
  attempt (the cohort holds 10,027 entrants), whatever their
  later enrollment. Parentheses hold 95\% intervals:
  a Wald interval for a difference, in percentage points, and the
  log (Woolf) method for the odds ratio. Subjects with at least
  400 later-observed entrants, ordered by the number with
  a DFW. An entrant can appear in several rows. The comparisons are
  unadjusted and descriptive, they were computed after the analysis
  plans were frozen, and they do not establish the order of the
  course outcome and the change of major.}
\label{tab:majorchangedfw}
\resizebox{\textwidth}{!}{%
\begin{tabular}{@{}lrrrrccc@{}}
\toprule
 & \multicolumn{6}{c}{Observed enrolled at least one year after entry} & All entrants \\
\cmidrule(lr){2-7}\cmidrule(l){8-8}
 & & & \multicolumn{2}{c}{Changed major (\%)} & Difference & & Difference \\
\cmidrule(lr){4-5}
Subject & Entrants & With DFW & with DFW & without & (points) & Odds ratio & (points) \\
\midrule
Mathematics & 6,390 & 2,196 & 65.2 & 54.6 & $+10.6$ $(8.1,\,13.1)$ & 1.56 $(1.40,\,1.73)$ & $+2.8$ $(0.7,\,5.0)$ \\
Political science & 4,301 & 1,133 & 60.1 & 57.2 & $+2.9$ $(-0.5,\,6.2)$ & 1.13 $(0.98,\,1.29)$ & $-2.0$ $(-4.7,\,0.8)$ \\
Writing composition & 4,937 & 976 & 57.0 & 58.4 & $-1.4$ $(-4.9,\,2.0)$ & 0.94 $(0.82,\,1.09)$ & $-10.2$ $(-12.9,\,-7.6)$ \\
History & 4,020 & 634 & 58.8 & 59.2 & $-0.4$ $(-4.6,\,3.8)$ & 0.98 $(0.83,\,1.17)$ & $-9.6$ $(-12.8,\,-6.4)$ \\
Chemistry & 2,982 & 605 & 69.3 & 56.4 & $+12.8$ $(8.7,\,17.0)$ & 1.74 $(1.44,\,2.11)$ & $+0.3$ $(-3.4,\,4.1)$ \\
Biology & 2,335 & 543 & 63.5 & 55.9 & $+7.6$ $(3.0,\,12.3)$ & 1.37 $(1.13,\,1.67)$ & $+1.7$ $(-2.2,\,5.7)$ \\
First-year seminar & 6,846 & 531 & 52.4 & 57.2 & $-4.8$ $(-9.2,\,-0.4)$ & 0.82 $(0.69,\,0.98)$ & $-16.0$ $(-19.0,\,-13.1)$ \\
Psychology & 2,174 & 395 & 60.3 & 55.5 & $+4.8$ $(-0.6,\,10.1)$ & 1.22 $(0.97,\,1.52)$ & $-3.9$ $(-8.1,\,0.3)$ \\
Sociology & 1,803 & 219 & 57.5 & 57.9 & $-0.4$ $(-7.3,\,6.6)$ & 0.99 $(0.74,\,1.31)$ & $-7.9$ $(-13.0,\,-2.9)$ \\
Computer science & 534 & 130 & 63.8 & 46.5 & $+17.3$ $(7.7,\,26.9)$ & 2.03 $(1.35,\,3.05)$ & $+7.3$ $(-1.5,\,16.0)$ \\
Physics & 550 & 127 & 55.1 & 44.4 & $+10.7$ $(0.8,\,20.5)$ & 1.54 $(1.03,\,2.29)$ & $+12.9$ $(4.5,\,21.3)$ \\
Geology & 901 & 124 & 58.1 & 55.2 & $+2.9$ $(-6.5,\,12.2)$ & 1.12 $(0.77,\,1.65)$ & $-11.0$ $(-18.0,\,-4.1)$ \\
Statistics & 783 & 97 & 60.8 & 51.0 & $+9.8$ $(-0.6,\,20.2)$ & 1.49 $(0.97,\,2.30)$ & $+2.3$ $(-6.4,\,11.1)$ \\
Spanish & 606 & 79 & 64.6 & 63.6 & $+1.0$ $(-10.3,\,12.3)$ & 1.04 $(0.64,\,1.71)$ & $-3.2$ $(-12.4,\,5.9)$ \\
Information systems & 1,162 & 78 & 75.6 & 64.9 & $+10.7$ $(0.8,\,20.6)$ & 1.68 $(0.98,\,2.85)$ & $-11.0$ $(-19.5,\,-2.4)$ \\
Art history & 428 & 60 & 46.7 & 55.2 & $-8.5$ $(-22.1,\,5.1)$ & 0.71 $(0.41,\,1.23)$ & $-16.1$ $(-24.9,\,-7.4)$ \\
Art & 644 & 52 & 59.6 & 57.6 & $+2.0$ $(-11.9,\,15.9)$ & 1.09 $(0.61,\,1.94)$ & $-15.5$ $(-25.5,\,-5.5)$ \\
\bottomrule
\end{tabular}}
\end{table}

\subsection{Section size: full results and sensitivity
  checks}\label{app:results:classsize}

\begin{figure}[tbp]
  \centering
  \includegraphics[width=0.95\textwidth]{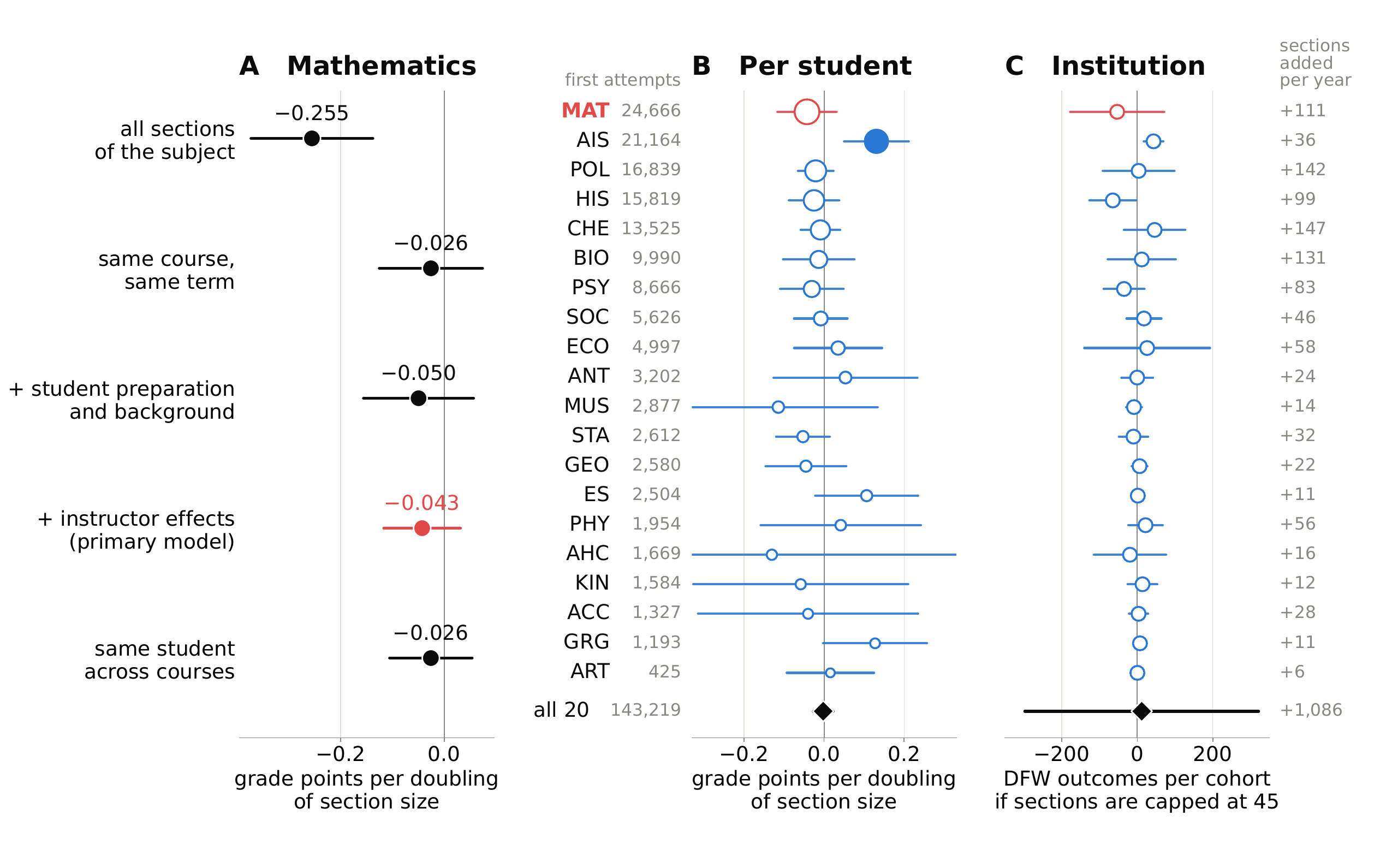}
  \caption{Section size and first-attempt outcomes among
    first-time fall entrants of Fall 2014 to Fall 2018 in
    face-to-face lecture sections of more than 10 students.
    Intervals are 95\%, from a wild cluster bootstrap over
    instructors (the same-student row uses instructor-clustered
    standard errors). (A) Mathematics
    (24,666 first attempts, 18,878
    students, 1,040 sections,
    73 instructors): the difference in
    grade points that goes with a doubling of section size as the
    comparison tightens. Across all sections of the subject the
    coefficient is $-0.255$; with course-by-term effects, the
    student covariates, and instructor effects (the prespecified
    primary model, red), it is $-0.043$
    ($-0.117$ to $+0.031$; $p=0.258$), and the
    matching difference in the DFW rate is $+1.0$
    percentage points ($-1.3$ to
    $+3.3$). The last row compares each student
    with the same student in other courses. (B) The primary
    coefficient in each of 20 subjects, ordered by the
    first attempts each serves (printed beside the code, from
    24,666 in MAT to 425 in ART), with
    marker area proportional to that count; filled markers pass
    false-discovery control at $q=0.05$
    (1 of 20). The diamond is the random-effects
    pooled estimate, $-0.002$ ($-0.026$ to $+0.022$), with an interval that includes the covariance between subject estimates. The axis is set by the 90th percentile of interval reach; 2 wider intervals run past it.
    (C) The same subjects on the scale of the institution: the
    projected change in DFW outcomes per entry cohort if every
    lecture section above 45 students were split (the
    scenario of Table~\ref{tab:classsizescenario}), which
    multiplies each subject's DFW coefficient by the first
    attempts affected and by the change in their section sizes;
    the lecture sections the split would add per year are printed
    at right.
    Mathematics: $-53$
    ($-178$ to
    $+71$); all 20
    subjects (diamond): $+12$
    ($-297$ to $+321$). No subject passes
    false-discovery control on the DFW coefficient. Every
    estimate is an association conditional on entry. Section
    assignment is administrative, and the projection of panel C
    applies only if the association is the effect of size. Subject
    codes: MAT, mathematics; AIS, the first-year seminar; POL, political science; HIS, history; CHE, chemistry; BIO, biology; PSY, psychology; SOC, sociology; ECO, economics; ANT, anthropology; MUS, music; STA, statistics; GEO, geology; ES, environmental science; PHY, physics; AHC, art history and criticism; KIN, kinesiology; ACC, accounting; GRG, geography; ART, art.}
  \label{figClassSizeLadder}
\end{figure}

This subsection supports Section~\ref{sec:baseline:classsize}, which
states the results for mathematics. It reports the comparison across
models and subjects (Figure~\ref{figClassSizeLadder}), the scenario
(Table~\ref{tab:classsizescenario}), and the sensitivity checks.

\paragraph{From all sections to the primary model.}
Panel~A of Figure~\ref{figClassSizeLadder} follows the mathematics
coefficient as the comparison tightens. It falls from
$\SizeMatAllSectionsGrade$ grade points per doubling of section size
across all sections to $\SizeMatCourseTermGrade$ when course-by-term
effects enter, and it is $\SizeMatPrimaryGrade$ in the primary model.
The differences between successive models are descriptive
and are not exact shares attributable to courses, students, or
instructors. Comparing each
student with the same student in other courses gives $\SizeMatSameStudentGrade$. Within
a mathematics course and term, the entry mathematics score differs by
$\SizeMatBalanceMathScore$ standard deviations per doubling of section size. That
balance on one measured score is compatible with assignment by room
availability; it does not establish that assignment was as good as
random, and differences by schedule, format, major, or repeat status
remain possible.

\paragraph{Subjects.}
The pooled interval of the \SizeSubjectsModeled{} subjects
(Section~\ref{sec:baseline:classsize}) includes the covariance
between subject estimates, which is close to zero here. \SizeSubjectsPassingCapitalWord{} subject passes
false-discovery control, with the
opposite sign: in the first-year seminar (AIS), larger sections go
with higher grades ($\SizeAisPrimaryGrade$). Its section sizes have a median of
\SizeAisSizeMedian{} students and a 90th percentile of \SizeAisSizeNinetieth{}, which suggests that size
there marks a format.

A coefficient is an association per student, and the subjects differ
widely in the entrants they serve, from \SizeMatGradeFirstAttempts{} first attempts in
mathematics to \SizeFewestGradeFirstAttempts{} in ART. Panel~B of
Figure~\ref{figClassSizeLadder} therefore orders the subjects by
first attempts and sizes each marker by that count, and panel~C
places each subject on the scale of the institution: the projected
change in DFW outcomes per entry cohort under the scenario of
Table~\ref{tab:classsizescenario}, which multiplies the DFW
coefficient of the subject by the first attempts affected and by the
change in their section sizes. On that scale the largest projected
reductions are in history ($\SizeHisDfwPerCohort$ DFW outcomes per cohort, $\SizeHisDfwPerCohortLow$ to
$\SizeHisDfwPerCohortHigh$) and mathematics ($\SizeMatDfwPerCohort$, $\SizeMatDfwPerCohortLow$ to $\SizeMatDfwPerCohortHigh$), and the largest
projected increases are in chemistry ($\SizeCheDfwPerCohort$, $\SizeCheDfwPerCohortLow$ to $\SizeCheDfwPerCohortHigh$) and
the first-year seminar ($\SizeAisDfwPerCohort$, $\SizeAisDfwPerCohortLow$ to $\SizeAisDfwPerCohortHigh$). A subject with a
large coefficient and few entrants weighs
little: music has the second-largest negative coefficient per student
($\SizeMusPrimaryGrade$) and a projected change of $\SizeMusDfwPerCohort$ DFW outcomes per cohort
($\SizeMusDfwPerCohortLow$ to $\SizeMusDfwPerCohortHigh$). No subject passes false-discovery control on the
DFW coefficient. Weighting every first attempt equally, as the
same-student model does, gives $\SizeSameStudentGrade$ grade points per doubling
($\SizeSameStudentGradeLow$ to $\SizeSameStudentGradeHigh$).

\paragraph{Scenario.}
Table~\ref{tab:classsizescenario} reports the scenario in which every
lecture section above \SizeCap{} students is split
into equal sections of \SizeCap{} or fewer. The calculation applies the
fitted size association to the proposed reductions while holding
staffing and other conditions constant
(Section~\ref{sec:data:link} states the conditions under which it
would describe an intervention). In mathematics, the
affected first attempts sit in sections with a median of \SizeAffectedMedianSize{}
students, and the split would place them in sections with a median
of \SizeAffectedMedianSizeAfterSplit{}. Of these first attempts, \SizeAffectedShareWithSmallSection{}\% occurred in a course and term
that also had a section of \SizeCap{} or fewer students. This establishes
that smaller sections existed in those offerings; it does not
establish that students and instructors were comparable across
section sizes. The affected first attempts number \SizeMatAffectedPerCohort{} per entry cohort
and are two thirds of the first-year first attempts in mathematics.
Over all of them the projection is \SizeMatDfwPerCohortFewer{} fewer DFW outcomes per cohort, with an interval from \SizeMatDfwPerCohortLowFewer{} fewer
to \SizeMatDfwPerCohortHighMore{} more. Across the \SizeSubjectsModeled{} subjects the same threshold would add \SizeAllSectionsAddedPerYear{}
lecture sections to the \SizeAllSectionsPerYear{} taught per year, with a projected
change of $\SizeAllDfwPerCohort$ DFW outcomes per cohort ($\SizeAllDfwPerCohortLow$ to $\SizeAllDfwPerCohortHigh$).

The intervals of the scenario come from the wild cluster bootstrap of
each contrast. For the third of first attempts with the lowest entry
scores, the projection of \SizeLowThirdDfwFewer{} fewer DFW outcomes per 100 affected
first attempts has an interval of $\SizeLowThirdDfwLow$ to $\SizeLowThirdDfwHigh$; inverting the
bootstrap test gives $\SizeLowThirdDfwInvertedLow$ to $\SizeLowThirdDfwInvertedHigh$, and the clustered-$t$ interval
that the plan specified was narrower ($\SizeLowThirdDfwClusterLow$ to $\SizeLowThirdDfwClusterHigh$).

\begin{table}[tbp]
  \centering
  \caption{Scenario: every face-to-face lecture section above
    45 students is split into equal sections of
    45 or fewer. Projected changes apply the primary
    within-course coefficient of each subject to the first attempts
    of first-time fall entrants in the affected sections (annual
    means over the Fall 2014 to Fall
    2018 entry cohorts), with 95\% intervals from
    the wild cluster bootstrap of the coefficient. Sections per
    year count every lower-division lecture section of the
    subject.
    $^\dagger$Rows by third of the entry mathematics score use
    the model with the size-by-score interaction; each is a
    linear contrast of the size coefficient and the interaction
    with its own wild cluster bootstrap interval, computed after
    the plan was frozen in place of the clustered-$t$ intervals
    the plan specified. The projection describes an intervention
    only if section size causes the estimated differences, the
    association is linear in the logarithm of size over the
    split, and added sections are taught as existing ones are;
    the observational models do not establish these conditions.
    The interval of the total includes the covariance between
    subject estimates. The threshold of
    45 students is a scenario value: the record holds
    no enrollment-cap field.}
  \label{tab:classsizescenario}
  \footnotesize
  \setlength{\tabcolsep}{2.5pt}
  \begin{tabular}{lrrrrrr}
    \toprule
    & First attempts & Grade & DFW rate, & DFW outcomes & \multicolumn{2}{c}{Sections per year} \\
    & per cohort & points & pct.\ points & per cohort & now & added \\
    \midrule
    Mathematics & 3,708 & $+0.06$ {\scriptsize($-0.05$ to $+0.17$)} & $-1.4$ {\scriptsize($-4.8$ to $+1.9$)} & $-53$ {\scriptsize($-178$ to $+71$)} & 246 & 111 \\
    \quad lowest third$^\dagger$ & 1,181 & $+0.12$ {\scriptsize($-0.01$ to $+0.25$)} & $-4.6$ {\scriptsize($-8.9$ to $-0.4$)} & $-55$ {\scriptsize($-105$ to $-5$)} & & \\
    \quad middle third$^\dagger$ & 1,409 & $+0.06$ {\scriptsize($-0.05$ to $+0.17$)} & $-1.6$ {\scriptsize($-5.1$ to $+2.0$)} & $-22$ {\scriptsize($-72$ to $+28$)} & & \\
    \quad highest third$^\dagger$ & 1,111 & $-0.01$ {\scriptsize($-0.11$ to $+0.10$)} & $+2.0$ {\scriptsize($-1.3$ to $+5.2$)} & $+22$ {\scriptsize($-14$ to $+58$)} & & \\
    \midrule
    All 20 subjects & 23,622 & & & $+12$ {\scriptsize($-297$ to $+321$)} & 1,086 & 1,086 \\
    \bottomrule
  \end{tabular}
\end{table}

\paragraph{Sensitivity checks.}
Two sensitivity checks bear on the mathematics result. The first
concerns grading standards. Of the \SizeNextOriginFirstAttempts{} mathematics first attempts
in the entry fall, \SizeNextContinuers{} are followed by another mathematics course
within a year, and \SizeNextBothGrades{} have graded records in both courses. Among
these continuers, the next-course grade given the first grade has a
coefficient of $\SizeNextGrade$ grade points per doubling of the first
section's size, with an interval that includes zero ($\SizeNextGradeLow$ to
$\SizeNextGradeHigh$). For \SizeNextDifferentInstructor{} of them the next course has a different
instructor, and on that subset the coefficient is $\SizeNextGradeDifferentInstructor$ ($\SizeNextGradeDifferentInstructorLow$
to $\SizeNextGradeDifferentInstructorHigh$). This check uses a selected sample, the entrants who
continue, and it conditions on the first grade. It gave no clear
evidence of a relationship, and it does not rule out differences in
grading standards or in learning between small and large sections.
The second check concerns shape: the banded coefficients decline in
order across size bands
($\SizeBandSmallGrade$ for \SizeBandSmallLow{} to \SizeBandSmallHigh{} students, $\SizeBandMiddleGrade$ for \SizeBandMiddleLow{} to \SizeBandMiddleHigh{}, $\SizeBandLargeGrade$ for
\SizeBandLargeLow{} to \SizeBandLargeHigh{}, and $\SizeBandLargestGrade$ above \SizeBandLargeHigh{}, against sections of \SizeBandReferenceLow{} to \SizeBandReferenceHigh{}),
with none distinguishable from zero. One further quantity is
descriptive, because it was seen before the plan was frozen: entrants
in larger first-term mathematics sections are less likely to take
another mathematics course within a year ($\SizeContinuationPerDoubling$ percentage points
per doubling, $\SizeContinuationPerDoublingLow$ to $\SizeContinuationPerDoublingHigh$).

\subsection{Reference-cohort progression}\label{sec:baseline:cohort}

Table~\ref{tab:cohortbaseline} reports the progression quantities of
the reference cohort. Of the \InstCohortN{} first-time freshmen whose first
term fell between Fall 2014 and Fall 2015, \InstSeniorPct{}\% reached
senior-level coursework (an enrollment record in a course numbered
above 4000)
within the observation window (the entry term through the same term
\InstHorizonYearsWord{} years later, Section~\ref{sec:data:cohort}), and \InstMajorChangePct{}\% changed
major within the same window. Exposure to the gateway subjects within
the window is nearly universal for MAT and AIS (\InstMatTakersPct{}\% and \InstAisTakersPct{}\%) and
high for WRC (\InstWrcTakersPct{}\%). The cohort-restricted association between each
gateway subject and progression appears in
Table~\ref{tab:oddsnaivecohort} (Appendix~\ref{sec:results:design}):
taking MAT and WRC associates with higher odds of having no
senior-level enrollment in the window (odds ratios \InstOddsMat{} and \InstOddsWrc{}),
while AIS reverses to
strongly favorable (\InstOddsAis{}) once the design is restricted to the entry
cohort. The outcome is an enrollment record; a student without one
may have transferred, as CAP entrants plan to do, so the outcome
measures progression at this institution and gives no measure of
attrition from higher education. One caveat applies to all three: taking a
subject within the horizon requires persisting long enough to take
it, so the exposed groups tilt toward survivors, while preparation,
credited requirements, major, and timing also differ between the
groups. The direction of the net bias is unknown. Exposure here is
any course under the subject code at any point in the window, and the
cohort includes CAP entrants who transfer by design
(Section~\ref{sec:data:cohort}); the mathematics odds ratio is \InstNoCapOddsMat{}
among students without the CAP flag. The \InstOddsMat{} odds ratio corresponds
to non-progression of \InstMatNoSeniorPct{}\% among students with a mathematics course
against \InstNoMatNoSeniorPct{}\% among those without one. These are associations within
a selected entry cohort. The linked models of Appendix~\ref{sec:link}
measure first-year performance against preparation and context
directly.

\begin{table}[tbp]
  \centering
  \caption{Baseline quantities of the reference cohort: first-time freshmen entering Fall 2014 through Fall 2015, outcomes measured over the declared observation window (the entry term through the same term four years later, inclusive; Section~\ref{sec:data:cohort}), so that every measurement closed before Fall 2020. Generated from the institutional database by \texttt{cohort\_tables.py}; nothing is transcribed by hand.}
  \label{tab:cohortbaseline}
  \begin{tabular}{@{}lrr@{}}
    \toprule
    Quantity & $N$ & Share of cohort \\
    \midrule
    First-time freshmen entering Fall 2014--Fall 2015 & 10,027 & 100.0\% \\
    Reached a course numbered above 4000 in window & 5,062 & 50.5\% \\
    Changed major within the window & 4,311 & 43.0\% \\
    Took MAT within the window & 9,126 & 91.0\% \\
    Took WRC within the window & 7,058 & 70.4\% \\
    Took AIS within the window & 9,283 & 92.6\% \\
    \bottomrule
  \end{tabular}
\end{table}

\subsection{Reference quantities of the
  baseline}\label{sec:declaration}

The principal quantities of the baseline are collected below. Each is
generated by a reproducible pipeline from hash-verified sources, each
names its population and vintage, and together they constitute the
pre-2020 reference state against which post-disruption comparisons
can be measured.

\begin{enumerate}
\item \textbf{Cohort progression.} Of the $N=\InstCohortN$ first-time
  freshmen whose first term fell between Fall 2014 and Fall 2015,
  \InstSeniorPct{}\% reached senior-level
  coursework within the observation window (the entry term through
  the same term \InstHorizonYearsWord{} years later, Section~\ref{sec:data:cohort}),
  \InstMajorChangePct{}\% changed major, and exposure to the gateway subjects was
  \InstMatTakersPct{}\% (MAT), \InstWrcTakersPct{}\% (WRC), and \InstAisTakersPct{}\% (AIS)
  (Table~\ref{tab:cohortbaseline}).
\item \textbf{Gateway associations.} Cohort-restricted odds ratios of
  having no senior-level enrollment in the window: MAT \InstOddsMat{}, WRC
  \InstOddsWrc{}, AIS \InstOddsAis{},
  for exposure at any point in the window, with the caveats of
  Appendix~\ref{sec:baseline:cohort}
  (Table~\ref{tab:oddsnaivecohort}).
\item \textbf{Relative-grading structure.} The matrix of grade
  deviations by subject and initial major among the cohort's
  major-changers. Below movers'
  norms: chemistry per mover ($\InstHeatChePerMover$ GPA points), then mathematics
  and political science ($\InstHeatMatPolPerMover$ each), which are the largest in
  total across movers ($\InstHeatMatTotal$ and $\InstHeatPolTotal$ GPA points). Above
  movers' norms: information systems per mover ($\InstHeatIsPerMover$) and the
  first-year seminar in total ($\InstHeatAisTotal$)
  (Figure~\ref{figHeatmap}).
\item \textbf{National position.} College Scorecard coordinates at
  the 2019--20 vintage: \ScoreBaselineUndergraduates{} undergraduates, completion within
  150\% of normal time \ScoreBaselineCompletion{}, Pell share \ScoreBaselinePellShare{}, residual $\ScoreBaselineResidual$
  against the Pell-conditional national expectation; and at the
  current 2024--25 vintage: \ScoreCurrentUndergraduates{}, \ScoreCurrentCompletion{}, \ScoreCurrentPellShare{}, residual $\ScoreCurrentResidual$
  (Figure~\ref{figScorecard}).
\item \textbf{District and sending context.} Course enrollment per
  estimated student in the course's own grades across \GradCourseFamiliesWord{} course
  families: a near-zero association with district child poverty for
  Algebra~II, Geometry, and Biology; a decline of \GradAdvancedRelDeclineMin{}\% to
  \GradAdvancedRelDeclineMax{}\% of the national mean per ten percentage points of child
  poverty for Calculus and AP mathematics and science; and a point
  estimate of \GradEarlyAlgebraRelDecline{}\% for Algebra~I in grades 7--8, with an interval
  that includes zero in the primary cycle;
  the Calculus slope ($\CalcSlope$ per unit of the poverty rate,
  state-clustered SE \CalcSlopeSE{}) and its correlation with district income
  (\CalcIncomeCorr{}); the same contrast among the \GradBexarDistrictsWord{} districts of Bexar
  County, where Calculus is taken at \GradBexarGroupOneCalculusRatio{} and \GradBexarGroupTwoCalculusRatio{} times the national
  rate in the \GradBexarLowerGroupsWord{} lowest-poverty groups of districts and at \GradBexarUpperGroupsCalculusRatioMin{} to
  \GradBexarUpperGroupsCalculusRatioMax{} times in the other \GradBexarUpperGroupsWord{}, with a county gradient of $\GradBexarCalculusRel\%$
  ($\GradBexarCalculusRelLow$ to $\GradBexarCalculusRelHigh$); the
  sending-school provision profile with its opposite-signed
  credit-arrival shares, and the sending-district poverty
  distribution (student-weighted median \XwalkPovertyMedianStudents{}\% against a
  child-weighted Texas median of \XwalkPovertyMedianChildren{}\%)
  (Figures~\ref{figCalculusPanels}, \ref{figCourseGradients},
  \ref{figProvisionBands}, and~\ref{figSendingPoverty}).
\item \textbf{Preparation and first-year performance.} Among \LinkSample{}
  linked entrants of Fall 2014 through Fall 2019, a population chosen
  after the plan's first run on \LinkPlanCohortsWord{} cohorts: \LinkVarBetweenPct{}\% of the variation in
  the entry mathematics score lies between sending schools and
  districts; the score differs by $\LinkIndexPointsSigned$ points per standard
  deviation of
  the school's advanced-mathematics index and by $\LinkPovertyPointsSigned$ per standard
  deviation of district child poverty; per standard deviation of the
  mathematics score, first-year grades are higher by \LinkPooledMathOther{} points in
  the non-quantitative subjects (\LinkPooledMathOtherLow{} to \LinkPooledMathOtherHigh{}) and by \LinkPooledMathQuant{} in the
  quantitative subjects
  (\LinkWeightedMathOther{} and \LinkWeightedMathQuant{} when subjects are weighted by first attempts),
  net of the verbal score; and with sending schools held out, district
  poverty and the school block change the proportional reduction in
  squared error for relative grades by a
  median $\LinkPovertyAddedSigned$ and $\LinkSchoolAdded$ once the other blocks are included
  (Table~\ref{tab:subjectpooled},
  Figures~\ref{figSubjectPredictors}
  and~\ref{figPreparationVariance}).
\item \textbf{Section size.} Among first attempts of the Fall 2014
  through Fall 2018 entrants in face-to-face lecture sections of
  mathematics, larger sections are associated with more DFW outcomes
  among entrants with lower entry mathematics scores, which
  identifies a hypothesis for testing smaller sections: the
  interaction of section size with the entry
  mathematics score is $\SizeMatInteractionGradeSigned$ grade points per standard deviation
  (\SizeMatInteractionGradeLow{} to \SizeMatInteractionGradeHigh{}), and for the third with the lowest entry scores
  the \SizeCap{}-student scenario projects \SizeLowThirdDfwFewer{} fewer DFW outcomes per 100
  affected first attempts ($\SizeLowThirdDfwLow$ to $\SizeLowThirdDfwHigh$) and \SizeLowThirdGrade{} more grade
  points ($\SizeLowThirdGradeLow$ to $\SizeLowThirdGradeHigh$). Averaged over all entrants the
  association is $\SizeMatPrimaryGrade$
  grade points per doubling of section size with course-by-term and
  instructor effects ($\SizeMatPrimaryGradeLow$ to $\SizeMatPrimaryGradeHigh$), against $\SizeMatAllSectionsGrade$ across
  all
  sections of the subject, and $\SizePooledGrade$ pooled over \SizeSubjectsModeled{} subjects
  (Figure~\ref{figClassSizeLadder},
  Table~\ref{tab:classsizescenario}).
\end{enumerate}

The institutional course outcomes behind these quantities were
recorded before the Spring 2020 disruption. Three items use later or
undated sources and are labeled accordingly: the district analysis
includes a 2021--22 robustness cycle, the national position includes
the 2024--25 Scorecard data year, and the extract records no date for
the aid variables, the entry tests, or the extraction of the
per-student credit table. Table~\ref{tab:populations} lists the
window and vintage of each analysis. Every
cohort quantity uses the observation window of
Section~\ref{sec:data:cohort}, inclusive upper boundary included; a
comparison study must reproduce that boundary exactly. The generation
paths are recorded with the figure and table snippets themselves; the
source files sit behind SHA-256 hashes in the data manifest and the
analysis database's provenance layer (Section~\ref{sec:data:oir},
Appendix~\ref{app:data:repro}).

\section{Data Appendix: Materials, Linkage Validation, and Source
  Data Quality}\label{app:data}

This appendix documents the materials behind the paper's
measurements: the datasets and their vintages; the analysis database
and its provenance layers; the construction and validation of the
linkage from entrant to sending high school to district poverty and
neighborhood context; the inferential conventions; the selection
audits; and the disclosure-control regime under which student records
were processed. It also documents, with standing remedies, the
source-data flaws that pose risk for reproducible analysis. Several
of these flaws produce plausible but incorrect values without raising
any error or warning, and several live in public datasets (CRDC,
SAIPE) where they will affect any research team that joins them.
Section~\ref{app:data:defects} states each flaw, its consequence if
unhandled, and the rule that neutralizes it.
Table~\ref{tab:populations} lists the population, unit, and
observation window of each analysis in the paper.

\begin{table}[tbp]
\centering\footnotesize
\setlength{\tabcolsep}{4pt}
\caption{Populations and observation windows, by analysis. The
  institutional course outcomes of every row were recorded before the
  Spring 2020 disruption; the Scorecard comparison and one robustness
  cycle of the district analysis use later vintages, labeled as
  such. The extract holds no field dating the aid application, the
  adjusted gross income, or the entry tests
  (Section~\ref{sec:data:link}).}
\label{tab:populations}
\begin{tabular}{@{}>{\raggedright\arraybackslash}p{0.21\textwidth}>{\raggedright\arraybackslash}p{0.36\textwidth}>{\raggedright\arraybackslash}p{0.35\textwidth}@{}}
\toprule
Analysis (section) & Population and unit & Window or vintage \\
\midrule
District course enrollment (\ref{sec:context:calculus},
\ref{sec:context:courses}) &
\CalcDistricts{} school districts, and the \CalcBexarDistricts{} districts of Bexar County among
them; district totals of CRDC course enrollment and SAIPE child
poverty &
CRDC 2017--18 with SAIPE 2017; robustness on CRDC 2015--16 and
2021--22 \\[2pt]
Sending-school credit profile and sending-district poverty
(\ref{sec:context:provision}, \ref{sec:context:poverty}) &
\XwalkProvisionStudents{} students from \XwalkProvisionSchools{} sending schools, and \XwalkPovertyStudents{}
students from \XwalkPovertyDistricts{} Texas districts; aggregate counts by school band
or by district over the students active between Fall 2014 and Fall
2019 &
CRDC 2017--18 school snapshot; mean of the SAIPE 2010--2018
vintages \\[2pt]
Preparation at entry (\ref{sec:link}) &
\LinkSample{} linked first-time fall entrants; one row per student &
entry cohorts Fall 2014 to Fall 2019; school context from the CRDC
cycle nearest the senior year \\[2pt]
First-year performance (\ref{sec:link}) &
\LinkSampleGradeCohorts{} linked first-time fall entrants; first attempts in
lower-division courses &
entry cohorts Fall 2014 to Fall 2018; entry fall and the following
spring and summer \\[2pt]
Section size (\ref{sec:baseline:classsize}, models) &
\SizeEntrants{} first-time fall entrants, linked or not; \SizeFirstAttempts{} first
attempts in face-to-face lecture sections &
entry cohorts Fall 2014 to Fall 2018; entry fall and the following
spring and summer \\[2pt]
Reference cohort (\ref{sec:baseline:flows},
\ref{sec:baseline:barriers}, \ref{sec:results:design},
\ref{app:results:majorchange}, \ref{sec:baseline:cohort}) &
\InstCohortN{} first-time freshmen entering Fall 2014 to Fall 2015; one
row per student &
entry term through the same term four years later, inclusive \\[2pt]
Full-window descriptives (\ref{sec:baseline:flows}, curves of
\ref{sec:baseline:classsize}) &
all students enrolled, entrants and transfers alike; students or
sections &
Fall 2014 to Fall 2019 \\[2pt]
National position (\ref{sec:results:position}) &
predominantly-bachelor's Title IV institutions &
College Scorecard data years 2019--20 and 2024--25 \\
\bottomrule
\end{tabular}
\end{table}

\subsection{Sources and provenance}\label{app:data:sources}

Table~\ref{tab:sources} inventories every dataset in the analysis
database. The database (\textsc{gateway}, on PostgreSQL~16.2) is
layered by provenance: a \texttt{raw} layer holds source files
verbatim with their file name and SHA-256 hash and is never edited; a
staging layer types the data and quarantines sentinel codes; a
conformed core holds dimensions and facts; and analysis-ready marts
back each figure. Every raw load is registered in a provenance table
(dataset, vintage, path, SHA-256, byte count, source URL, load
timestamp, row count), so any published number can be traced to a
file and a hash. The archives acquired directly for this paper
additionally hold full manifest rows (identifier, agency, release,
acquisition URL and date, license, path, hash, size, schema, parser
version, lineage, geography, time coverage, unit, privacy policy,
revision policy). Bulk file downloads were preferred to API queries
throughout because files are hashable and reproducible offline.

\begin{table}[tbp]
\centering\footnotesize
\setlength{\tabcolsep}{4pt}
\caption{Data inventory. Rows are as loaded in the analysis database;
  student microdata (marked $\dagger$) is queried in place and never
  ingested (Section~\ref{app:data:governance}).}
\label{tab:sources}
\begin{tabular}{@{}>{\raggedright\arraybackslash}p{0.36\textwidth}>{\raggedright\arraybackslash}p{0.16\textwidth}>{\raggedright\arraybackslash}p{0.22\textwidth}>{\raggedright\arraybackslash}p{0.16\textwidth}@{}}
\toprule
Source & Unit & Window / vintage & Size as held \\
\midrule
SAIPE district poverty \citep{SAIPE2024} & district--year & 1995--2024, \DataSaipeVintages{} vintages & \DataSaipeRows{} rows \\
CRDC school files \citep{CRDC2018} & school--cycle & \DataCrdcCycles{} cycles, 2011--12 to 2021--22 & \DataCrdcSchoolCycleRows{} rows \\
\quad district aggregation & district--cycle & same & \DataCrdcDistrictCycleRows{} rows \\
\quad course-family enrollment (\DataCourseFamilies{} families; \DataCourseFamiliesEarliestCycle{} in 2013--14) & school--cycle $\times$course & 2013--14, 2015--16, 2017--18, 2021--22 & \DataCourseFamilyRows{} rows \\
NCES EDGE geocodes \citep{EDGE2018} & school--cycle & 2017--18, 2021--22 & \DataEdgeSchoolRows{} rows \\
NCES CCD directory \citep{CCD2014} & school & 2010--11, 2013--14 & recovery pass only \\
Opportunity Atlas \citep{chetty2018opportunity} & census tract & 2010 tract definitions & \DataAtlasTracts{} tracts \\
\quad aggregated to ZCTA & ZCTA & 2010 definitions & \DataAtlasZctas{} ZCTAs \\
ZCTA--tract relationship \citep{CensusZCTA2010} & ZCTA$\times$tract & 2010 definitions & \DataZctaTractWeights{} weights \\
NCES EDGE GRF tract--district relationship \citep{EDGEGRF2017} & LEA$\times$tract & GRF 2017 (SY 2015--16 bounds, 2010 tracts) & \DataDistrictTractShares{} shares \\
UT San Antonio records$^\dagger$ & student$\times$term $\times$course & terms Fall 2009--Fall 2019 & \PopExtractRecords{} rows \\
\quad students active in the window$^\dagger$ & student & Fall 2014--Fall 2019 & \PopActiveStudents{} students \\
\quad per-student credit table$^\dagger$ & student & (see text) & \PopActiveTableStudents{} of them \\
\bottomrule
\end{tabular}
\end{table}

Four source-level facts matter for interpretation. First, the SAIPE
district series exists for 1995, 1997, 1999, and 2000--2024; 1996 and
1998 were never produced, and the 1989 and 1993 estimates are
county-only, so no district poverty series can reach further back
than 1995. Second, the \DataCrdcCyclesWord{} CRDC cycles arrive in four different
package shapes and two different column vocabularies; the 2011--12
cycle differs from every later one in column vocabulary, in
missing-value convention, and in per-district reporting rate (roughly
\DataReportingRateExcessPct{}\% higher). All \DataCrdcCyclesWord{} ingested cycles are universe collections:
NCES records the CRDC as a universe collection from 2011--12 onward
(the sampled cycles were 2004, 2006, and 2009--10), so cross-cycle
differences here are coverage and convention differences among
universes. Consequences are developed in
Section~\ref{app:data:defects}. Third, calendar alignment between a
CRDC school year and a SAIPE income year is written down once, in a
dimension table pairing each CRDC cycle with the SAIPE vintage of its
fall year (2017--18 $\rightarrow$ 2017), and every join goes through
that table instead of re-deriving the convention per query. Fourth,
school districts change over thirty years: \DataSaipeDistrictsEver{} district
identifiers appear at least once in SAIPE and only \DataSaipeDistrictsEveryVintage{} (\DataSaipeDistrictsEveryVintagePct{}\%)
appear in every vintage, so a lifespan dimension records each
identifier's first and last year, and rank statistics are computed on
the balanced panel.

The institutional records are held in the university's SQL Server and
are non-normalized by design (one row per student, term, and course
section, with student attributes repeated). The extract holds
\PopExtractStudents{} distinct students, and \PopActiveStudents{} of
them hold a course record between Fall 2014 and Fall 2019; every
institutional count of this paper is of these active students. Key
coverage is high: \PopActiveWithNamePct{}\% of them record a sending
high-school name, \PopActiveTexasNamePct{}\% a Texas high school, and
\PopActiveWithZipPct{}\% a home ZIP code. A companion per-student
table holds pre-college credit in gateway mathematics courses (AP,
dual-enrollment, and College-Level Examination Program (CLEP) credit)
for \PopActiveTableStudents{} of the active students; its extraction
window is not independently attested in the source system, and the
absence of \PopTableShortfall{} active students from it is a
documented unknown, so the two tables are never used interchangeably
here. The linked entrant models
(Section~\ref{sec:data:link}) additionally read, from the course-row
table, the aid-application flag, the adjusted gross income of aid
applicants, the highest mathematics and verbal entry scores, and the
CAP flag; income enters models as a within-cohort rank and is never
tabulated or exported, and student rows are held in memory only. Its AP grade fields record only the
literal \texttt{CR}, while its dual-enrollment fields record letter
grades, an asymmetry recorded here because it constrains any future
credit-grade analysis. We did not reuse the project's earlier
database, because we could not reconstruct how its tables had been
produced. We rebuilt the public-data tables from source files and
recorded the hashes of those files.

\subsection{Linkage construction and validation}\label{app:data:linkage}

Every contextual measurement reaches a student through one of two
branches. The \emph{district branch} resolves the student's free-text
high-school name to an NCES school identifier, whose first seven
digits are the district identifier (LEAID), which joins SAIPE poverty
exactly; nothing in this branch is probabilistic once the school is
resolved. The \emph{neighborhood branch} resolves the student's home
ZIP to Opportunity Atlas context. The two branches fail differently
and were validated separately.

\paragraph{Name resolution.}
The crosswalk was built for the \XwalkNameRowsAll{} distinct
\emph{Texas} high-school name--city--ZIP combinations of the extract;
\XwalkNameRows{} of them are recorded for the \PopActiveTexasName{}
active students with a Texas high-school name. The combinations were
matched against the EDGE 2017--18 directory using the score of
Equation~\eqref{eq:linkscore}. Whole-string similarity fails
structurally here, because the institution records full honorific
names while NCES records short forms (\emph{Sandra Day O'Connor High
Sch} vs.\ \emph{O'CONNOR H S}), which is why the score is
containment-first. Auto-acceptance requires a score of at least \XwalkAutoScore{}.
Within the statewide block it additionally requires a margin of at
least \XwalkAutoMargin{} over the runner-up, and a near-tie between two real
schools goes to human review; within city and ZIP blocks the margin
requirement is waived because the block itself disambiguates, a rule
stated here exactly because it means the accepted set includes
low-margin matches whose error rate has not been measured against an
adjudicated gold standard. Candidates scoring below \XwalkMinCandidateScore{} are
suppressed entirely, on the principle that a wrong district
identifier is worse than a missing one.

\begin{table}[tbp]
\centering\small
\caption{Linkage validation among the students active between Fall
  2014 and Fall 2019: staged coverage of the school crosswalk, and the
  final disposition of the \PopActiveTableStudents{} active students
  of the per-student table. Percentages in the left panel are of the
  \PopActiveTexasName{} active students with a Texas high-school name
  (the crosswalk base); the right panel partitions the students of
  the per-student table, each counted once.}
\label{tab:linkage}
\begin{tabular}{@{}lr@{\qquad}lr@{}}
\toprule
\multicolumn{2}{@{}l}{Crosswalk stage (usable students)} &
\multicolumn{2}{l@{}}{Final disposition (all students)} \\
\midrule
automatic match            & \XwalkStageAutoPct{}\%  & matched                & \XwalkMatched{} \\
+ hand adjudication (\XwalkReviewRows{} rows) & \XwalkStageAdjudicatedPct{}\%  & high school outside Texas & \XwalkOutsideTexas{} \\
+ vintage recovery         & \XwalkStageVintagePct{}\%  & unresolved             & \XwalkUnresolved{} \\
+ CCD directory recovery   & \XwalkCoveragePct{}\%  & private (outside NCES) & \XwalkPrivate{} \\
                           &         & no school recorded     & \XwalkNoSchool{} \\
                           &         & rejected on review     & \XwalkRejected{} \\
\bottomrule
\end{tabular}
\end{table}

Table~\ref{tab:linkage} shows the staged coverage. Three validation
episodes shaped the final design, and each replaced a naive approach
with a stated rule. (i)~Low-threshold matching yields high-scoring
false matches: a recovery pass blocked on state alone at a score
threshold of \XwalkFirstRunScore{} accepted a sixth-grade school, an entire school
district, a ninth-grade campus, and an alternative-education center
as high schools; it was revoked and re-run at \XwalkRecoveryScore{} with an extended
exclusion list, and the pass now retracts anything a previous run
promoted. The standing rule is that name matching without city-level
blocking is unreliable even at nominally high score thresholds.
(ii)~Private schools are structural absences: \XwalkPrivateNames{} names
(\XwalkPrivateNamedStudents{} students, \XwalkPrivateNamedPct{}\% of the crosswalk base; the right panel's
\XwalkPrivate{} counts the same group within the per-student table) are
outside the NCES public universe by construction, and counting them
as match failures had overstated the error rate by that share.
(iii)~A name pattern is the wrong high-school classifier: deciding
``is this a high school'' from the name silently dropped combined
middle/high campuses, which are disproportionately rural and small,
hence non-random. A later recovery pass using CCD grade spans showed
that \XwalkRecoveredInDirectoryPct{}\% of the students it recovered had been reachable in the
original directory all along; the correct classifier is the grade
span (highest grade in 9--12), recorded as the standing rule for any
rebuild. The rebuild itself was left unperformed at \XwalkCoveragePct{}\% coverage,
because it would invalidate the completed hand adjudication; instead
the residual is audited (Section~\ref{app:data:selection}).

The crosswalk was subsequently validated against an external source:
a school-level table in the institutional database whose provenance
was established by exact-match joins against the CRDC ingest of this
project (its Algebra~II column matches CRDC 2017--18 on \XwalkTableCycleSchools{}
schools at exactly \XwalkTableCycleEqualPct{}\%, against \XwalkTableOtherCyclesLow{}--\XwalkTableOtherCyclesHigh{}\% for every other
cycle). It covers \XwalkSchoolsInTablePct{}\% of the \XwalkMatchedSchools{} matched schools of the
active students, confirming that the
identifiers assigned by name matching are real schools in the
national universe. This is an independent check on the crosswalk and
on the CRDC ingest simultaneously; it leaves open whether the correct
school was chosen among similarly named candidates.

\paragraph{Neighborhood resolution.}
The home-ZIP branch was validated before use, because it rests on the
untested assumption that the first recorded ZIP is a childhood
address. We compared the locations of the recorded home ZIP and the
recorded high-school ZIP of the active students; only percentiles and
shares are written. The 90th percentile of the distance between the
two is \XwalkDistanceNinetieth{} miles, and the 95th percentile is
\XwalkDistanceNinetyFifth{} miles. Among students more than
\XwalkFarMiles{} miles from their recorded high school,
\XwalkFarLocalHomePct{}\% have a San Antonio home ZIP. This pattern
is consistent with the recording of a local address after a student
moved to the city; the geography alone does not date the address
field. We therefore do not read the full distance distribution as
commuting distance, and distance was retained only as a linkage gate.
The primary sample is students within \XwalkPrimaryMiles{} miles (\XwalkWithinPrimaryPct{}\%
of students on the ZIP-centroid measure); the point-coordinate
re-check below confirms the threshold at \XwalkPointWithinPrimaryPct{}\% on its subsample, and
a strict sensitivity variant uses \XwalkStrictMiles{} miles. District assignment always
goes through
the resolved school: Bexar County alone has \GradBexarDistrictsWord{}
school districts, so a displacement of a few miles
routinely crosses a district line. An initial version of this
diagnostic also mismeasured the school end, using ZIP centroids: a
zero-mile spike containing \XwalkZipZeroPct{}\% of students fell to
\XwalkPointZeroPct{}\% when schools were
measured as points, without moving any threshold. The spike was an
artifact of ZIP-centroid geometry.

Assigning students \emph{down} to a census tract was tested and
rejected. Weighted by where students actually live, the modal-tract
assignment is wrong roughly four times in five (the median student's
best tract holds \XwalkBestTractSharePct{}\% of their ZIP's population; the average
student's ZIP spans \XwalkTractsPerZip{} tracts). The unweighted ZCTA-level statistic
looks acceptable (median \XwalkBestTractShareAllZips{}) because it describes ZIPs; weighted
by students it fails. The design was therefore inverted: Opportunity
Atlas values are aggregated \emph{up} to the ZCTA as a
population-weighted mean over its tracts, using the Census 2010
relationship file (chosen over the HUD crosswalk because it is a
hashable download, records population weights instead of address
overlap, and is on 2010 tract definitions, exactly the Atlas
vintage). Coverage is \XwalkContextCoveragePct{}\% of students. The resulting variable is
doubly ecological, and the constraint is enforced in the schema
itself: context tables live in schemas named for their level, every
row holds an \texttt{is\_ecological} flag and a tract-vintage column,
and the table comment (queryable at join time) states that no value
describes an individual.

\subsection{Source-data flaws and standing
  remedies}\label{app:data:defects}

The source files hold eight properties material enough to change
results if handled naively. Table~\ref{tab:defects} summarizes each
with its remedy; the five with the largest consequences are described
below together with the checks that reveal them. The manner of
detection is itself a methods result: each flaw surfaced from a
robustness or consistency check whose stated purpose was something
else, and in two cases the flaw was invisible under the specification
originally chosen. Those checks are therefore part of the method.

\begin{table}[tbp]
\centering\footnotesize
\caption{Source-data flaws that pose risk for reproducible analysis.
  ``Silent'' means the flaw produces plausible output with no error or
  warning.}
\label{tab:defects}
\begin{tabular}{@{}>{\raggedright\arraybackslash}p{0.30\textwidth}>{\raggedright\arraybackslash}p{0.30\textwidth}>{\raggedright\arraybackslash}p{0.32\textwidth}@{}}
\toprule
Flaw & Consequence if unhandled & Remedy \\
\midrule
CRDC reserve codes (discrete values such as $-3$, $-5$, $-6$, $-8$,
$-9$, $-11$; cycle-specific) stored inside count columns &
silent: \texttt{SUM()} absorbs $-9$ sentinels into enrollment totals &
sentinel $\rightarrow$ NULL, code kept in a parallel column; suppression
stays reportable \\[2pt]
CRDC 2015--16 ships numerically damaged school keys &
silent: \DataStoredKeySchoolsLost{} of \DataStoredKeySchools{} schools fold onto duplicate keys;
\DataStoredKeyEnrollmentLostPct{}\% of the cycle's enrollment vanishes &
reconstruct the key as LEAID(7)$\,\|\,$SCHID(5) in every cycle; the
stored key is unreliable \\[2pt]
CRDC 2011--12 records explicit zeros in enrollment columns where
later cycles record a skip code &
silent: \DataAlgTwoZeroPctEarliestCycle{}\% of schools hold an explicit zero for Algebra~II
(vs.\ \DataAlgTwoZeroPctLaterMin{}--\DataAlgTwoZeroPctLaterMax{}\% in every later cycle), a difference of coding
convention that a naive offering indicator reads as a change in
schools &
no offering comparison spans 2011--12 until zeros and skip codes are
harmonized; rank statistics are read with that caveat \\[2pt]
\DataMismatchDistrictsCapitalWord{} districts with more Algebra~II enrollees than children 5--17
(ratio up to \DataAlgTwoPerChildMax{}) &
silent: inverts the sign of the unweighted national gradient &
require $\geq \DataMinChildren$ children 5--17 (a pre-treatment filter on the
denominator) \\[2pt]
Course titles are truncated free text with drifting spellings &
silent: one course fragments into up to four; course-level cells
understate & canonicalization on a punctuation-free key before any
grouping \\[2pt]
SAIPE 2024 reordered its columns under machine names &
silent: poverty read as population if parsed positionally &
resolve columns by name; positional parsing is forbidden \\[2pt]
Demographic fields are label strings & silent: numeric coercion NaNs
entire columns, which vanish from output &
explicit label comparison; no numeric coercion \\[2pt]
School ZIP stored one digit short (\XwalkMisreadZipStudents{} students) &
a Houston school reads as New Jersey once the ZIP is padded on the
left & flagged; excluded by the distance gate \\
\bottomrule
\end{tabular}
\end{table}

\paragraph{Damaged school keys (2015--16).}
The 2015--16 CRDC wide file ships the school key in a numerically
damaged form consistent with spreadsheet or scientific-notation
coercion somewhere upstream; the exact pathway is unverified, and the
observed fact is nonuniqueness. Read as stored, \DataStoredKeySchoolsLost{} of \DataStoredKeySchools{}
schools fold onto duplicate keys and the cycle's Algebra~II
enrollment drops from \DataAlgTwoEnrolleesRebuiltKey{} to \DataAlgTwoEnrolleesStoredKey{}, a \DataStoredKeyEnrollmentLostPct{}\% loss that
produces a plausible-looking panel with a mysterious dip. The key is
now reconstructed from its components (LEAID and SCHID) in every
cycle, with a uniqueness assertion at load time.

\paragraph{Explicit zeros and skip codes (2011--12).}
In 2011--12, \DataAlgTwoZeroPctEarliestCycle{}\% of schools record an explicit zero for
Algebra~II enrollment; in every later cycle the figure is
\DataAlgTwoZeroPctLaterMin{}--\DataAlgTwoZeroPctLaterMax{}\%. The difference coincides with a change of coding
convention. The 2011--12 cycle differs from the later ones in column
vocabulary, in marking missing values with blanks where later cycles
use reserve codes, and in a per-district reporting rate about \DataReportingRateExcessPct{}\%
higher. In the later cycles a school that reports no classes in a
course holds the skip code on the enrollment item (verified exactly
for 2017--18, below), which is consistent with the 2011--12 zeros
being schools without the course; that reading has not been checked
against the 2011--12 documentation. The parser used here coerces
blanks and
reserve codes to missing and keeps all-missing sums missing, so the
zeros are explicit values in the archived source file. An indicator
built as ``reports a non-negative Algebra~II count'' is therefore
nearly constant from 2013--14 onward for a reason of coding: schools
without the course leave the universe of reported counts. It says
little about how many schools offer the course. The problem surfaced
in a stability
analysis whose first version reported \DataOfferAgreePctMin{}--\DataOfferAgreePctMax{}\% cross-cycle agreement
on that indicator; Cohen's $\kappa \approx \DataOfferKappa$ alongside it showed the
indicator to be degenerate, and rebuilding on enrollment rates
exposed the difference in zeros. The finding is a harmonization
problem: a comparison of offering across cycles needs equivalent
offering flags and a common treatment of zeros and skip codes, which
this paper does not build. Consequences here: no offering comparison
spans 2011--12, and cross-cycle rank statistics that span it are
read with the caveat that a block of tied zeros covering a quarter of
schools perturbs ranks.

\paragraph{The frame-mismatch ratios.}
Algebra~II enrollment per resident child divides a CRDC numerator by
a SAIPE denominator, and in very small districts the two frames
disagree:
\DataMismatchDistrictsWord{} districts in 2017--18 record more Algebra~II enrollees than
children aged 5--17 (maximum ratio \DataAlgTwoPerChildMax{}; median offending district
holds \DataMismatchMedianChildren{} children), which happens when a school serves a wider area
than its district of record. Those \DataMismatchDistrictsWord{} districts, of \DataAlgTwoDistrictsNoFloor{},
invert the sign of the unweighted national regression: $\DataUnweightedSlopeNoFloor$ with
them included, against $\DataUnweightedSlopeWithoutMismatch$ once they are excluded and
$\DataUnweightedSlopeFloor$ above the standing floor of \DataMinChildren{} children (the
weighted slope is $\DataWeightedSlope$ in all three fits). The flaw had been
invisible because the published specification weighted by district
size, which suppresses \DataMismatchMedianChildren{}-child districts by construction; it
surfaced only when the unweighted robustness fit was finally run. The
remedy filters the denominator (clipping the ratio would condition on
the dependent variable), and at a floor of \DataMinChildren{} children the maximum
ratio becomes physically plausible (\DataAlgTwoPerChildMaxFloor{}) and the weighted and
unweighted slopes agree to within \DataSlopeGapFloor{}, with the weighted slope
invariant across the entire filter ladder ($\DataWeightedSlopeLadderFlattest$ to $\DataWeightedSlopeLadderSteepest$).

\paragraph{The fragmented courses.}
Course titles in the institutional records are free text truncated to
roughly thirty characters, with spelling drifting across the
2012--2019 window (\emph{Algebra with Calculus for Bus} appears under
four spellings; \emph{Algebra-Scientists \& Engineers} under four
more). Grouped on the raw title, one course fragments into pieces,
and a fragment is indistinguishable from a genuinely small course: a
first-version analysis selected its courses ``by volume'' over
fragments, which is how a cell below the disclosure minimum reached a
results table. Canonicalization on a punctuation-free lowercase key
corrected the counts without moving any calculus-track number, and
all downstream analyses reproduce identically.

\paragraph{The sentinel codes.}
CRDC stores discrete reserve codes inside its count columns ($-3$,
$-5$, $-6$, $-8$, $-9$, and $-11$ in the 2017--18 public-use file,
with cycle-specific variants; $-11$ marks data-quality suppression),
and its own published totals treat them as zeros. A natural
cast-and-sum silently adds $-9$ sentinels into enrollment totals,
for roughly \DataAlgTwoSkipCodeSchoolsRounded{} schools per cycle in the Algebra~II field alone. The
staging layer replaces every sentinel with NULL and preserves the
code in a parallel column, which keeps suppression \emph{reportable}:
the decomposition it enabled resolved an apparent coverage crisis
(only $\sim$\DataAlgTwoSchoolsWithCountPct{}\% of schools report Algebra~II) into structural
missingness (elementary and middle schools, code $-9$), and thereby
closed a suspected downward bias in the district sums. One limit on
that reading is recorded with it: the public-use manual applies
disclosure-avoidance adjustments that perturb published counts
without leaving a sentinel, so absence of a code establishes absence
of a coded suppression only, and the counts read as protected
published values rather than exact tallies.

The course-family comparison of Section~\ref{sec:context:courses}
applies one further reading that the parallel column makes possible.
On a course item, code $-9$ marks either a school without high-school
grades or a high school reporting zero classes in the subject: in
2017--18 the \DataAlgTwoSchoolsWithCount{} schools reporting at least one Algebra~II class
are exactly the schools holding a non-negative Algebra~II enrollment.
Within a district that serves high-school grades, a skipped course
item is therefore a structural zero and enters the district sum as
zero, while every other code stays missing. A district total is then
in one of three states: a count, a structural zero, or an incomplete
total in which some school holds a missing code. \DataIncompleteDistrictsCapitalWord{} districts of
the Algebra~II frame hold an incomplete total in some course family
in 2017--18 (\DataIncompleteDistrictsEarlierWord{} in 2015--16, \DataIncompleteDistrictsLaterWord{} in 2021--22); they leave the
common frame, and keeping them changes no conclusion. Two further
groups of districts fall outside this reading and are handled by the
frame. In 2017--18,
\DataNoCoreDistricts{} districts serving high-school grades (\DataNoCoreChildren{} children aged
5--17) report no enrollment in any of the \DataCoreCoursesWord{} core high-school
mathematics and science courses, a reporting gap that the Algebra~II
frame excludes; and \DataNoAlgTwoDistricts{} districts (\DataNoAlgTwoChildren{} children) report no
Algebra~II alongside positive enrollment in other core courses. The
neutral-frame variant admits the second group (\DataNeutralFrameDistricts{} districts
against \CalcDistricts{}) and leaves the contrast among course families
unchanged.

All five flaws described above were silent: each analysis completed
without error and produced plausible magnitudes. The checks that
reveal such flaws are therefore standing practice: running the stated
robustness check even when the primary estimate appears stable,
computing agreement statistics such as $\kappa$ alongside raw
agreement, and testing a ratio's numerator and denominator frames
against each other.

\subsection{Inferential practice}\label{app:data:inference}

Uncertainty is computed at the level the exposure varies. District
poverty varies by district and state policy by state, however many
students the cells contain; ignoring that made the district
gradient's standard errors roughly five times too narrow
(state-clustered \DataSlopeSeStateClustered{} vs.\ naive \DataSlopeSeNaive{} on identical point
estimates). Resampling and clustering follow the same principle: the
course-family comparison resamples states, the level at which
education policy varies, and the linked entrant models cluster on
sending school, the level at which school context varies. Summaries
across subjects use the covariance between subject estimates, because
the same students and schools contribute to several subject models.
The section-size models take instructors as clusters and a wild
cluster bootstrap for every size term and every contrast of size
terms, after dry runs on shuffled grades showed that the clustered
$t$ test rejects too often with few, unequal instructor clusters. Where
families of subjects are tested, the
Benjamini--Hochberg procedure controls the false discovery rate at
the stated level (Appendix~\ref{sec:data:quant}), and every reported
association names its design and window.

\subsection{Selection audits}\label{app:data:selection}

Three selection margins could bias the linked results, and each was
audited directly.

\paragraph{The unresolved residual.}
The \XwalkUnresolved{} students whose Texas high school no method
could resolve differ from the \XwalkMatched{} matched students in the
sample: they hold dual-enrollment credit more often
(\XwalkUnresolvedDualPct{}\% vs.\ \XwalkMatchedDualPct{}\%), AP credit
less often (\XwalkUnresolvedApPct{}\% vs.\ \XwalkMatchedApPct{}\%),
have lower mean mathematics entry scores (\XwalkUnresolvedScore{}
vs.\ \XwalkMatchedScore{}; the highest SAT or concorded ACT
mathematics score, on the 200--800 scale), and receive Pell grants
more often (\XwalkUnresolvedPellPct{}\% vs.\
\XwalkMatchedPellPct{}\%). With students treated as independent,
these contrasts are large against their standard errors
($z=\XwalkDualZ$, $\XwalkApZ$, $\XwalkScoreZ$, and $\XwalkPellZ$, in
that order). Students of one school share its credit offerings and
its community, and the unresolved students come from
\XwalkUnresolvedSchools{} school names. With sending schools as
clusters the same contrasts give $z=\XwalkDualZClustered$,
$\XwalkApZClustered$, $\XwalkScoreZClustered$, and
$\XwalkPellZClustered$. The differences are therefore within the
variation between schools, and the audit leaves open whether linkage
is selective on these markers. The
direction of any resulting bias in district-linked results is
unknown: the district poverty of the excluded students is unobserved,
and a similar marginal outcome (first-course failure
\XwalkUnresolvedFirstCoursePct{}\% vs.\
\XwalkMatchedFirstCoursePct{}\%; $z=\XwalkFirstCourseZ$ with
independent students, $\XwalkFirstCourseZClustered$ with school
clusters) does not determine how context and outcomes are
jointly distributed among them. The linked models of
Appendix~\ref{sec:link} address this margin in two ways, with
inverse-probability-of-linkage weights and with a comparison of
coefficients between the linked sample and all entrants; neither
changes the results.

\paragraph{Students with no Texas school to resolve.}
\XwalkOutsideTexas{} students record a high school outside Texas,
which the crosswalk does not cover, and \XwalkNoSchool{} record no
high-school name. Both groups are mostly transfer students
(\XwalkOutsideTexasTransferPct{}\% and \XwalkNoSchoolTransferPct{}\%,
against \XwalkAllTransferPct{}\% of all students of the table), whose
pre-college credit was recorded by their first institution and is
largely invisible here: AP credit is recorded for
\XwalkOutsideTexasApPct{}\% and \XwalkNoSchoolApPct{}\% of them and
dual-enrollment credit for \XwalkOutsideTexasDualPct{}\% and
\XwalkNoSchoolDualPct{}\%, against \XwalkAllApPct{}\% and
\XwalkAllDualPct{}\% of all students of the table. They are excluded
from all linked analyses and must never be pooled with ``no credit.''

\paragraph{The private-school margin.}
\XwalkPrivate{} students attended schools outside the NCES public universe.
They are markedly advantaged (Pell \XwalkPrivatePellPct{}\%, first-generation \XwalkPrivateFirstGenPct{}\%,
against \XwalkMatchedPellPct{}\% and \XwalkMatchedFirstGenPct{}\% among matched), so their exclusion removes
an advantaged group from every district-linked result; its effect on
any gradient is unmeasured.

\subsection{Governance and disclosure control}\label{app:data:governance}

Student-level records are processed under the institutional review
board protocol and data request identified in the Ethics
approval statement, with a governance rule stronger than the formal
requirement:
individual student records may be queried and processed locally but
never ingested into the analysis database. Student rows exist only
in memory during computation, and in a local cache for the older
figure programs; what
enters the database is aggregate cells, each supported by at least
\InstDisclosureMinCellWord{} students, published through a code path that refuses smaller
cells and records how many were withheld alongside how many were
published. File outputs and legacy tables produced outside that
code path are checked and brought under the same minimum-cell rule as
they are found, and an automated scanner enumerates every
figure-companion export whose count cells fall below \InstDisclosureMinCellWord{}; its
findings are treated as release blockers for any package leaving the
authorized team. The repository that accompanies the paper holds only
the outputs on an explicit list: the files this manuscript reads and
the backing file of each figure and table. The scanner is applied to
every listed file, including the statistics files, before release,
and the working exports of the linkage and distance diagnostics
remain with the authorized team. A figure can also state a count
through its geometry. In the flow diagram of
Figure~\ref{figChordShort} the width of a ribbon is a number of
students, so the diagram is drawn from the flows of at least
\InstChordMinFlowWord{} students only, its sectors are sized by the
drawn flows, and the release scan recovers the count of every ribbon
from the PDF and holds it to the same minimum. One early analysis
table violated the ingestion rule
(a one-row-per-student context table, built before the rule was
articulated) and was dropped in its entirety and rebuilt as local
processing plus aggregate publication; the event is recorded because
the audit trail is part of the method. The local cache itself remains
restricted material: it holds row-level frames and is kept outside
the repository and outside any release package. Ecological variables
state their level in the schema
itself (Section~\ref{app:data:linkage}); income is banded or ranked
before any use and is never tabulated; and the distance diagnostic
reads ZIP codes only and writes percentiles and shares. In the
section-size models the instructor identifier is
replaced by a surrogate inside the query, instructor names are never
read, and no output is indexed by section or by instructor.

\section{Angular Index of Majors}\label{app:chords}

The chord diagram in the body of the paper (Figure~\ref{figChordShort})
labels each sector with the name of its major.
Table~\ref{tab:majorangles} lists the majors in alphabetical order
with the angular position and size of each sector, so that a major
can be located on the ring.

\begin{table}[hbp]
  \centering
  \footnotesize
  \setlength{\tabcolsep}{4pt}
  \caption{Angular index to the chord diagram of Figure~\ref{figChordShort}: every major in alphabetical order, the approximate angular position of its sector center (degrees counter-clockwise from the 3 o'clock direction), and $N$, the number of students it donated plus received in the flows the figure draws (flows of at least 10 students). The majors named General are the college-wide undeclared or general-studies majors of the college in parentheses: Business (COB), Engineering (COE), Sciences (COS), Liberal and Fine Arts (COLFA), Architecture, Construction and Planning (COA), and Education \& Human Development (COEHD). Other collects the 14 majors with the smallest total flow, whose sectors would be too small to see: Music; General (COA); Anthropology; Non Degree Seeking; General (COEHD); Geological Sciences; Public Administration; Modern Languages \& Literatures; Race, Ethnicity, Gender and Sexuality Studies; Undeclared - GATES; COPP (Admin Unit); Counseling; Social Work; and Bicultural/\allowbreak{}Bilingual Studies.}
  \label{tab:majorangles}
  \begin{tabular}{>{\raggedright\arraybackslash}p{5.0cm}rr@{\quad}>{\raggedright\arraybackslash}p{5.0cm}rr}
    \toprule
    \textbf{Major} & \textbf{Angle} & \textbf{N} & \textbf{Major} & \textbf{Angle} & \textbf{N} \\
    \midrule
    Accounting & 230$^\circ$ & 972 & General (COLFA) & 119$^\circ$ & 239 \\
    Architecture & 124$^\circ$ & 273 & General (COS) & 152$^\circ$ & 372 \\
    Art \& Art History & 96$^\circ$ & 109 & History & 105$^\circ$ & 169 \\
    Biology & 19$^\circ$ & 3,648 & Information Systems \& Cyber Security & 210$^\circ$ & 933 \\
    Biomedical Engineering and Chemical Engineering & 135$^\circ$ & 314 & Interdisciplinary Learning \& Teaching & 250$^\circ$ & 1,006 \\
    Chemistry & 185$^\circ$ & 588 & Kinesiology, Health \& Nutrition & 355$^\circ$ & 2,348 \\
    Civil \& Environmental Engineering & 171$^\circ$ & 503 & Management & 337$^\circ$ & 1,834 \\
    Communication & 283$^\circ$ & 1,199 & Management Science \& Statistics & 101$^\circ$ & 160 \\
    Computer Science & 295$^\circ$ & 1,291 & Marketing & 308$^\circ$ & 1,322 \\
    Construction Science & 110$^\circ$ & 176 & Mathematics & 146$^\circ$ & 350 \\
    Criminal Justice & 193$^\circ$ & 673 & Mechanical Engineering & 272$^\circ$ & 1,148 \\
    Economics & 114$^\circ$ & 213 & Philosophy and Classics & 178$^\circ$ & 566 \\
    Electrical \& Computer Engineering & 201$^\circ$ & 689 & Physics \& Astronomy & 92$^\circ$ & 85 \\
    English & 130$^\circ$ & 279 & Political Science \& Geography & 164$^\circ$ & 420 \\
    Environmental Science and Ecology & 141$^\circ$ & 315 & Psychology & 321$^\circ$ & 1,585 \\
    Finance & 261$^\circ$ & 1,014 & Sociology & 240$^\circ$ & 978 \\
    General (COB) & 220$^\circ$ & 946 & University College & 55$^\circ$ & 5,818 \\
    General (COE) & 158$^\circ$ & 397 & Other & 84$^\circ$ & 1,396 \\
    \bottomrule
  \end{tabular}
\end{table}

\end{document}